\documentclass{aa}  
\usepackage{aalongtable,lscape}
\usepackage[super]{nth}
\usepackage{graphicx}
\usepackage{txfonts}
\usepackage{float}

\usepackage{xcolor}

\usepackage{natbib,twoopt}
\usepackage[breaklinks=true]{hyperref} %% to avoid \citeads line fills
\bibpunct{(}{)}{;}{a}{}{,} %% natbib format for A&A and ApJ
\makeatletter
\newcommandtwoopt{\citeads}[3][][]{\href{http://adsabs.harvard.edu/abs/#3}%
{\def\hyper@linkstart##1##2{}%
\let\hyper@linkend\@empty\citealp[#1][#2]{#3}}}
\newcommandtwoopt{\citepads}[3][][]{\href{http://adsabs.harvard.edu/abs/#3}%
{\def\hyper@linkstart##1##2{}%
\let\hyper@linkend\@empty\citep[#1][#2]{#3}}}
\newcommandtwoopt{\citetads}[3][][]{\href{http://adsabs.harvard.edu/abs/#3}%
{\def\hyper@linkstart##1##2{}%
\let\hyper@linkend\@empty\citet[#1][#2]{#3}}}
\newcommandtwoopt{\citeyearads}[3][][]%
{\href{http://adsabs.harvard.edu/abs/#3}
{\def\hyper@linkstart##1##2{}%
\let\hyper@linkend\@empty\citeyear[#1][#2]{#3}}}
\makeatother

\newcommand{\kms} {km\,s$^{-1}$}

\newcommand{\ioni}[2]{{#1\,\sc{#2}}}

\begin{document}

\title{High-resolution spectroscopic study of massive blue and red supergiants in Per~OB1} 
\subtitle{I. Definition of the sample, membership, and kinematics}
\titlerunning{Massive supergiant stars in the Per~OB1 association}
\author{de~Burgos, A.\inst{1,2,3}, Simon-D\'iaz, S.\inst{3,4}, Lennon, D.~J.\inst{3,4}, Dorda, R.\inst{3,4}, Negueruela, I.\inst{5}, Urbaneja, M.~A.\inst{6}, Patrick, L.~R.\inst{3,4,5}, Herrero, A.\inst{3,4}}
\authorrunning{A. de~Burgos et al.}
\institute{
Nordic Optical Telescope, Rambla Jos\'e Ana Fern\'andez P\'erez 7, E-38\,711 Bre\~na Baja, Spain
\and
Isaac Newton Group of Telescopes, E-38700, La Palma, Spain
\and
Universidad de La Laguna, Dpto. Astrof\'isica, E-38206 La Laguna, Tenerife, Spain
\and
Instituto de Astrof\'isica de Canarias, Avenida V\'ia L\'actea, E-38205 La Laguna, Tenerife, Spain
\and
Departamento de F\'{\i}sica Aplicada, Facultad de Ciencias, Universidad de Alicante, Carretera de San Vicente s/n, E03690, San Vicente del Raspeig, Spain
\and
Institut für Astro- und Teilchenphysik, Universität Innsbruck, Technikerstr. 25/8, A-6020 Innsbruck, Austria
}
\date{Received --- / Accepted ---}
% \abstract{}{}{}{}{} 
% 5 {} token are mandatory
\abstract 
% context heading (optional), leave it empty if necessary 
{The Perseus OB1 association, including the $h$~and~$\chi$~Persei double cluster, is an interesting laboratory for the investigation of massive star evolution as it hosts one of the most populous groupings of blue and red supergiants (Sgs) in the Galaxy at a moderate distance and extinction.}
{We discuss whether the massive O-type, and blue and red Sg stars located in the Per~OB1 region are members of the same population, and examine their binary and runaway status.}
{We gathered a total of 405 high-resolution spectra for 88 suitable candidates around 4.5\,deg from the center of the association, and compiled astrometric information from {\em Gaia} DR2 for all of them. This was used to investigate membership and identify runaway stars. By obtaining high-precision radial velocity (RV) estimates for all available spectra, we investigated the RV distribution of the global sample (as well as different subsamples) and identified spectroscopic binaries (SBs).}
{Most of the investigated stars belong to a physically linked population located at d\,=\,2.5$\,\pm$\,0.4~kpc. We identify 79 confirmed or likely members, and 5 member candidates. No important differences are detected in the distribution of parallaxes when stars in $h$~and~$\chi$~Persei or the full sample are considered. In contrast, most O-type stars seem to be part of a differentiated population in terms of kinematical properties. In particular, the percentage of runaways among them (45\%) is considerable higher than for the more evolved targets (which is lower than $\sim$5\% in all cases). A similar tendency is also found for the percentage of clearly detected SBs, which already decreases from 15\% to 10\% when the O star and B Sg samples are compared, respectively, and practically vanishes in the cooler Sgs. Concerning this latter result, our study illustrates the importance of taking the effect of the ubiquitous presence of intrinsic variability in the blue-to-red Sg domain into account to avoid the spurious identification of pulsating stars as SBs.}
{All but 4 stars in our working sample (including 10 O~giants/Sgs, 36 B~Sgs, 9 B~giants, 11 A/F~Sgs, and 18 red Sgs) can be considered as part of the same (interrelated) population. However, any further attempt to describe the empirical properties of this sample of massive stars in an evolutionary context must take into account that an important fraction of the O stars is or likely has been part of a binary/multiple system. In addition, some of the other more evolved targets may have also been affected by binary evolution. In this line of argument, it is also interesting to note that the percentage of spectroscopic binaries within the evolved population of massive stars in Per~OB1 is lower by a factor 4\,--\,5 than in the case of dedicated surveys of O-type stars in other environments that include a much younger population of massive stars.}
 %maximum 6
\maketitle

%========================================================================== 

\section{Introduction}
\label{section.1_intro}

The study of the physical properties and evolution of massive stars (M\,>\,8-9~M$_{\odot}$) is crucial for many aspects of our understanding of the Universe. They play an important role in the chemodynamical evolution of the galaxies \citepads{2012ceg..book.....M} and were key players in the epoch of reionization of the Universe (\citeads{1999ApJ...527L...5B}; \citeads{2000ApJ...540...39A}). They are the precursors of hyperenergetic supernovae, long-duration $\gamma$-ray burst (see \citeads{2012ARA&A..50..107L}, and references therein), and the recently detected gravitational wave events (e.g., \citeads{2016PhRvL.116f1102A}; \citeads{2017PhRvL.119p1101A}; \citeads{2020arXiv200201950A}). Their high luminosities make them observable individually at large distances, and they are thus optimal tools for access to invaluable information about abundances and distances in galaxies at up to several megaparsec (e.g., \citeads{2003ApJ...584L..73U}; \citeads{2008A&A...485...41C}; \citeads{2013ApJ...779L..20K}). Moreover, through their feedback into the interstellar medium in the form of ultraviolet radiation and stellar winds, massive stars critically affect the star formation process by both triggering the formation of new generations of stars and stopping mass accretion in the surrounding forming stars.

Most massive stars are found within or are linked to young open clusters and the so-called OB associations (\citeads{2003ARA&A..41...57L}; \citeads{2010ARA&A..48..431P}). These stellar groupings are therefore perfect laboratories to study them. 

\citetads{1978ApJS...38..309H} compiled a catalog of all known supergiants (Sgs) and O stars in associations and clusters of the Milky Way, including over 1000 objects of this type. Among the list of associations quoted in that paper, Per~OB1, which also includes the famous $h$~and~$\chi$~Persei double cluster, clearly stands out as one of the richest. In particular, it is one of the few Galactic OB associations in which, given its age ($\sim$\,13\,--\,14\,Myr, \citeads{2002ApJ...576..880S}; \citeads{2019ApJ...876...65L}), a massive star population covering a wide range of evolutionary stages can be found (e.g., it harbors 23 red Sgs and several dozen blue Sgs). In addition, it is also relatively close to us (d\,$\sim$\,2.2\,--\,2.4\,kpc, \citeads{2018A&A...616A..10G}; \citeads{2019MNRAS.486L..10D}) and is characterized by a moderate extinction (E(B-V)$\sim$0.6, \citeads{2002ApJ...576..880S}). This unique combination of characteristics makes Per~OB1 a very interesting testbed for the study of a large interrelated population of evolved massive stars from an evolutionary point of view.

Per~OB1 has attracted the attention of the astrophysical community for many years and has been the subject of studies from many different fronts. We highlight the investigation of how the association could have been formed (\citeads{2008ApJ...679.1352L}); the membership of stars to the association (\citeads{1970ApJ...160.1149H}; \citeads{1978ApJS...38..309H}; \citeads{1992A&AS...94..211G}; \citeads{2008ApJ...679.1352L}; \citeads{2009MNRAS.400..518M}) and, in particular, to $h$~and~$\chi$~Persei (\citeads{2002PASP..114..233U}; \citeads{2010ApJS..186..191C}); the characterization of the kinematics of the region (\citeads{2017MNRAS.472.3887M}; \citeads{2019A&A...624A..34Z}); the identification of blue Sg binaries (\citeads{1973ApJ...184..167A}); or the spectroscopic characterization of different samples of blue stars in the region (including the determination of rotational velocities, stellar parameters and surface abundances \citeads{1968ApJ...154..933S};  \citeads{1988A&A...195..208L}; \citeads{1995A&A...298..489K}; \citeads{1996A&A...310..564K}; \citeads{2005AJ....129..809S}; \citeads{2019ApJ...876...65L}), also reaching the red Sg domain (\citeads{2014ApJ...788...58G}).

Despite all the information compiled about the Per~OB1 association, and particularly, $h$~and~$\chi$~Persei, we still lack a complete homogeneous empirical characterization (that also takes environmental and kinematical information into account) of the physical and evolutionary properties of its massive star population. This is the main objective of this series of papers, which is based on a set of high-quality observations including high-resolution, multi-epoch spectroscopy (mostly gathered in the framework of the IACOB project, see \citeads{2015hsa8.conf..576S} and references therein), and astrometric information delivered by the {\em Gaia} mission (\citeads{2018A&A...616A...1G}; \citeads{2018A&A...616A...2L}). The compiled empirical information resulting from the analysis of this observational dataset will allow us to proceed in our understanding of massive star evolution, and also investigate some long-standing and new open questions in this important field of stellar astrophysics. These questions include the evolutionary status of the blue supergiants, or the effect that binarity and rotation have on the evolution of massive stars.

In this first paper, we carry out a membership analysis of a sample of 88 blue and red Sgs located within 4.5\,deg from the center of the Per~OB1 association, and we also investigate some of its kinematical properties. In Sect.~\ref{section.2_samobs} we present the sample of stars and the main characteristics of the compiled observations. 
In Sect.~\ref{section.3_rv} we describe the strategy we followed to derive reliable radial velocities (RVs). Sect.~\ref{section.4_results} presents the results extracted from the analysis of the observations, mainly referring to parallaxes, proper motions, RV measurements, and the identification of spectroscopic signatures of binarity and other types of spectroscopic variability phenomena.
In Sect.~\ref{section.5_discus} we use all these results to establish and apply our membership criteria to all stars in the sample, and we also identify outliers for each of the considered quantities, in particular, binary and runaway stars. We also analyze some global features of Per~OB1, and discuss some individual cases of interest. The main conclusions of this work and some future prospects are provided in Sect.~\ref{section.6_summary}.

\begin{figure}[t!]
\centering
\includegraphics[width=.47\textwidth]{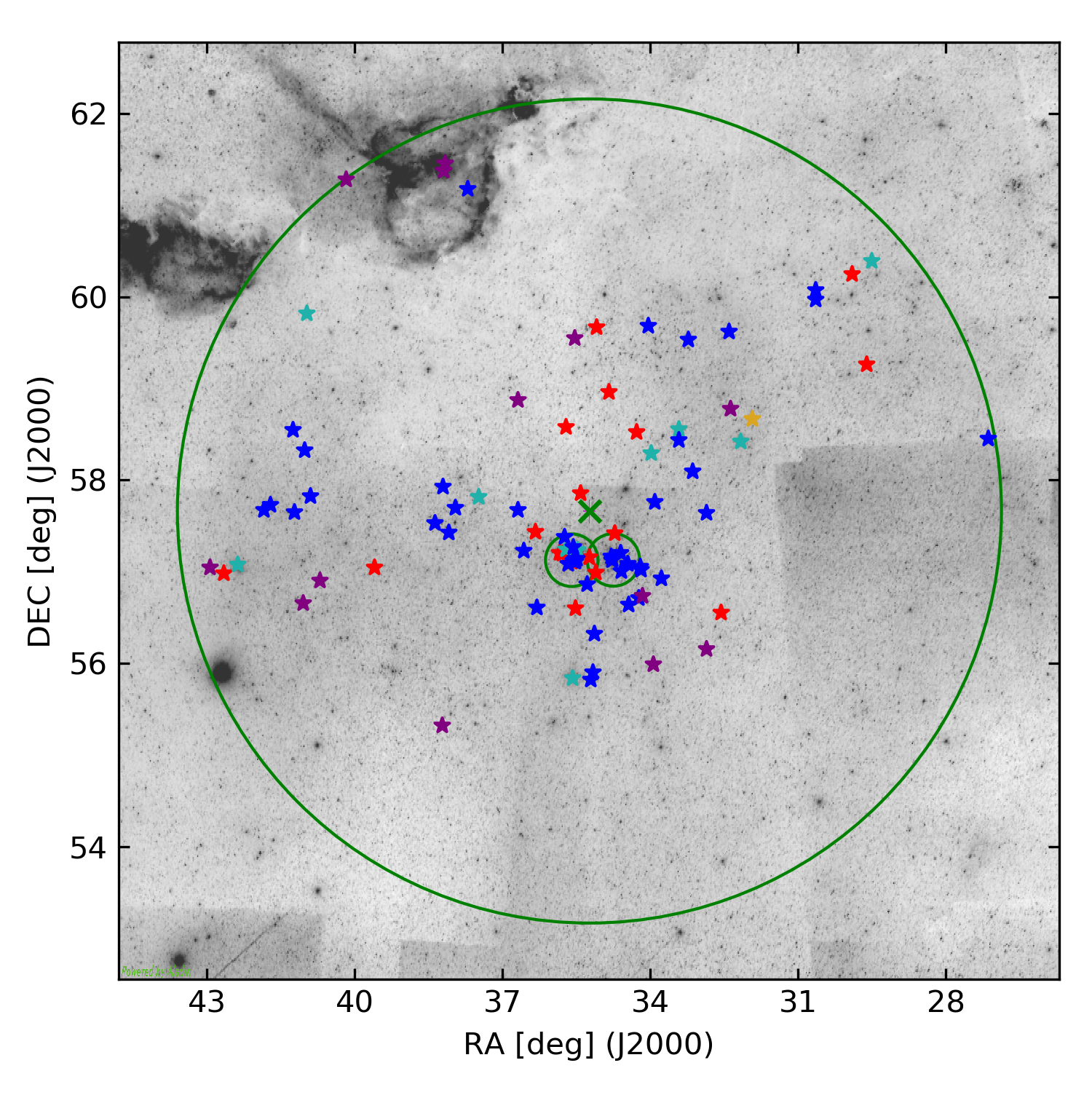}
\caption{Sky map with the complete sample of stars. Purple, blue, cyan, golden, and red symbols represent the O-, B-, A-, F-, and K- and M-type stars, respectively. This color code is the same in all the plots unless otherwise specified. The central green cross denotes the center of the Per~OB1 association taken from \citetads{2017MNRAS.472.3887M}. The large green circle indicates a 4.5-degree circle around the center. The small green circles show the positions of $h$~and~$\chi$~Persei. The background image, used for reference, was taken from DSS-red.}
\label{figure.fig1}
\end{figure}

%========================================================================== 

\section{Sample definition and observations}
\label{section.2_samobs}

In this section, we describe the process we have followed to build the sample under study, and to compile the associated observations. The latter mainly refers to high-quality spectroscopy obtained with the FIbre-fed Echelle Spectrograph (FIES), \citepads{2014AN....335...41T} and the High Efficiency and Resolution Mercator Echelle Spectrograph (HERMES) \citepads{2011A&A...526A..69R} high-resolution
spectrographs attached to the 2.56~m Nordic Optical Telescope (NOT) and the 1.2~m Mercator telescope, respectively, and astrometric and photometric data delivered by the {\em Gaia} mission in the second data release \citepads[DR2, ][]{2018A&A...616A...1G, 2018A&A...616A...2L, 2018A&A...616A...4E}.

%__________________________________________________ 

\subsection{Sample definition}
\label{subsection.21_sample}
The final sample of targets considered for this work comprises 88 blue and red massive stars located within 4.5\,deg from the center of the Per~OB1 association (as defined in \citeads{2017MNRAS.472.3887M}). 
To restrict the sample to the most massive stars, the luminosity classes (LCs) were limited to bright giants (Gs) and Sgs (LC II and I, respectively) in the case of the O- and B-type stars, and to Sgs when we refer to A- and later-type stars. In addition, the sample includes a few O and early-B Gs for which we already had available observations in the IACOB spectroscopic database (see Sect.~\ref{subsection.22_specobs}).

\begin{figure}[t!]
\centering
\includegraphics[width=.47\textwidth]{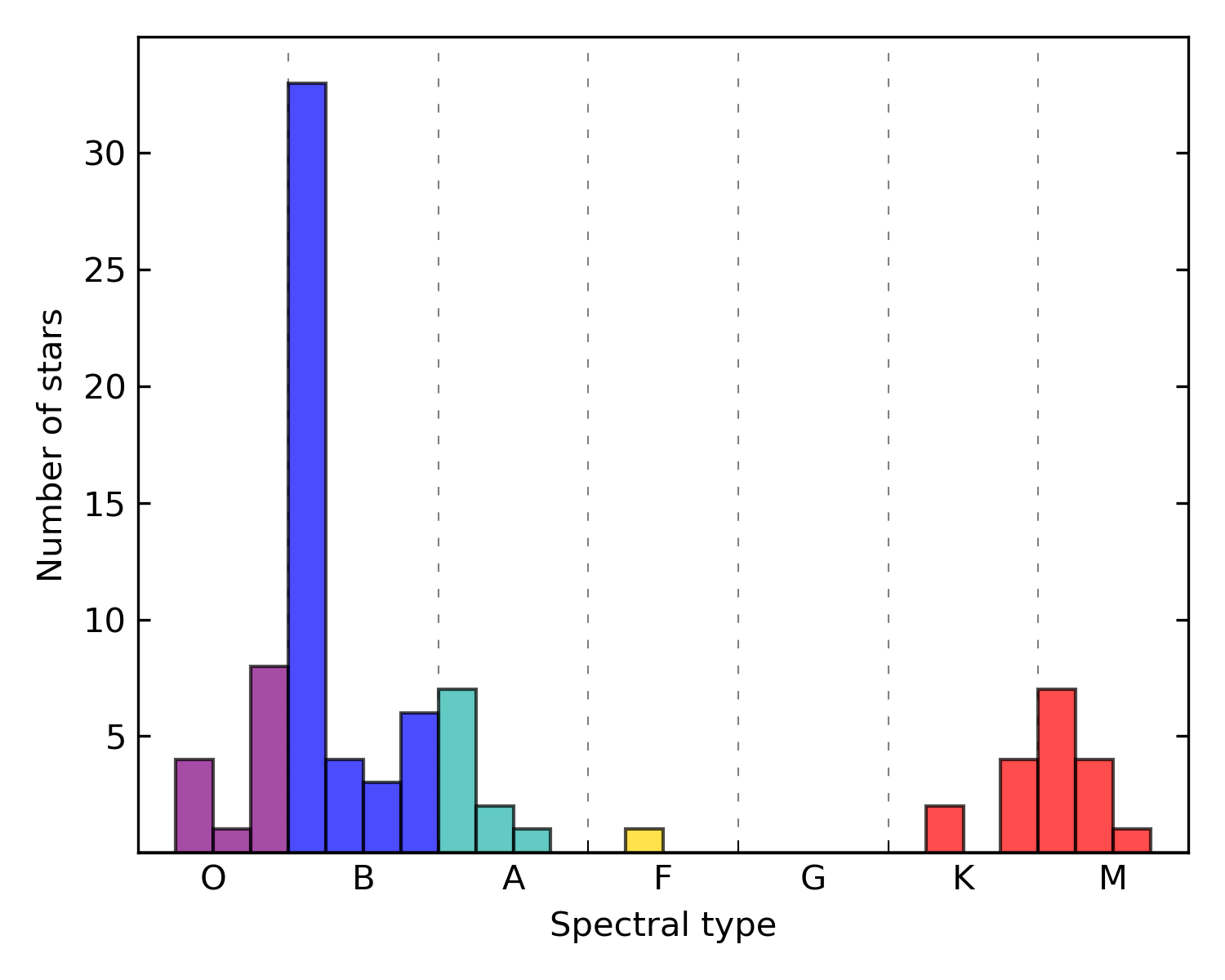}
\includegraphics[width=.47\textwidth]{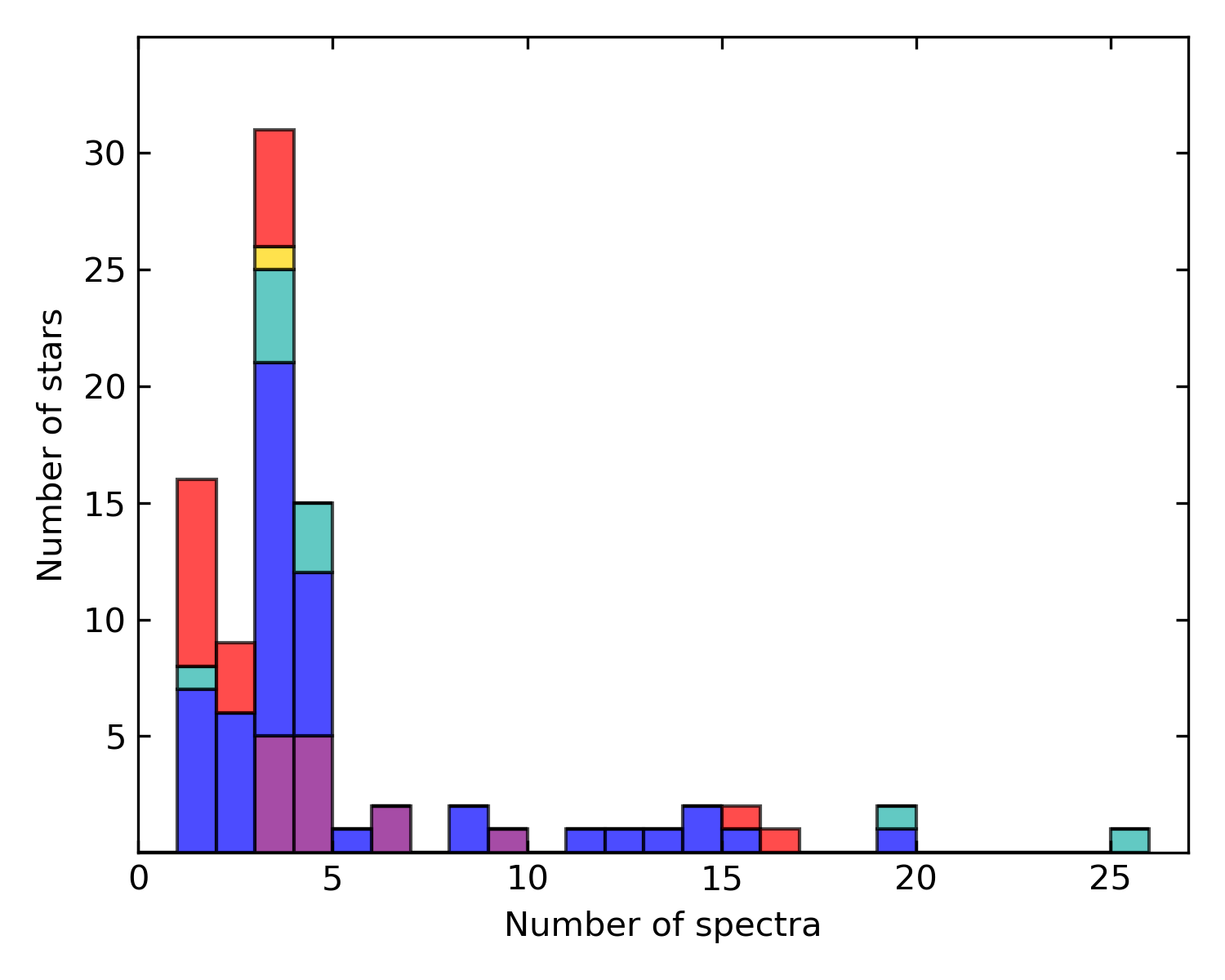}
\caption{Histograms by SpT separated with colors (top) and number of spectra separated by SpT and stacked (bottom).} 
\label{figure.fig2}
\end{figure}

Table~\ref{table.A1} summarizes the list of targets, separated and ordered by spectral type (SpT). We note that the quoted spectral classifications were carefully revised using the spectra with the best signal-to-noise ratio (S/N) of our own spectroscopic observations (see Sect.~\ref{subsection.22_specobs}) following the criteria explained in \cite{Negueruela2020}, in prep.) and \citeads{2018A&A...618A.137D}, for the case of the blue and red Sg samples, respectively.
In addition, Fig.~\ref{figure.fig1} shows their location on the sky. We also indicate as a large green circle the search area of 4.5\,deg around the center of Per~OB1, marked as a green cross. Most stars, including those from the $h$~and~$\chi$~Persei double cluster (indicated as two small green circles), are concentrated along the diagonal of the image. In addition, our sample includes four stars lying within one degree from the center of IC\,1805 (the Heart nebula, located in the top left corner of the figure). The top panel in Fig.~\ref{figure.fig2} depicts the histogram of SpT of the sample, which shows that the majority of stars are B~Sgs.

To assemble this sample of stars, we considered several bibliographic sources, including the works by \citetads{1978ApJS...38..309H}, \citetads{1992A&AS...94..211G}, \citetads{2010ApJS..186..191C}, and \citetads{2014ApJ...788...58G}. In a first step, we used the Topcat\footnote{\href{Topcat}{http://www.star.bris.ac.uk/~mbt/topcat/}} Virtual Observatory tool to cross-match all the stars that are quoted in these four papers and fulfilled the criteria indicated above and the list of targets with spectra available in the IACOB spectroscopic database. In a second step, we tried to obtain new spectra of as many of the missing stars as possible using the NOT or Mercator telescopes (see Sect.~\ref{subsection.22_specobs}).

From the original lists of luminous stars in Galactic OB associations quoted in \citetads{1978ApJS...38..309H} and \citetads{1992A&AS...94..211G}, we found 207 targets that are located within 4.5\,deg of the center of Per~OB1. Only 109 of these fulfill our luminosity class criteria; the rest are either dwarfs, (sub)giants, or do not have a defined luminosity class. Our sample includes 82 of these, but we miss spectra for another 12 (7 B and 5 M Sgs).

We also used the list of targets quoted in the extensive study of the stellar population of $h$~and~$\chi$~Persei by \citetads{2010ApJS..186..191C} to find suitable candidates. From the complete list of several ten thousand stars, only 23 were found to have luminosity classes I or II. We currently have spectra for 17 of them. Of the remaining 6 (all of them B~Sgs), one was identified previously when we cross-matched our observations with the list of targets in \citetads{1978ApJS...38..309H} and \citetads{1992A&AS...94..211G}. This means that we lack spectra for another 5 blue Sgs at the time of writing.

Last, our sample includes all the red Sgs of those listed in \citetads{2014ApJ...788...58G}. In summary, the sample of stars we discuss here comprises all the blue and red Sgs (except for 12 B and five M Sgs, listed at the end of Table~\ref{table.A1} for future reference) that are quoted in the abovementioned papers and are located within 4.5\,deg around the center of Per~OB1. Further notes on the actual completeness of our sample can be found in Sect.~\ref{subsection.43_compl}.

%__________________________________________________ 

\subsection{Spectroscopic observations}
\label{subsection.22_specobs}

The spectroscopic observations of the stars in the sample come from different observing runs performed between November 2010 and December 2019 using either the FIES (NOT) or the HERMES (Mercator) instruments.

The first observations, comprising an initial sample of B, A, and M Sgs in Per~OB1 selected from \citetads{1978ApJS...38..309H}, were obtained in 2010 during an observing run of three nights with Mercator (PI. M.A. Urbaneja). The O stars in the sample were targeted by the IACOB project (P.I. S. Sim\'on-D\'iaz) as part of a more general objective of observing all O stars in the Northern Hemisphere up to $V_{mag}$ = 9. These observations, obtained with both HERMES and FIES, include a minimum of three epochs per target (see more details in \citeads{2018A&A...613A..65H}; \citeads{Holgado2019}; \citeads{2020arXiv200505446H}). We also benefit from the multi-epoch observations available for a subsample of O and B~Sgs as gathered by the IACOB project as part of a subproject aimed at investigating line-profile variability phenomena in the OBA Sg domain and its relation with pulsational-type phenomena (see, e.g., \citeads{2010ApJ...720L.174S}; \citeads{2017A&A...597A..22S}; \citeads{2018A&A...612A..40S}; \citeads{2017A&A...602A..32A}; \citeads{2018MNRAS.476.1234A}). The time span of these observations covers several years. We also count on multi-epoch observations of red Sgs obtained during several of our observing runs with HERMES. Last, all these observations have more recently been complemented by FIES spectroscopy obtained as part of the time granted to A. de Burgos in 2018 by the Spanish time-allocation committee, and through internal service observations performed in 2019 and 2020 by A. de Burgos. In addition, we were able to obtain a new epoch for a large fraction of stars in the sample during an observing run with Mercator in December 2019. 

FIES is a cross-dispersed high-resolution \'echelle spectrograph mounted at the 2.56~m Nordic Optical Telescope (NOT), located at the Observatorio del Roque de los Muchachos on La Palma, Canary Islands, Spain. The observations made with FIES were taken with different fibers/resolutions from R\,$\sim$\,25000 to R\,$\sim$\,67000, and with a wavelength coverage of 370-830\,nm. 

HERMES is a fibre-fed prism cross-dispersed \'echelle spectrograph mounted at the 1.2~m Mercator Telescope, also located at the Observatorio del Roque de los Muchachos. It provides a spectral resolution of R\,$\sim$\,85000 and wavelength coverage of 377-900 nm, similar to FIES. 

The FIES and HERMES spectrographs provide good mechanical and thermal stability that allows for a good precision in RV measurements. For FIES, the RV y accuracy\footnote{http://www.not.iac.es/instruments/fies/fies-commD.html} for the high-resolution fiber has been proved to be 5-10\,m/s, regardless of the atmospheric conditions. For the medium-resolution fiber under poor conditions, the precision reaches 150\,m/s. In the case of HERMES, the precision obtained for the low- and high-resolution fibers is 2.5 and 2\,m/s, respectively (\citeads{2011A&A...526A..69R}). In both cases this precision is well above the precision required for this work, as we expect variations of several \kms\ for the blue Gs/Sgs, and a few \kms\ for the red Sgs. 

All the spectra were reduced using the FIESTool (\citeads{2017ascl.soft08009S}) and HermesDRS\footnote{http://www.mercator.iac.es/instruments/hermes/drs/} dedicated pipelines. Both pipelines provide merged wavelength-calibrated spectra. In addition, we used our own programs, implemented in IDL, to normalize the spectra and compute the heliocentric velocity to be applied to each spectrum before the associated RV was measured (see Sect.~\ref{section.3_rv}).

As indicated above, we have multi-epoch spectroscopy for a large fraction of the stars in our sample. The bottom panel in Fig.~\ref{figure.fig2} summarizes this characteristic of our observations, showing the histogram of the collected number of spectra per star. In addition, Table~\ref{table.A2} quotes all those stars for which we have five or more spectra. This table includes the time span covered by the spectra, together with the total number of spectra for each of these stars, separated by SpT. It is important to remark that the cadence of the spectra taken for each star is very inhomogeneous, as they were gathered during different observing runs, as described at the beginning of this section. 

%__________________________________________________ 

\subsection{Photometric and astrometric data}
\label{subsection.23_gaiaobs}

For all the stars in the sample, Table~\ref{table.A1} quotes the {\em Gaia} $G_{mag}$ and $BP_{mag}$\,-\,$RP_{mag}$, parallaxes ($\varpi$) and proper motions ($\mu_{\alpha}$,$\mu_{\delta}$), as well as associated errors, retrieved from {\em Gaia} DR2. Sources in the {\em Gaia} catalog were identified using Topcat, defining a radius threshold of 2~arcsec. 

We adopted a parallax zero-point offset of $-0.03$\,mas (see \citeads{2018A&A...616A...2L}), which is already applied to all values quoted in Table~\ref{table.A1} and used to generate the various figures in the paper. We note, however, that some other authors push this value up to $-0.08$\,mas (see \citeads{2018ApJ...862...61S}; \citeads{2019MNRAS.486L..10D}). 

The {\em Gaia} DR2 renormalized unit weight error (RUWE) is also included in the last column of Table~\ref{table.A1}. The value of this quantity is used to estimate the goodness of the {\em Gaia} astrometric solution for each individual target. Following recommendations by the {\em Gaia} team for the known issues\footnote{https://www.cosmos.esa.int/web/gaia/dr2-known-issues}, we decided to adopt a RUWE\,=\,1.4 to distinguish between good and bad solutions.

\begin{figure}[t!]
\centering
\includegraphics[width=.47\textwidth]{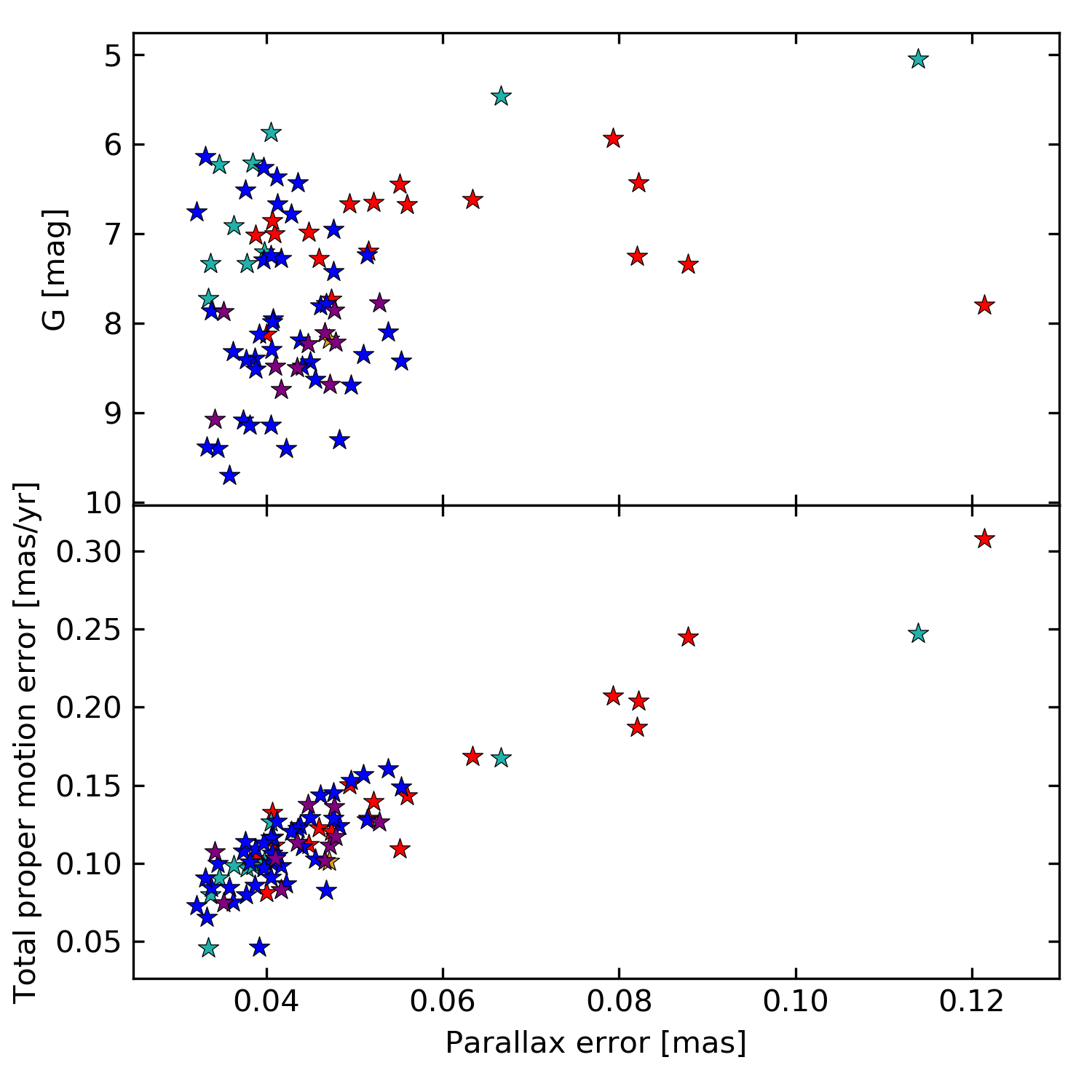}
\caption{(Top) $G_{mag}$ against the {\em Gaia} error in parallax. (Bottom) {\em Gaia} error in total proper motion against the error in parallax. Both panels include all stars in our sample except for the seven targets with {\em Gaia} RUWE > 1.4 (see Sect.~\ref{subsection.23_gaiaobs}).}
\label{figure.fig3}
\end{figure}

Seven stars (or 8$\%$ of the sample) have an associated RUWE higher than this value. Their parallaxes and proper motions are indicated in parentheses in Table~\ref{table.A1}. Hereafter, we call them stars with "unreliable astrometry" or "unreliable astrometric solution". For all the stars with a RUWE $<$ 1.4, the top panel of Fig.~\ref{figure.fig3} shows the $G_{mag}$ against the {\em Gaia} error in parallax, and the bottom panel shows the {\em Gaia} error in total proper motion against the {\em Gaia} error in parallax. 

The $G_{mag}$ of the stars in our working sample ranges between 5.1 and 9.7\,mag. It has been shown that bright sources ($G_{mag}$ < 6) also result in unreliable astrometric solutions because of uncalibrated CCD saturation (\citeads{2018A&A...616A...2L}). In the sample, four stars have magnitudes lower than 6, and they are discussed in detail in Sect.~\ref{subsection.51_memb}. In order to verify the {\em Gaia} DR2 parallaxes and proper motions for the brightest stars in the sample, we also retrieved the values provided in the {Hipparcos} (\citeads{2007A&A...474..653V}), and TGAS (\citeads{2015A&A...574A.115M}) catalogs; however, the results were not better. 

The {\em Gaia} errors in parallax range between 0.032, and 0.121\,mas, while the errors in total proper motion range between 0.046, and 0.308\,mas/yr. Six stars have uncertainties in parallax $\sim$0.08\,mas or larger. The same have uncertainties in total proper motion larger than 0.18\,mas/yr. They are all red Sgs except for HD\,14489 (the A~Sg in the upper right corner). The explanation for their large errors lies in the combined effect of large size and variability for the red Sgs, and the high brightness for HD\,14489 ($G_{mag}$ = 5.1). In both cases, the {\em Gaia} astrometric solution is affected (see Sect.~\ref{subsection.51_memb}). 
Of the red Sgs, HD\,14528 (in the upper right corner) has the largest errors and also a relatively high RUWE value (1.25), followed by HD\,14489, which in comparison has a RUWE = 0.81. In particular, for HD\,14528, we adopt the results from \citetads{2010ApJ...721..267A} from this point on, who used the very long baseline interferometry (VLBI) technique to derive the astrometric parameters.

For the stars with RUWE $>$ 1.4, errors in parallax range between 0.086, and 0.384\,mas with a mean of 0.170\,mas, and errors in total proper motion range between 0.171, and 0.501\,mas/yr, with a mean of 0.347\,mas/yr. We note that as expected, all these stars have larger errors than those associated with the main concentration of stars in the bottom panel of Fig.~\ref{figure.fig3}. 

%========================================================================== 

\section{Radial velocity measurements}
\label{section.3_rv}

We first generated various suitable lists of spectral lines, optimized for the different SpT, using information available in the Atomic Line List interface\footnote{https://www.pa.uky.edu/~peter/newpage/} \citep{2018Galax...6...63V}, and the SpectroWeb\footnote{\href{SpectroWeb}{http://spectra.freeshell.org/whyspectroweb.html}} database. Each line list comprises a few to several dozen strong (log(gf) > --0.5), unblended lines covering the full 390\,--\,650~nm spectral window (or 510\,--\,870~nm in the case of the red Sgs). 

For early and mid O-type stars, a few lines of N~{\sc iii-v} and O~{\sc iii} were used. In addition, we also included some He~{\sc i} lines to compensate for the lower number of available metal lines. For the late O-type stars, we added some lines of Si~{\sc iv} and O~{\sc ii}. The situation improves for the B and A Sgs, were a much larger sample of lines is available, including lines from Si~{\sc ii-iv}, N~{\sc ii-iii}, O~{\sc ii-iii}, S~{\sc ii-iii}, C~{\sc ii}, Mg~{\sc ii}, and Fe~{\sc ii}. Last, in the case of red Sgs, we mostly used lines from Mg~{\sc i}, Ti~{\sc i}, Fe~{\sc i}, Ca~{\sc i}, Cr~{\sc i}, Ni~{\sc i}, and V~{\sc i}.

We then used our own tool (developed in Python\,3.6) to perform a RV analysis. For each star, the corresponding list of lines was selected based on its SpT. For each line, an iterative normalization of the surrounding local continuum was made. Then, each line was fit to either a Gaussian or a Gaussian plus a rotational profile, depending on the first estimate of the full width at half maximum (FWHM) of the line. The measured central wavelength was then used to calculate the RV of each individual line in the initial line list (see above). From all the identified lines we removed those with equivalent widths lower than 25\,{m\AA} directly before we carried out an iterative sigma clipping (using a threshold of 2\,$\sigma$) to remove potential poorly fit lines or incorrect identifications. The RVs of the surviving lines were then averaged, and we calculated the standard deviation of the final RV. This process was repeated for each spectrum and for each star in the sample. 

The measurements of the individual RVs, together with the number of lines used for each spectrum, are listed in Table~\ref{table.rvstable}. For O-type stars, the average number of lines is 12, the final average number of lines after sigma clipping is 6, and the typical uncertainties associated with the dispersion of RV measurements obtained after sigma clipping is $\sim$3.9\,\kms. For the B-type stars, these values are 37 and 22 lines and $\sim$0.9\,\kms , respectively. For A/F-type stars, they are 42 and 32 lines, and $\sim$0.26\,\kms. Finally, for the K/M-type stars, they are 31 and 24 lines, and $\sim$0.17\,\kms. This error is larger for the O-type stars for two main reasons: the first is that fewer lines are available, and the second reason is related to the broadening of the diagnostic lines, which is much larger for the O-type stars than in the cooler B, A, and red Sgs.

The RV results for the spectra with the best S/N are shown in the last column of Table~\ref{table.A1}. For each star, we also searched for double-lined spectroscopic binaries (SB2) by looking at different key diagnostic lines (e.g., \ioni{He}{i}~$\lambda$5875, \ioni{Si}{iii}~$\lambda$4552, \ioni{O}{iii}~$\lambda$5592, \ioni{C}{ii}~$\lambda$4267, \ioni{and Mg}{ii}~$\lambda$4481). 

We were able to measure individual RVs for the two components in three of the five SB2. We used the spectrum of maximum separation between them. Their values are listed in Table~\ref{table.A1}.

For each star with four or more spectra, an average RV was calculated as the mean of the RVs obtained for each individual spectrum. In addition to the associated standard deviation, the peak-to-peak amplitude of variability in RV (RV$_{\rm PP}$) was calculated as the difference between the highest and lowest individual RVs, and its error was calculated as the square root of the sum of the squares of the their individual uncertainties. The results for the stars for which multi-epoch spectroscopy is available are listed in Table~\ref{table.A2}. 

Last, we also visually inspected the line-profile variability in each star with available multi-epoch spectroscopy. By doing this we were able to identify those cases in which any detected variability is more likely due to stellar oscillations than to (single-line) spectroscopic binarity (see Sect.~\ref{subsubsection.442_multi})

%========================================================================== 

\section{Results}
\label{section.4_results}

\begin{figure*}[t!]
\centering
\includegraphics[width=.72\textwidth]{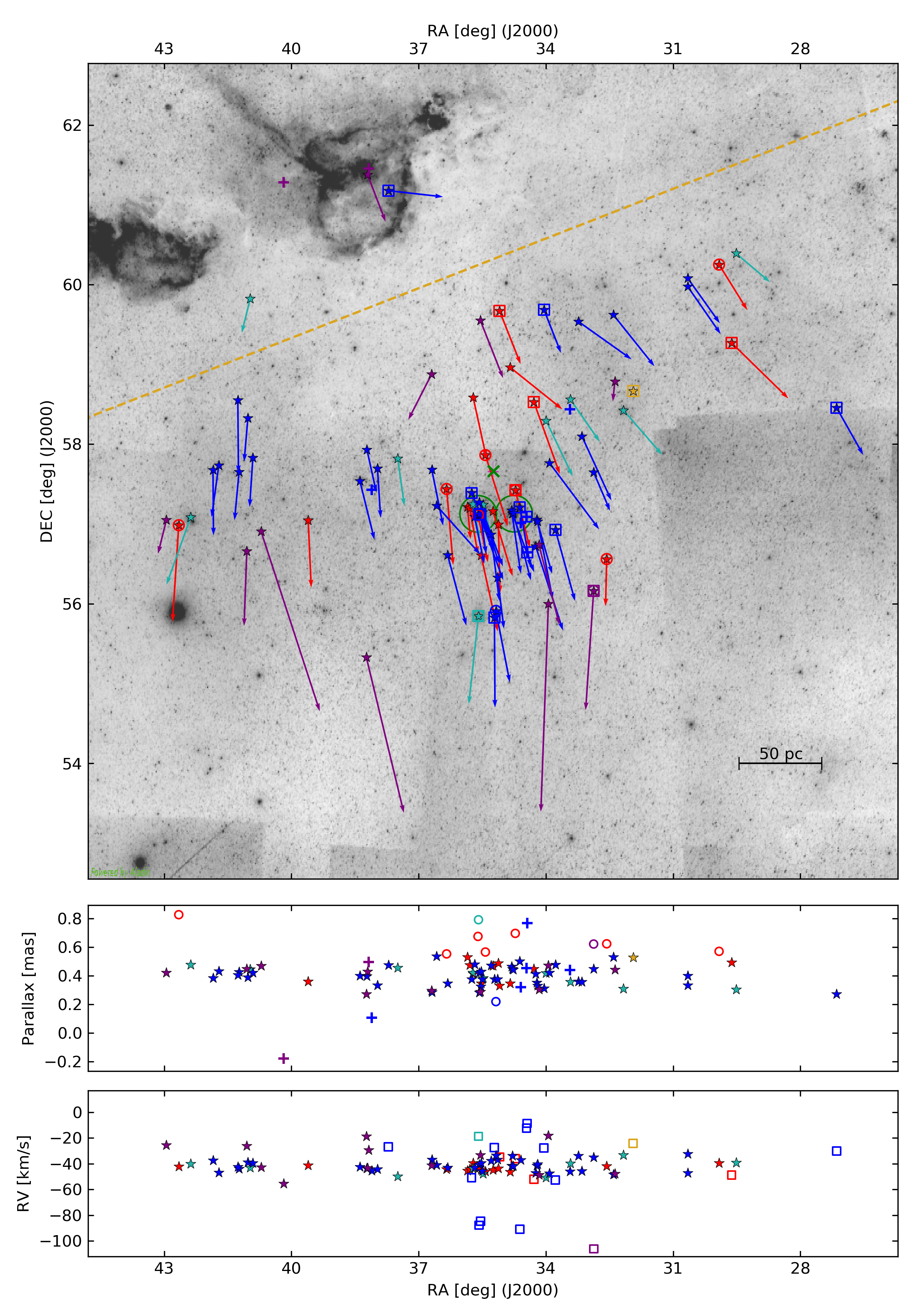}
\caption{{\em Top pannel}: Sky map of all the stars in the sample overplotted over a DSS-red image of the region. Dashed yellow line indicates the galactic plane, and the central green cross marks the center of Per~OB1 (as defined in \citetads{2017MNRAS.472.3887M}). Green circles indicated the location of the $h$~and~$\chi$ Persei double cluster. Colored vectors indicate the individual proper motion of each star. {\em Middle and bottom panels}: Parallax and RV of the spectrum with the highest S/N, respectively, for each star in the sample against their position in right ascension. Open circles and square symbols indicate stars that deviate more than 2$\sigma$ from the mean of the distribution of parallaxes and RVs, respectively (see Sects.~\ref{subsection.41_plx&pm} and \ref{subsubsection.441_bestsnr}). Stars with bad astrometry (see Sect.~\ref{subsection.23_gaiaobs}) are indicated with a plus, and no proper motion vectors are overplotted.}
\label{figure.fig4}
\end{figure*}

Fig.~\ref{figure.fig4} summarizes all the compiled information on astrometry and RVs (except for the information we extracted from the multi-epoch spectroscopy, which is presented in Fig.~\ref{figure.fig9}). The top panel of the figure shows the position of the stars in the sky, and the corresponding proper motions are indicated with arrows. For reference, we also indicate the location of the $h$~and~$\chi$~Persei double cluster (green circles at the center of the image) and the Galactic plane (dashed yellow line).

This image is complemented with another two panels, in which the distribution of parallaxes and RVs (as derived from the best S/N spectrum of each star) is plotted against the right ascension (middle and bottom panels, respectively). These two panels allow us to better identify the location in the sky of the outliers of both distributions, and to easily connect the information of the three investigated quantities.

From a first visual inspection of this summary figure, it becomes clear that generally speaking, the stars in our sample (including those located in the $h$~and~$\chi$~double cluster) belong to a connected population in terms of proper motions, parallaxes, and RVs. In addition, there is a non-negligible number of outliers that we discuss in detail in the next sections. They are potential nonmembers of the Per~OB1 association, and/or runaway stars and binary systems.

%__________________________________________________ 

\subsection{Parallaxes and proper motions}
\label{subsection.41_plx&pm}

\begin{figure}[t!]
\centering
\includegraphics[width=.49\textwidth]{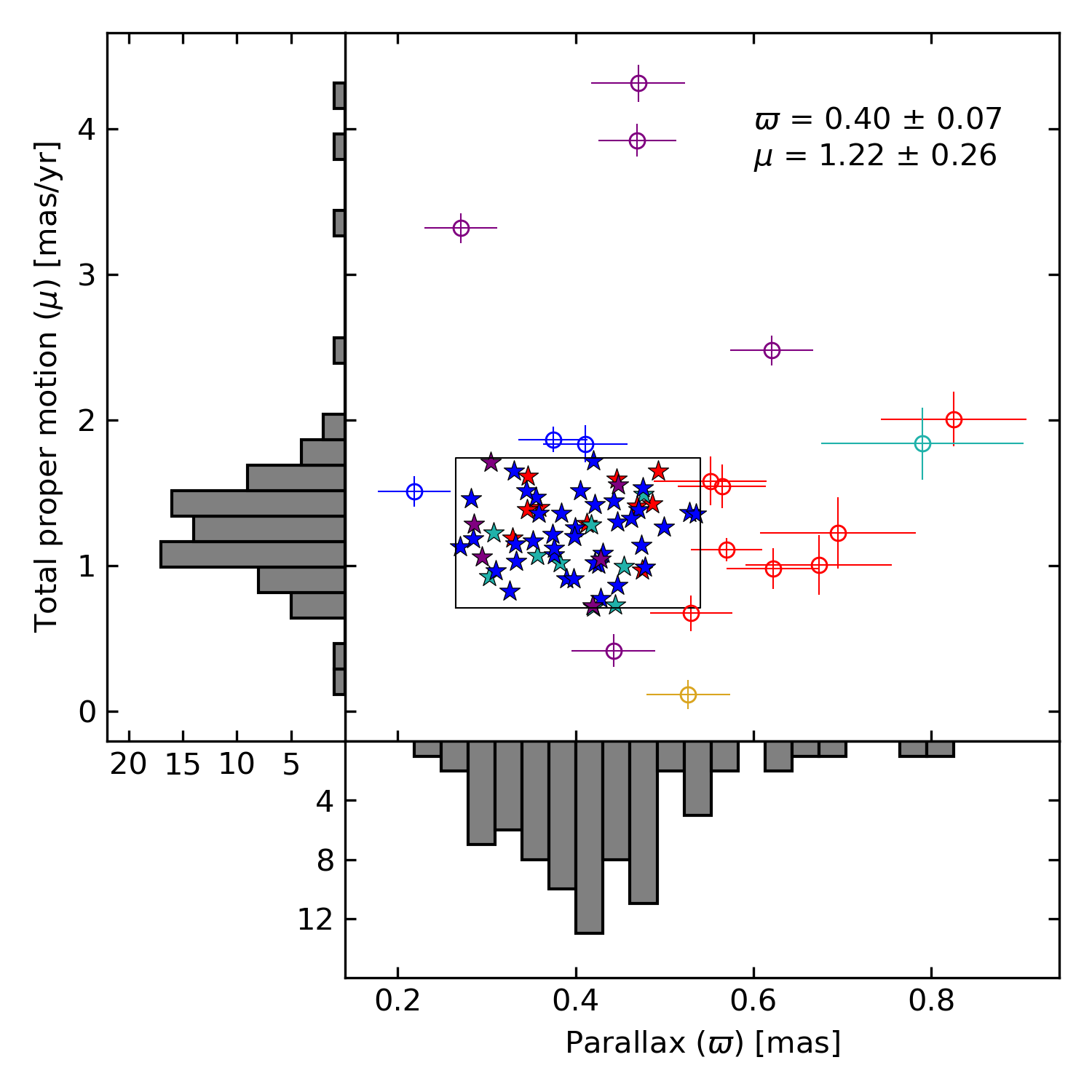}
\caption{Total proper motions against parallax for the sample of stars except for those labeled "unreliable astrometry" (see Sect.~\ref{subsection.23_gaiaobs}). The 2$\sigma$ boundaries of the distribution are shown as a rectangle. Empty colored circles show outliers of the distribution of any of the two quantities, and the associated uncertainties are overplotted. The mean and standard deviation obtained from the stars within the 2$\sigma$ box are shown in the top right corner.}
\label{figure.fig5}
\end{figure}

Figure~\ref{figure.fig5} shows again the results for proper motions and parallaxes ($\varpi$) from a different perspective. The central panel of the figure depicts the combined distribution of these two quantities, this time using the modulus of the proper motion ($\mu$), defined as the square root of the sum of the squares of the proper motion in right ascension and declination. Stars labeled "unreliable astrometric solution" (see Sect.~\ref{subsection.23_gaiaobs}) are excluded from this figure. 

Most of the stars are grouped together around $\varpi$\,$\approx$\,0.4~mas and $\mu$\,$\approx$\,1.2~mas\,yr$^{-1}$. This is also shown in the left and bottom panels of Fig.~\ref{figure.fig5}, where histograms of both parallax and total proper motion are shown. 

An iterative 2$\sigma$ clipping of these distributions results in $\varpi$ = 0.40 $\pm$ 0.07~mas, and $\mu$ = 1.22 $\pm$ 0.26~mas\,yr$^{-1}$, and the identification of a total of 18 outliers (i.e., deviating more than 2$\sigma$ from the mean of the distribution). The 2$\sigma$ boundaries of the distribution (0.265 < $\varpi$ < 0.540~mas, and 0.706 < $\mu$ < 1.740~mas\,yr$^{-1}$, respectively) and the outliers are highlighted in Fig.~\ref{figure.fig5}. The latter are also indicated in the second and third columns of Table~\ref{table.A4} and are discussed in Sect.~\ref{subsection.51_memb}.

\begin{figure}[t!]
\centering
\includegraphics[width=.49\textwidth]{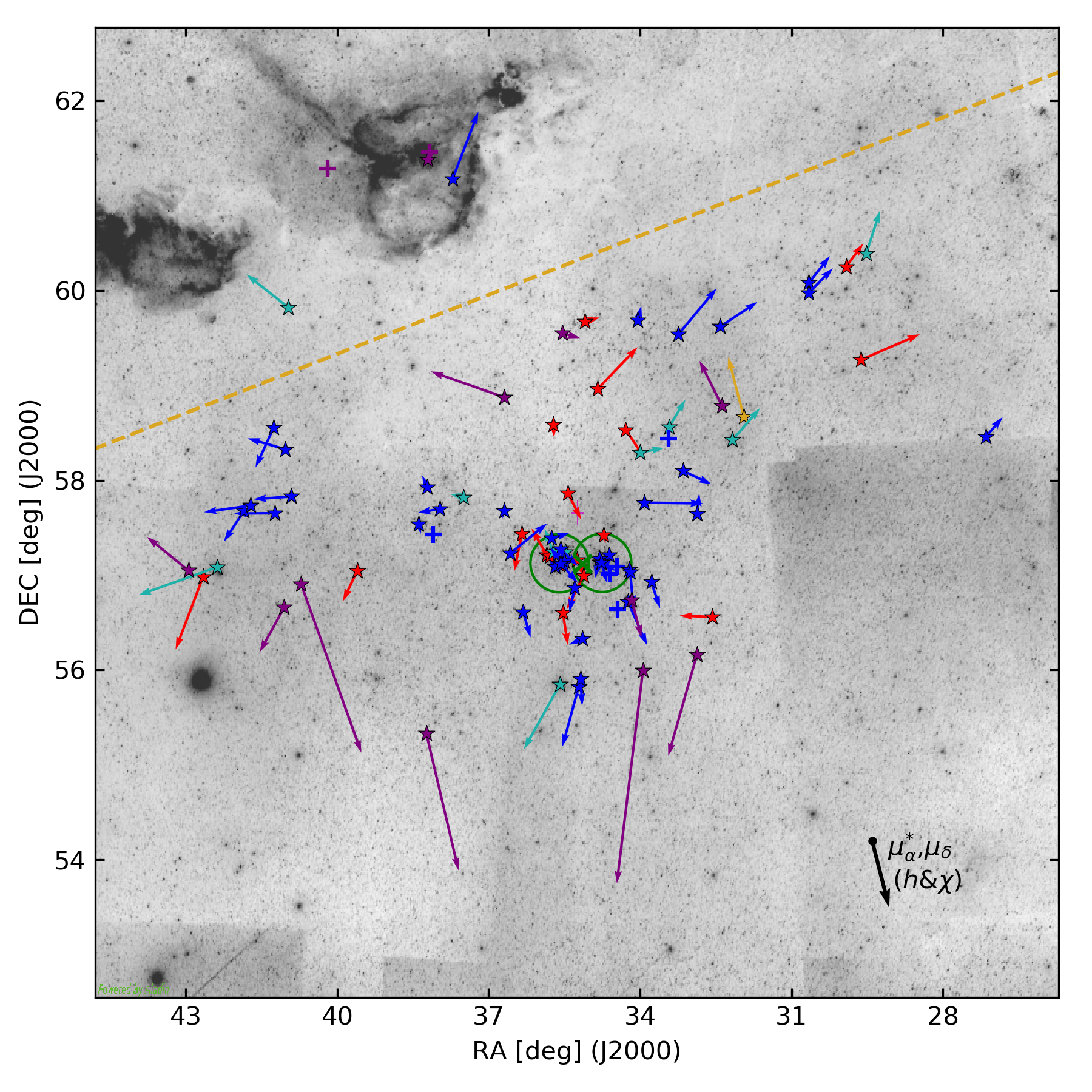}
\caption{Same as the top panel of Fig.~\ref{figure.fig4}, but this time, the individual proper motion of each star is referred to the mean proper motion of 16 stars in $h$~and~$\chi$~Persei with good astrometric solution (black arrow in the bottom right corner of the top panel).}
\label{figure.fig6}
\end{figure}

These results assume that no different local substructures exist in the region, especially in terms of parallax. To investigate this statement further, we show again in Fig.~\ref{figure.fig6} an image of the region with the proper motions overplotted, but this time using the mean proper motion obtained by considering the 16 stars located within 15\,arcmin from the center of $h$~and~$\chi$~Persei, respectively, and having reliable astrometry (see the black arrow in the bottom right corner of the figure, corresponding to $\mu_{\alpha}\cos{\delta}$ = $-0.47$ and $\mu_{\delta}$ = $-0.99$\,mas\,yr$^{-1}$). 

This figure is complemented with the information provided in Table~\ref{table.radiplxpm}, where we summarize the resulting means and standard deviations of parallaxes and proper motions when the sample is divided into circular regions around the center of $h$~and~$\chi$~Persei. The first region only includes the double cluster. The other regions extend outward by one degree each, starting at a distance of 30 arcmin from the center of the double cluster.

\begin{table}
  \centering
  \caption[]{Mean and standard deviation of parallaxes and proper motions for different groups of stars located at increased distance from the center of $h$~and~$\chi$~Persei. Proper motions are referred to the mean of the proper motions of stars within 15 arcmin of each of the clusters and with good astrometric solution.} 
    \label{table.radiplxpm}
      \begin{tabular}{cccc}
\hline
\hline
\noalign{\smallskip}
Radius [deg]  & N$_{\rm stars}$  & $\varpi$ [mas] &$\mu$ [mas/yr] \\  %
\noalign{\smallskip}
\hline
\noalign{\smallskip}
$h$~and~$\chi$ Persei & 16 & 0.43 $\pm$ 0.06 & 0.31 $\pm$ 0.13 \\ % 
0.5 < R < 1.5  &19  &0.39 $\pm$ 0.07  & 0.43 $\pm$ 0.22 \\ %
1.5 < R < 2.5  & 9  &0.34 $\pm$ 0.04  & 0.54 $\pm$ 0.33 \\ %
2.5 < R < 3.5  & 8  &0.40 $\pm$ 0.07  & 0.60 $\pm$ 0.23 \\ %
3.5 < R < 4.5  &14  &0.40 $\pm$ 0.07  & 0.74 $\pm$ 0.29 \\ % 
\hline
      \end{tabular}
\end{table}

Based on the results presented in this section, we conclude the following: There is some empirical evidence of the existence local substructures in the spacial distribution of proper motions (see further discussion in Sect.~\ref{subsubsection.521_pm}). These subgroups of stars have a compatible distribution of parallaxes and proper motions. As a result, this justifies the decision to use the whole sample of stars to obtain the mean values and standard deviations of these two quantities to characterize this population of stars, as well as to identify potential outliers in parallax (i.e., nonmembers) and proper motion (i.e., runaway stars).

%__________________________________________________ 

\subsection{Comparison with previous works}
\label{subsection.compa}

%_ _ _ _ _ _ _ _ _ _ _ _ _ _ _ _ _ _ _ _ _ _ _ _ _ _

\subsubsection{Distance}
\label{subsection.compa_plx}

We have obtained an average value for the parallax of $\varpi$ = 0.398 $\pm$ 0.066\,mas (adopting a zero offset of $-0.03$~mas). This value represents the mean of all stars in the sample with good astrometric solution that are not outliers in parallax and total proper motion.

Based on the corrected computed distances to these stars from \citetads{2018AJ....156...58B}, we obtain an average of $d$ = 2566 $\pm$ 432\,pc. This is compatible with the distance obtained using the inverse of our derived parallax: $d$ = 2510 $\pm$ 415\,pc. When we assume this distance, the projected distance extends up to $\sim$180\,pc for the furthest stars in the association. In particular for the stars in the double cluster used in Table~\ref{table.radiplxpm}, we obtain a distance of $d$ = 2340 $\pm$ 328\,pc using the inverse of the derived parallax. 

This result agrees well with previous estimates for $h$~and~$\chi$~Persei using different approaches. In addition, it also indicates that the parallax zero-point offset correction proposed by \citetads{2018A&A...616A...2L} is adequate.

To give some examples, \citetads{2002PASP..114..233U} obtained an average double cluster distance of $d$ = 2014 $\pm$ 46\,pc using the ZAMS fitting approach. \citetads{2010ApJS..186..191C} used main-sequence stars with a very large sample and obtained a distance to each cluster of $d_{h}$ = 2290$^{+87}_{-82}$\,pc and $d_{\chi}$ = 2344$^{+88}_{-85}$\,pc. The previously mentioned work by \citetads{2010ApJ...721..267A} estimated a distance to HD\,14528 of $d_{h}$ = 2420$^{+110}_{-90}$\,pc using high-precision interferometric observations. More recently, \citetads{2018A&A...616A..10G} published mean parallaxes for a broad selection of open clusters using {\em Gaia} DR2 including $h$~and~$\chi$~Persei. By applying a --0.03\,mas zero-point offset, they obtained $d_{h}$ = 2239\,pc and $d_{\chi}$ = 2357$^{+88}_{-85}$\,pc. Finally, \citetads{2019MNRAS.486L..10D} estimated the distance to $h$ Persei in $d_{h}$ = 2250$^{+160}_{-140}$\,pc, adopting an offset of $-0.05$\,mas for the {\em Gaia} parallaxes. 

The aim of this work is not to provide a better estimate, but to ensure that the stars selected here based on their parallax belong to the association. Only a few works provide distances to the Per~OB1 association. For instance, \citetads{2019ApJ...882..180S} used the photometric distance and {\em Gaia} parallaxes for a selection of O-type stars to derive a distance to the association of $d$ = 2.99\,$\pm$\,0.14\,kpc and $d$ = 2.47\,$\pm$\,0.57\,kpc, respectively.

%_ _ _ _ _ _ _ _ _ _ _ _ _ _ _ _ _ _ _ _ _ _ _ _ _ _

\subsubsection{Proper motions}
\label{subsection.compa_proper}

For the stars that are not outliers in proper motion and parallax, we obtain mean values and standard deviation for the individual components of the proper motion of $\mu_{\alpha}\cos{\delta}$ = $-0.51$ $\pm$ 0.48\,mas/yr, $\mu_{\delta}$ = $-1.00$ $\pm$ 0.31\,mas/yr. This result agrees quite well with previous results obtained in the literature by other authors and different samples of stars. For example, \citetads{2019A&A...624A..34Z} investigated a sample of more than 2100 stars (covering a much wider range in mass than our study) located within 7.5 degrees around the $h$~and~$\chi$~Persei double cluster. They found for each cluster $\mu_{\alpha}\cos{\delta}$ = $-0.71$ $\pm$ 0.18\,mas/yr and $\mu_{\delta}$ = $-1.12$ $\pm$ 0.17\,mas/yr, respectively. Similar results were also obtained by \citetads{2017MNRAS.472.3887M} and \citetads{2019ApJ...876...65L}.

%__________________________________________________ 

\subsection{Completeness of the sample}
\label{subsection.43_compl}

\begin{figure}[t!]
\centering
\includegraphics[width=.49\textwidth]{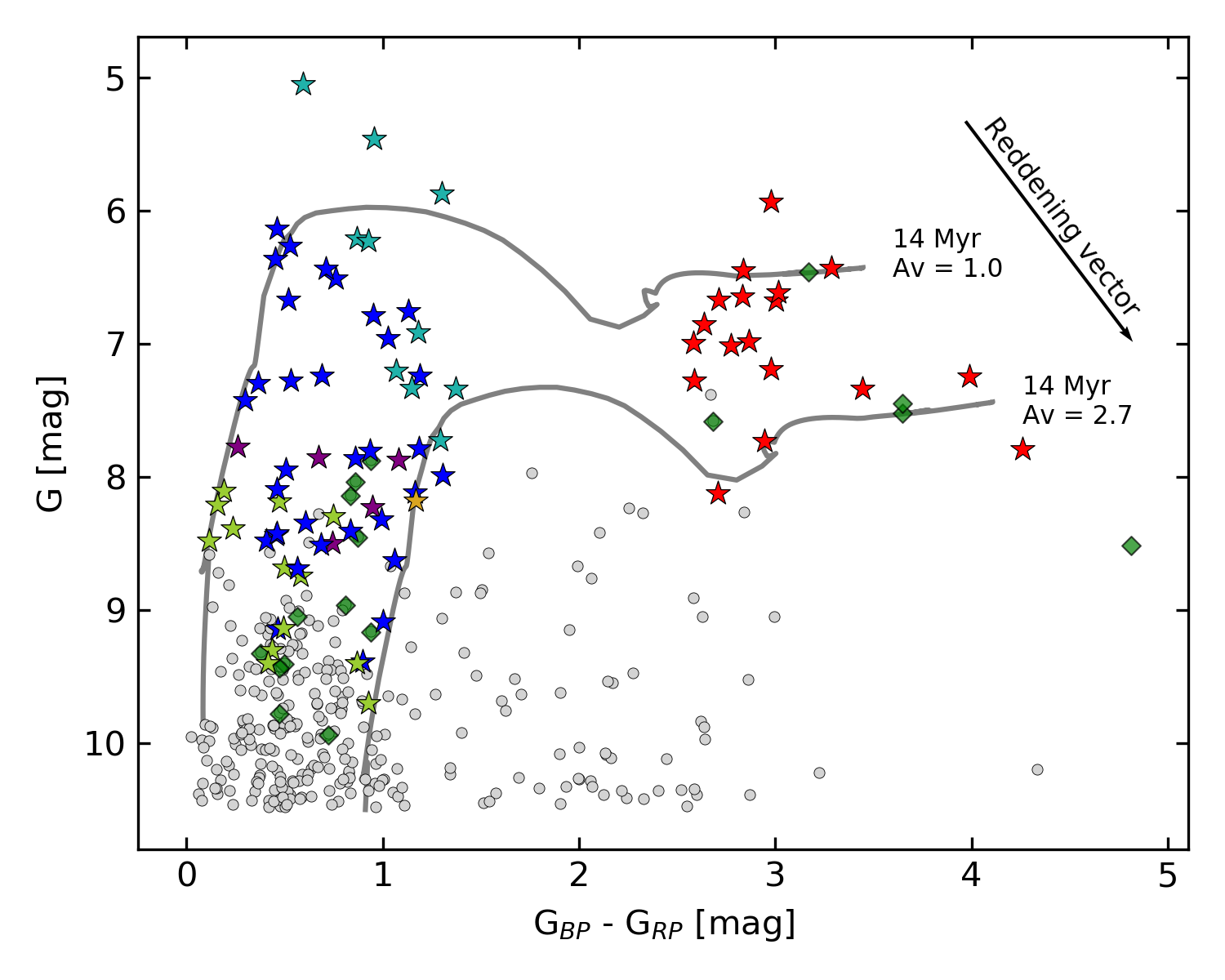}
\caption{Color-magnitude diagram (using {\em Gaia} photometry) of stars located within 4.5~degrees from the center of the Per~OB1 association. Colored stars shows the stars in our sample, gray circles represent the remaining stars from {\em Gaia}, green diamonds show 17 blue and red supergiants quoted in the literature for which we lack spectra (see the last part of Table~\ref{table.A1}). Two isochrones and a reddening vector are also included for reference purposes. See Sect.~\ref{subsection.43_compl} for explanation.}
\label{figure.fig7}
\end{figure}

As indicated in Sect.~\ref{subsection.21_sample}, our sample of 88 stars includes almost all blue and red Sgs (LC I and II) quoted in \citetads{1978ApJS...38..309H}, \citetads{1992A&AS...94..211G}, \citetads{2010ApJS..186..191C}, and \citetads{2014ApJ...788...58G}, plus a few LC III objects (Gs) with late-O and early-B spectral types. In particular, from a total of 107 targets quoted in these four papers that meet our selection criteria, we only lack spectra for 12 B and 5 M Sgs.

To further evaluate the completeness of our sample, we benefit from photometry provided by {\em Gaia} and the results about parallaxes and proper motions described in Sect.~\ref{subsection.41_plx&pm}. To this aim, we retrieved all the stars in the {\em Gaia} DR2 catalog with $G_{mag}$ brighter than 10.5 whose parallaxes and total proper motions lie within 2$\sigma$ of the distributions depicted in Fig.~\ref{figure.fig5}. We then removed all stars with RUWE larger than 1.4, and those classified by the SIMBAD Astronomical Database\footnote{\href{SIMBAD}{http://simbad.u-strasbg.fr/simbad/}} as dwarfs or subgiants (luminosity classes V and IV). 

The results are presented in a color-magnitude diagram (CMD) in Fig.~\ref{figure.fig7}, where we use the same color-code as in previous figures for the stars in our sample, but this time, we also highlight the 17 stars that are classified as LC III or II-III stars in light green. 
For reference purposes, we also include a $A_{\rm v}$ = 1.7\,mag reddening vector and two reddened 14\,Myr isochrones\footnote{Downloaded from the {\em Mesa Iscochrones and Stellar Tracks} interface, MIST (\citeads{2016ApJS..222....8D,2016ApJ...823..102C}).} (solid lines) shifted to a distance of 2.5\,kpc (or, equivalently, a distance modulus of 12\,mag.). The values of reddening for the isochrones (A$_{\rm v}$ = 1.0 and 2.7, respectively) were selected to embrace the main-sequence band, corresponding to the region of the CMD with higher density of gray points in the bottom left corner. 

From inspection of this figure we can conclude that the level of completeness in our sample is very high, specially when we concentrate on the region of the CMD where the blue and red Sgs are located (purple, dark blue, cyan, and red stars). Interestingly, we also find that a high percentage of the 12 B~Sgs quoted in \citetads{1978ApJS...38..309H}, \citetads{1992A&AS...94..211G}, and \citetads{2010ApJS..186..191C} are likely B~Gs, instead of B~Sgs. These refer to all green diamonds with $G_{mag}$ < 9, most of them classified as B Sgs in \citetads{2010ApJS..186..191C} (see the last rows of Table~\ref{table.A1}).

In addition, Fig.~\ref{figure.fig7} allows us to conclude that the blue and red Sg population of Per~OB1 is affected by a variable reddening that ranges from $A_{\rm v}$~$\sim$~1.0 to 2.7\,mag (in agreement with previous findings by \citeads{2019ApJ...876...65L}), and that the age associated with the blue and red Sg population is not compatible because the higher mass present in the 14~Myr isochrone is $\sim$14\,$M_{\odot}$, while all O, B, and A Gs/Sgs included in our sample are expected to have masses higher than 20\,$M_{\odot}$. This latter result will be further investigated in the next paper of this series, after information about the stellar parameters of the full working sample is included.

%__________________________________________________ 

\subsection{Radial velocities}
\label{subsection.44_rv}

By following the strategy described in Sect.~\ref{section.3_rv}, we obtained RV estimates for all the available spectra in our sample of stars. These measurements are used (1) to investigate the RV distributions resulting from the best S/N spectra, (2) to provide empirical constraints on intrinsic spectroscopic variability typically associated with the various types of stars, and (3) to identify spectroscopic binaries and runaway candidates.

%_ _ _ _ _ _ _ _ _ _ _ _ _ _ _ _ _ _ _ _ _ _ _ _ _ _

\subsubsection{Best S/N spectra}
\label{subsubsection.441_bestsnr}

\begin{figure}[t!]
\centering
\includegraphics[width=.4\textwidth]{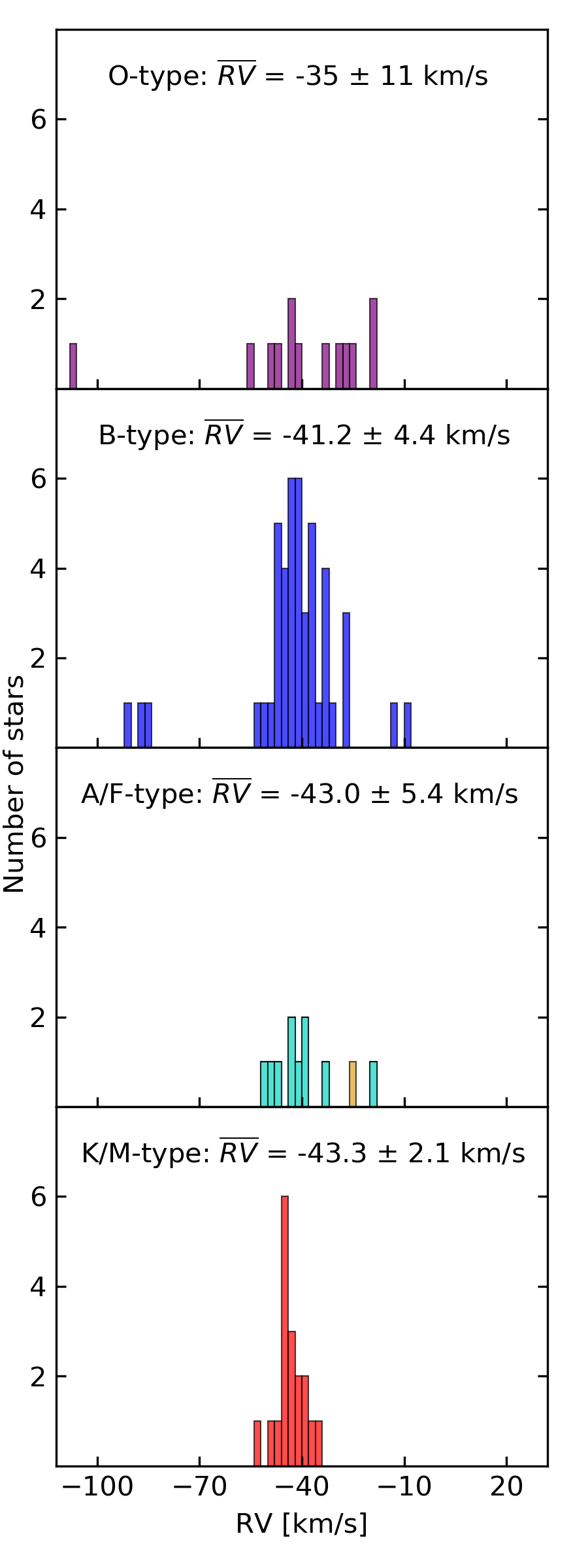}
\caption{RV distributions associated with the different SpT groups resulting from the analysis of the best S/N spectra. The orange bin in the second panel from the bottom is HD\,12842, the F Sg.}
\label{figure.fig8}
\end{figure}

The bottom panel of Fig.~\ref{figure.fig4} shows the RVs of all stars in the sample, obtained from the best S/N spectra, as a function of the position of the stars in right ascension. The associated distributions, this time separated by SpTs, are depicted in the form of histograms in Fig.~\ref{figure.fig8}, with the mean and standard deviation associated with each RV distribution (after performing an iterative 2$\sigma$ clipping) indicated at the top of the various panels. The corresponding outliers in each distribution are indicated as open squares in the bottom panel of Fig.~\ref{figure.fig4} and listed in the fourth column of Table~\ref{table.A4}.

From a visual inspection of Fig.~\ref{figure.fig8} we can conclude that except for the case of O-type stars, which has a flatter and more scattered distribution, the other three distributions are quite similar (when the outliers are eliminated), following a more or less clear Gaussian shape. (For the A/F-type stars, only the two situated on the right-most side of Fig.~\ref{figure.fig8} are outliers. The consequence of having fewer stars than for the B- and K/M-type stars results in a poorer Gaussian shape.) The mean values of these three distributions are compatible within the uncertainties, with a difference smaller than 2~\kms. Interestingly, the standard deviation of the distributions significantly drops from O- to B- and A-type stars, and continues to decrease to the K/M-type stars (see further notes in Sects.~\ref{subsubsection.442_multi} and \ref{subsubsection.443_binaries}).

%_ _ _ _ _ _ _ _ _ _ _ _ _ _ _ _ _ _ _ _ _ _ _ _ _ _

\subsubsection{Multi-epoch spectra: intrinsic variability}
\label{subsubsection.442_multi}

\begin{figure}[t!]
\centering
\includegraphics[width=.49\textwidth]{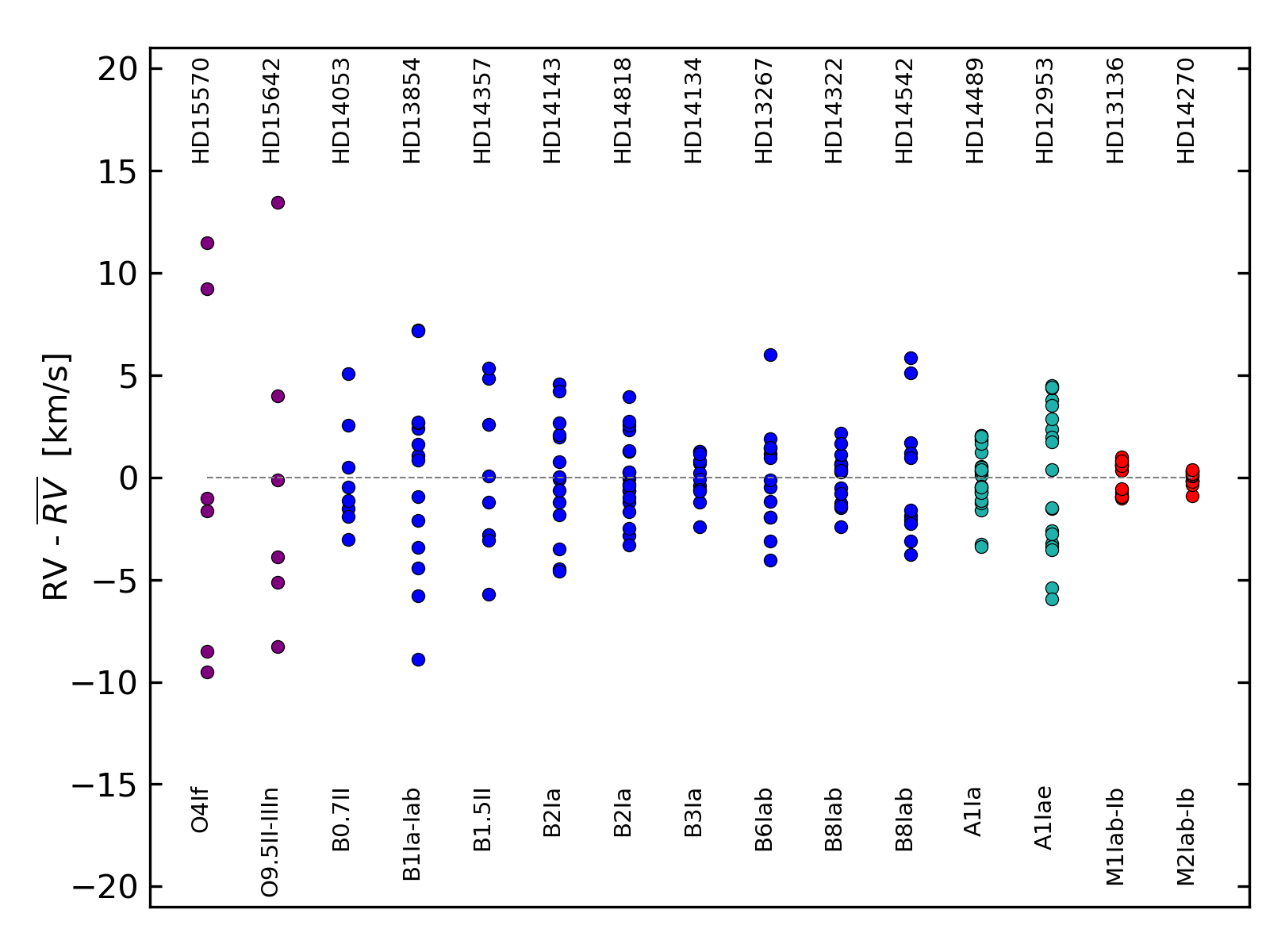}
\caption{Measured RVs (subtracted from their mean) for a sample of 15 stars (ordered by SpT) for which we have five or more spectra, and whose detected variability in RV is more likely produced by intrinsic variability than by the orbital motion in a binary system.}
\label{figure.fig9}
\end{figure}

As indicated in the bottom panel of Fig.~\ref{figure.fig2}, we have more than one spectrum for 73 of the stars in the sample. These observations can be used to identify binaries; however, as extensively discussed in \cite{SimonDiaz2020}, in prep.), the effect of intrinsic variability also needs to be taken into account to minimize the spurious detection of single-line spectroscopic binaries (SB1) in the blue supergiant domain (see also further notes regarding the red supergiant domain in \citeads{2019A&A...624A.129P,2020A&A...635A..29P}).

Some examples of the type of spectroscopic variability phenomena produced by stellar oscillations or the effect of a variable stellar wind in the OBA Sg domain can be found in \citetads{1996ApJS..103..475F, 2004A&A...418..727P, 2006A&A...457..987P, 2006A&A...447..325K, 2015A&A...581A..75K, 2017A&A...597A..22S,2018A&A...612A..40S, 2017A&A...602A..32A, 2018MNRAS.476.1234A}, for example. This effect is also illustrated in Fig.~\ref{figure.fig9} using a subsample of 15 stars in PerOB1 for which five or more spectra are available, and whose detected variability in RV is more likely produced by intrinsic variability than by the orbital motion in a binary system (see Table~\ref{table.A2} and further notes in Sect.~\ref{subsubsection.443_binaries}). 

These results warn us about the dangers of using a single snapshot observation to associate the outliers detected in the RV distributions shown in Fig.~\ref{figure.fig8} with potential runaway stars or spectroscopic binaries. Some of these cases might even correspond to a single measurement in a specific phase of the intrinsic variability of the star instead of being associated with the orbital motion in a binary system or with a single star with an anomalous RV due to an ejection event. They also partially explain why the standard deviation of the RV distributions presented in Fig.~\ref{figure.fig8} becomes smaller when moving from the blue to the red Sgs. This is just a consequence of the behavior of the characteristic amplitude of spectroscopic variability with SpT (see Table~\ref{table.variab} and \cite{SimonDiaz2020}, in prep.). Last, it also affects the fraction of detected SB1 stars using multi-epoch observations, or the final sample of outliers in RV (see further notes in Sects.~\ref{subsubsection.443_binaries} and \ref{subsubsection.523_rw&bin}, respectively).

To evaluate the effect that including information about multi-epoch spectroscopy has on the identification of outliers in the RV distribution, we have repeated the same exercise as in the case of the single-snapshot observations (Sect.~\ref{subsubsection.441_bestsnr}), but modifying the individual measurements (obtained from the analysis of the best S/N spectra) of stars for which four or more spectra are available by the mean of the multi-epoch RV measurements. Results of this exercise are presented in the "RV multi" column of Table~\ref{table.A4}. Although the number of stars with a modified outlier status in RV is small in this specific example (only HD\,13402 and HD\,12953), the results presented in Fig.~\ref{figure.fig9} indicate that it could have been larger if other epochs of the time series had been selected as single-snapshot observations.

\begin{table}
  \centering
  \caption[]{Summary of detected variability (mean and maximum of peak-to-peak amplitude of RV in each SpT group) for the sample of 15 stars depicted in Fig.~\ref{figure.fig9}. RVs in \kms.} 
    \label{table.variab}
      \begin{tabular}{lcccc}
\hline
\hline
\noalign{\smallskip}
SpT group & N$_{\rm stars}$ & $\overline{\rm N}_{\rm spectra}$ & $\overline{RV}_{\rm PP}$ & RV$_{\rm PP, max}$ \\ %
\noalign{\smallskip}
\hline
\noalign{\smallskip}
O-type   &  2 &  6 & 21.3 $\pm$ 0.4  & 21.7 \\  % 
B-type   &  9 & 13 &  8.8 $\pm$ 3.5  & 16.1 \\  %
A/F-type &  2 & 22 &  7.9 $\pm$ 2.5  & 10.4 \\  %
K/M-type &  2 & 16 &  1.6 $\pm$ 0.4  &  2.0 \\  %
\noalign{\smallskip}
\hline
      \end{tabular}
\end{table}

%_ _ _ _ _ _ _ _ _ _ _ _ _ _ _ _ _ _ _ _ _ _ _ _ _ _

\subsubsection{Multi-epoch spectra: spectroscopic binaries}
\label{subsubsection.443_binaries}

Given the stability of the FIES and HERMES instruments, and the accuracy reached in the RV measurements for most of the stars in the sample with multi-epoch spectroscopy, it might be tempted to assign the SB1 status to all stars showing a RV$_{\rm PP}$ above a few \kms. However, as indicated in Sect.~\ref{subsubsection.442_multi}, the intrinsic variability in single blue supergiants can reach amplitudes of a few dozen \kms; hence, many of these identification may lead to spurious results.   

To avoid this situation as much as possible, and in order to identify the most secure candidates to be SB1, we performed a careful inspection of the type of line-profile variability detected in each of the stars with more than one spectrum. To this aim, we mainly considered the following diagnostic lines, whenever available: \ioni{He}{i}~$\lambda$5875, \ioni{Si}{iii}~$\lambda$4552, \ioni{O}{iii}~$\lambda$5592, \ioni{C}{ii}~$\lambda$4267, \ioni{and Mg}{ii}~$\lambda$4481. For the case of the two red supergiants with multi-epoch observations, we found that the measured RV$_{\rm PP}$ is lower than 2\,\kms, which we directly attribute to intrinsic variability.

The list of clearly identified SB1 is presented in Tables~\ref{table.A3}. In addition to the four SB1 stars quoted there, we found five SB2 systems (some of them directly detected from a single-snapshot observation) and labeled "LPV/SB1?" another five cases in which we are not entirely sure if the detected variability is due to binarity or intrinsic variability. All this information is also added to \ref{table.A4} (column "Spec. variability"). 

We also performed a bibliographic search for previously identified binaries in our sample of blue and red supergiants. We mainly concentrated in the works by \citetads{2018A&A...613A..65H, 2020arXiv200505446H} and \citetads{2019A&A...626A..20M} for the case of O-type stars, and \citetads{1973ApJ...184..167A,1985AbaOB..58..313Z}, and \citetads{2017A&A...598A.108L} for the B supergiant sample. In addition, we made use of \textit{The International Variable Star Index (VSI)}\footnote{\href{VSI}{https://www.aavso.org/vsx/index.php}}. 

In total, we found that six out of our sample of ten detected SB1 or SB2 systems from this work were previously identified in any of these references as such (HD\,16429 is actually a triple system \citepads{2003ApJ...595.1124M}). This implies four newly detected binaries: HD\,13969, HD\,14476, HD\,17378 (all SB1), and HD\,13402 (SB2).
We also found three binaries in the literature that were not detected from our available spectra because of short time-coverage: BD\,+56578, an eclipsing binary \citepads{2016AstL...42..674T, 2017A&A...598A.108L}, plus HD\,17603 and HD\,14956, identified as SB1 by \citetads{2018A&A...613A..65H} and \citetads{1973ApJ...184..167A}, respectively. All of them are labeled "(lit.)" in the corresponding column of Table~\ref{table.A4}.

The stars classified as "LPV/SB1?" are HD\,13036, HD\,13854, HD\,13267, HD\,14542, HD\,12953, and HD\,17378.
HD\,13036 (B0.2~III), HD\,13267 (B6~Iab) and HD\,14542 (B8~Iab) have RV$_{\rm PP}$ = 10 -- 14\,\kms; although this value is at the boundary of the expected variability due to pulsations, which may indicate an SB1 classification, we cannot conclude after visual inspection of their line-profile variability. HD\,13854 (B1~Ia-Iab), mostly looks like a pulsational variable, but we do not discard the possibility entirely that  this star might be a SB1. We note that \citetads{1973ApJ...184..167A} provide RV$_{\rm PP}$ = 24.8\,\kms; however, they did not consider it as a binary. For HD\,12953 (A1~Iae), we measured RV$_{\rm PP}$ = 10.4\,\kms, the largest variability in the A supergiant sample; however, after visual inspection of its line profile variability, we cannot conclude whether this is a SB1 system. \citetads{1973ApJ...184..167A} found RV$_{\rm PP}$ = 15.8\,\kms for this star, which would favor that it is an SB1. HD\,17378 (A6~Ia) has RV$_{\rm PP}$ = 8\,\kms, which is large enough to consider it as potential binary. However, we only have three spectra. 

Last, we found that although HD\,14956 (B2~Ia) was classified as an SB1 with a period of $P$ = 175\,days, and ${RV}_{\rm PP}$ = 27.0\,\kms \citepads{1973ApJ...184..167A} , and \citetads{2017A&A...598A.108L} classified this star as $\alpha$\,Cygni variable, we do not see such signs of SB1 variations, as we measure RV$_{\rm PP}$ = 5.5\,\kms. However, we do not have enough spectra (three) to discard this possibility.

These results about detected spectroscopic binaries, along with the RV distributions obtained from the analysis of the best S/N spectra (i.e., obtained from a single-snapshot observation), allow us to evaluate the extent to which these distributions can be used to identify spectroscopic binaries among the outliers. We find that only four out of all the SB1/SB2 systems detected by means of multi-epoch spectroscopy are outliers in the abovementioned distributions. In addition, some outliers have not been detected as spectroscopic binaries although more than four spectra are available for them (e.g., HD\,13268, O8.5~IIIn, RV = $-106.2$\,\kms). These results can be explained when we take into account (1) that the best S/N of some of the spectroscopic binaries correspond to an orbital phase in which the RV is close to the systemic velocity, and (2) that some outliers in RV might be runaways and not necessarily binaries. The latter situation is the case of HD\,13268, a well-known runaway star (see also Sect.~\ref{subsection.51_memb}). This means that if a given star is an outlier in RV, it is useful to first investigate its runaway nature (by means of its proper motion) before marking it as a potential spectroscopic binary, and vice versa; for example, although the measured RV of the B1~Ib-II star HD\,14052 ($-90.8$\,\kms) deviates more than 3$\sigma$ from the mean, this star is not an outlier in proper motion, and so we may conclude that it is more likely a spectroscopic binary than a runaway. This is confirmed through access to multi-epoch spectroscopy. We further discuss the percentage of spectroscopic binaries in our sample of stars in Sect.~\ref{subsubsection.523_rw&bin}.

%========================================================================== 

\section{Discussion}
\label{section.5_discus}

Table~\ref{table.A4} compiles and summarizes some information of interest for the discussion about membership and final identification of spectroscopic binaries and runaway stars. Columns "$\varpi$" and "$\mu$" indicate if a given star is part of the bulk distribution of parallaxes and proper motions, respectively, or if it is detected as an outlier of these distributions (Sect.~\ref{subsection.41_plx&pm}). Columns "RV best" to "RV final" provide similar information for the case of RV estimates obtained from the best S/N spectra (Sect.~\ref{subsubsection.441_bestsnr}) for stars with four or more spectra (Sect.~\ref{subsubsection.442_multi}), or the final distribution of RVs (Sect.~\ref{subsubsection.521_pm}), respectively. In all these cases, different symbols are used to identify secure or doubtful cases.

For completeness, we also add to Table~\ref{table.A4} information about confirmed spectroscopic binaries, our final decision on cluster membership status (columns "Spec. variability" and "Member"), as well as some other comments of interest for the final interpretation of results (column "Comments").

%__________________________________________________ 

\subsection{Cluster membership}
\label{subsection.51_memb}

As discussed in Sect.~\ref{subsection.41_plx&pm}, most of the stars in the sample with good astrometry (81 stars) are grouped together in the proper motion versus parallax diagram. The mean and standard deviation of the distribution of these two quantities are $\varpi$ = 0.40 $\pm$ 0.07\,mas, and $\mu$ = 1.22 $\pm$ 0.26\,mas\,yr$^{-1}$ , respectively. All the stars that are located within the 2$\sigma$ boundaries of the distribution (64 in total) are directly considered as members and labeled with filled circles in columns "$\varpi$" and "$\mu$" of Table~\ref{table.A4}. The remaining 17 stars are marked with an open circle or a cross in Table~\ref{table.A4} depending on whether they deviate by 2\,--\,3$\sigma$ or more than 3$\sigma$, respectively. We note that in this case, columns "$\varpi$" and "$\mu$" include information about the remaining seven stars that were not included in Fig.~\ref{figure.fig5}: those labeled "unreliable astrometry" (or RUWE > 1.4). Because the information about parallaxes and proper motions is uncertain for them, we exclude these stars for the moment and mark them using brackets surrounding the corresponding symbols in columns "$\varpi$" and "$\mu$" of Table~\ref{table.A4}.

These are not the only stars with unreliable parallaxes. Figs.~\ref{figure.fig3} and \ref{figure.fig5} include a small sample of 6 K/M-type supergiants that despite a RUWE value well below 1.4 have larger errors than the rest of stars in the sample, and interestingly, all of them are systematically shifted to larger parallaxes (although except for one, all have total proper motions within the 2$\sigma$ boundaries and proper motion vectors compatible with the bulk of member stars, Fig.~\ref{figure.fig4}). They are also all marked with brackets in Table~\ref{table.A4}.

This is likely connected to an already known problem that affects the reliability of the {\em Gaia} DR2 astrometric solution. In brief, as pointed out by \citetads{2011A&A...532A..13P} and \citetads{2018A&A...617L...1C}, the position of the centroid changes on timescales of several months or a few years because of the large size and strong intrinsic photocentric variability of red supergiants. This effect leads to unreliable parallaxes and errors. 

A particular example of interest regarding this issue with the astrometric solution of {\em Gaia} for the case of red supergiants is the the highly variable star HD~14528 (S~Per, $\varpi_{Gaia}$\,=\,0.25\,$\pm$ 0.12\,mas, $\mu_{total\,Gaia}$ = 2.57 $\pm$ 0.31\,mas/yr). This star has an average angular size of 6.6\,mas (\citeads{2012A&A...546A..16R}). It was monitored for six years by \citetads{2010ApJ...721..267A} with VLBI. The authors obtained an independent parallax of 0.413$\pm$0.017~mas, which is just at the center of the distribution. We therefore cannot discard completely  that these six K/M-type supergiants, which are outliers in parallax using data from {\em Gaia} DR2, are members
of Per~OB1.

The last star that we place in brackets is the A-type supergiant HD\,14489. This is the brightest stars in our sample, with $G_{mag}$ = 5.1. As shown in Fig.~\ref{figure.fig3}, this star also has much larger errors in parallax and proper motions than the bulk of stars in the sample. This may be related to the current limitation of {\em Gaia} DR2 regarding the reliability of the astrometric solutions for stars brighter than $G_{mag} \lesssim$ 6 \citepads{2018A&A...616A...2L}. Another three stars share this issue, but their associated astrometric errors are much smaller and their magnitudes are close to $G_{mag}$ = 6; therefore we decided to consider their astrometric solutions reliable.

Taking all this information into account, we decided to following strategy below to evaluate the membership to Per~OB1 of each star in our sample. Stars with reliable values on parallax and proper motion (i.e., not marked with parentheses in columns 2 and 3 of Table~\ref{table.A4}) are considered as {\em \textup{confirmed members}} if they do not deviate more than 2$\sigma$ from the mean of the distribution of parallaxes. For stars with unreliable values of parallax and proper motion (i.e., highlighted with brackets in columns "$\varpi$" and "$\mu$" of Table~\ref{table.A4}), we adopted the following: if they are not outliers in parallax, they are considered {\em \textup{likely members}}; if they are outliers in parallax, we consider them {\em \textup{candidate members}}, except for the K/M-type stars, which remain likely members because of the arguments provided above. Last, stars with reliable astrometry that are outliers in parallax (as well as those stars in IC\,1805, see below) are considered {\em \textup{nonmembers}}.

Most of the stars are properly classified using these criteria. However, a few cases deserve further attention.
\paragraph{HD~13022 (O9.7~III) and HD~12842 (F3~Ib):} These two stars are classified as members following the guidelines above, but they are outliers in proper motion (Fig~\ref{figure.fig5}). Interestingly, they have a very small proper motion compared to the rest of the stars in the sample (see in Fig.~\ref{figure.fig4} the two stars with very small vectors located at (RA, DEC)\,$\sim$\,(32, 58.5)\,deg). Awaiting a more detailed study of these two stars, we continue considering them members for the moment.
\paragraph{HD~16691 (O4~If), HD~15642 (O9.5~II-IIIn), HD~13745 (O9.7~II(n)), and HD\,13268 (ON8.5~IIIn):} These four O-type stars are clear outliers in proper motion (see Figs.~\ref{figure.fig5} and \ref{figure.fig4}). We consider the first three runaway members  because their parallaxes lie within the 2$\sigma$ boundaries. The fourth (HD~13268) is an interesting case; although this star has a somewhat larger parallax, it has a RV of $\sim$ 105\,\kms. Therefore, given its spectral classification and this high RV pointing to us, it can still be considered a runaway member of Per~OB1. This star is a well-known fast-rotating nitrogen-rich O-type runaway star (e.g., \citeads{1972AJ.....77..138A}; \citeads{1989A&AS...81..237M}; \citeads{2014A&A...562A.135S}; \citeads{2015A&A...578A.109M}; \citeads{2017A&A...603A..56C,2017A&A...604A.123C})
\paragraph{HD~14322 (B8~Iab):} This star is an outlier in parallax with a value of $\varpi$ = 0.21$\pm$0.04\,mas. Although the TGAS catalog provides a value for it of $\varpi$ = 0.44$\pm$0.38\,mas (within the boundaries of $\varpi$), the error is much larger. This inconsistency caused us to modify its status from nonmember to member candidate while awaiting {\em Gaia} DR3.

\paragraph{HD~14489 (A1~Ia):} This is a bright A-type star ($G_{mag}$ = 5.1), outlier in parallax, and with the largest parallax error. Although it has a RUWE = 0.81, we do not trust its {\em Gaia} astrometry, as explained before, because of its brightness. The result from TGAS provides a parallax of $\varpi$ = 0.45 $\pm $0.94\,mas, and although it is within the adopted boundaries of Per~OB1, the error is very large. This star is also an outlier in RV and close to the 2$\sigma$ boundary in proper motion. Therefore we decide to label it a runaway member candidate.

\paragraph{BD+56724 (M4-M5~Ia-Iab):} This star has the largest parallax in Fig.~\ref{figure.fig5}, and a RUWE = 0.93. Although the reliability of {\em Gaia} DR2 parallaxes for the K/M-type stars may be low, its large deviation from the mean of the distribution could mean that this star is not a member. It is also an outlier in proper motion, but its magnitude and RV are similar to other red supergiants in the sample. We therefore retain this star as member candidate for the moment.

\paragraph{HD\,15570 (O4~If), HD15558 (O4.5~III(f)), HD16429 (O9~II(n)), and BD~+60493 (B0.5~Ia):} All these stars are located within or in the surroundings of IC\,1805. Interestingly, all of them but one are O-type stars. Although they are located within the 2$\sigma$ boundaries of the parallax and proper motion distribution (except for HD~16429, but this is a triple system with a RUWE = 8.8), we decided to mark them nonmembers based on their separated location in the sky and their direct connection with the surrounding H~{\sc ii} region. They seem to be linked to a younger star-forming region located at higher galactic latitudes (but at the same distance). Most of them are also outliers in RV (see Table~\ref{table.A4}), but this is likely due to their binary nature. 

The final result of this classification, also taking into account the comments on some individual stars presented above, is summarized in column "Member" of Table~\ref{table.A4}. In total, we have 70 confirmed members, 9 likely members, 5 member candidates, and 4 nonmembers. Interestingly, only stars in IC\,1805 are finally classified nonmembers. The remaining 84 stars likely belong to the Per~OB1 association (although some of them are identified as runways, see Sect.~\ref{subsubsection.523_rw&bin}).

%__________________________________________________ 

\subsection{Kinematics.}
\label{subsection.52_kinem}

In Sect.~\ref{subsection.41_plx&pm} and \ref{subsection.44_rv} we provided a global overview of the results about proper motions and RVs for the complete sample of stars, also including some information about identified spectroscopic binaries. In this section we discuss these results more in detail. We also refer to  \citetads{2017MNRAS.472.3887M, 2019A&A...624A..34Z, 2020MNRAS.493.2339M} for complementary (and in some cases more detailed) information about the global and internal kinematical properties of stars in the Per~OB1 association.

\begin{figure*}[t!]
\centering
\includegraphics[width=.95\textwidth]{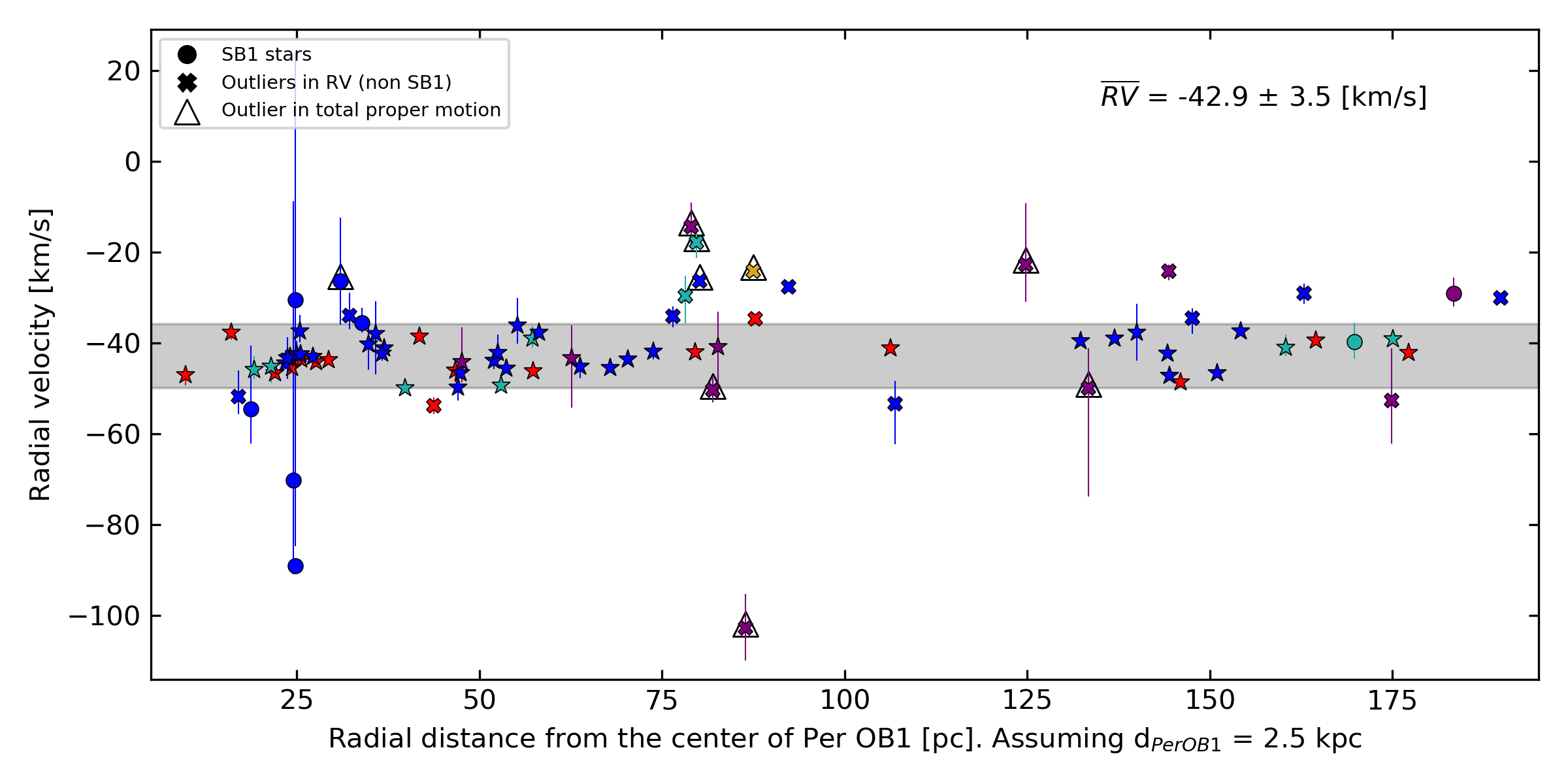}
\caption{Radial distance from the center of Per~OB1 against the mean RV for the stars with more than one spectrum, or RV for the stars with only one spectrum, excluding the SB2 binaries. The stars within 1\,degree from IC\,1805 were also excluded. The filled gray rectangle shows 2$\sigma$ of all the RVs, excluding those from stars identified as binaries (SB1 or SB2). The filled colored circles denote stars identified as SB1. The filled colored crosses denote starts outside 2$\sigma$ from the mean that are not identified as binaries. The colored stars show the remaining stars. The colored error bars present RV$_{\rm PP}$ for stars with multi-epoch data, except for SB1. The open triangles show stars that are outliers in proper motion and were therefore identified as runaways.} 
\label{figure.fig10}
\end{figure*}

%_ _ _ _ _ _ _ _ _ _ _ _ _ _ _ _ _ _ _ _ _ _ _ _ _ _

\subsubsection{Proper motions}
\label{subsubsection.521_pm}

Figure~\ref{figure.fig5} and the top panel of Fig.~\ref{figure.fig4} provide a global overview of the distribution of proper motions in the whole star sample. These figures show that (except for a few outliers) most of the stars in our sample that are located below the Galactic plane (among them, those in $h$~and~$\chi$ Persei) can be considered a dynamically connected population of stars. This result perfectly agrees with previous findings by \citetads{2008ApJ...679.1352L}. Using proper motions from the {\em Hipparcos} mission, these authors showed that the luminous members of the Per~OB1 association exhibit a bulk motion away from the Galactic plane, such that their average velocity increases with height above the Galactic plane.

Furthermore, inspection of the results for proper motions (relative to h\,and\,$\chi$~Persei) and parallaxes presented in Fig.~\ref{figure.fig6} and Table~\ref{table.radiplxpm} allows us to conclude that the distributions of parallaxes associated with stars located at increasing distances from the center of the double cluster are all compatible (at least we find no clear subgrouping in terms of parallax at least given the accuracy of {\em Gaia} DR2 astrometry -- except maybe the stars in $h$~and~$\chi$~Persei because their parallax is  somewhat larger or their distance is somewhat closer). We also conclude that the mean and standard deviation of the distribution of proper motions in $h$~and~$\chi$~Persei is much smaller than in the more extended population. 

Moreover, the spatial distribution of proper motions in the extended population of blue and red supergiants in Per~OB1 does not follow an expanding structure centered in the $h$~and~$\chi$~Persei double cluster. Instead, the local proper motions of most of the stars located north of these clusters seem to point outwards from an imagined center located at about 1 degree north of the double cluster (see also \citeads{2018ARep...62..998M}, \citeads{2020MNRAS.493.2339M}). These results from the proper motion, linked with the results by \citetads{2008ApJ...679.1352L} mentioned above, are compatible with a scenario in which the halo population of blue and red supergiants around the double cluster has been formed from a more diffuse region of interstellar material compared to the denser region associated with the clusters themselves. 

In addition, four O-type stars south of the region can be clearly considered runaways based on the size and direction of their proper motions (see also Fig.~\ref{figure.fig5}). Interestingly, their proper motion vectors do not point outward from $h$~and~$\chi$~Persei, but to a far more extended region of the Galactic plane (see also the discussion in Sect.~\ref{subsubsection.523_rw&bin}). Finally, as also indicated in Sect.~\ref{subsection.51_memb}, the stars located within or near IC~1805 likely belong to a younger population of stars that is not necessarily connected with the remaining stars in Per~OB1.

%_ _ _ _ _ _ _ _ _ _ _ _ _ _ _ _ _ _ _ _ _ _ _ _ _ _

\subsubsection{Radial velocities}
\label{subsubsection.522_rv}

Figure~\ref{figure.fig8} and the bottom panel of Fig.~\ref{figure.fig4} summarize the RV results obtained with the best S/N spectra. The analysis of these spectra has allowed us to characterize the RV distributions for the different SpT groups and to identify potential spectroscopic binary systems and runaway stars among the outliers of the distributions (see columns "RV best" and "RV multi" in Table~\ref{table.A4}). We then illustrated in Sect.~\ref{subsubsection.442_multi} (see also Fig.~\ref{figure.fig9}) the importance of incorporating information from the analysis of multi-epoch observations for the correct interpretation of the RV distribution, and in particular, to avoid the spurious identification of spectroscopic binaries (either from a single epoch or from multi-epoch observations) due to the effect on the measured RVs of the intrinsic variability caused by stellar oscillation and/or wind variability in the blue supergiant domain.

Last, we learned that after eliminating outliers associated with confirmed spectroscopic binaries (via multi-epoch spectroscopy) and runways (via proper motions), the RV distributions for the B, A/F, and K/M~Sgs are fairly compatible in terms of mean values and standard deviations. In addition, we found that most of the O-type stars in the sample are either (1) runaways, as detected from the proper motions, (2) spectroscopic binaries, or (3) are considered nonmembers because they are located nearby IC\,1805, far away from the main distribution of stars in Per~OB1. As a result, the RV distribution of the O-type sample is remarkably broader than for those associated with the other SpT.

We now take all these results into account to provide final information about RVs in Fig.~\ref{figure.fig10} and in the column "RV final" of Table~\ref{table.A4}. To do this, we first replaced the list of measurements obtained from the best S/N spectra by the mean value resulting from the analysis of the multi-epoch observations for those cases for which we have more than one spectrum. Then we used this list of values, except for all the SB2 binaries, and the stars identified as nonmembers (see Sect.~\ref{subsection.51_memb}), to obtain the mean and standard deviation by performing an iterative 2$\sigma$ clipping. 

\begin{table*}
  \centering
  \caption[]{Summary of the number of outliers in proper motion and RV that are used for the final identification of runaway stars. In the case of the proper motion, we indicate cases that deviate by more than 2$\sigma$ from the mean of the distribution for each individual component and the total proper motion. In the case of RV, we separate cases that deviate by more than 2$\sigma$ and 4$\sigma$, respectively. In parentheses, we indicate targets whose outlier characteristic is not entirely clear from the available data. The last column indicates the final number and percentage of clearly detected runaways for each SpT group.}
    \label{table.rw}
      \begin{tabular}{lccccccccc}
\hline
\hline
\noalign{\smallskip}
SpT & \multicolumn{3}{c}{PM} & & \multicolumn{2}{c}{RV} & & \multicolumn{2}{c}{Runaways} \\ %
\cline{2-4} \cline{6-7} \cline{9-10}
\noalign{\smallskip}
      &  $\mu_{\alpha}\cos{\delta}$      & $\mu_{\delta}$   & $\mu_{\rm Total}$ &  &   $>$2$\sigma$ &   $>$4$\sigma$ &  & \# & \% \\ %
\noalign{\smallskip}
\hline
\noalign{\smallskip}
O-type   &  3  &  5    & 5    & & 7~(+1) & 3 & & 5 & 45 \\ % 
B-type   &  1  &  2    & 2    & & 9~(+6) & 1 & & 2 &  5 \\ %
A/F-type &  1  &  1(+1)& 1(+1)& & 3 & 1 & & 1(+1) &  1(+1) \\ %
K/M-type & (1) & (1)   & (1)  & & 2 & 0 & & (1) &  (5) \\ %
\noalign{\smallskip}
\hline
      \end{tabular}
\end{table*}

The results of this process are presented in Fig.~\ref{figure.fig10}, where the RVs of all stars that were not excluded from the list are presented against the radial distance from the center of Per~OB1. The obtained mean and standard deviation are shown in the top right corner ($-42.9$ $\pm$ 3.5\,\kms) of the figure, and the horizontal gray band indicates the 2$\sigma$ boundaries. 

Although most of the stars in the sample are concentrated within the central 100\,pc, we observe that except for a few cases, the remaining stars also lie within the 2$\sigma$ boundaries. Therefore, once more, and as was suggested by \citetads{2010ApJS..186..191C} and \citetads{2019A&A...624A..34Z}, the extended population of blue and red Sgs in Per~OB1 (up to 200~pc, i.e., relatively far away from $h$~and~$\chi$~Persei) seems to have a common origin in terms of kinematics. No global gradient (as a function of distance to the center of the association) or local substructures are observed in the distribution of RVs. 

As indicated above, we obtain $\overline{RV}$ = $-42.9$ $\pm$ 3.5\,\kms\ using the whole sample of stars that are not excluded from the list. Regarding $h$~and~$\chi$~Persei, we obtained average values of RV$_{\rm \chi\,Per}$ = $-44.4$ $\pm$ 1.4\,\kms and RV$_{\rm h\,Per}$ = $-41.1$ $\pm$ 2.6\,\kms, respectively. These results agree well with those previously obtained by other authors. For the association as a whole, \citetads{2017MNRAS.472.3887M} provided a mean value of $\overline{RV}$ = $-43.2$ $\pm$ 7.0\,\kms\ using available information of member stars from the TGAS catalog. For the individual clusters, \citetads{1991AJ....102.1103L} provided RV$_{\rm \chi\,Per}$ = $-44.4$ $\pm$ 0.7\,\kms and RV$_{\rm h\,Per}$ = $-46.8$ $\pm$ 1.7\,\kms, respectively, using a sample of cluster stars (mainly early type, more specifically, B- and A-type stars). In the particular case of $h$\,Per, the fact that we have only three suitable stars to compute the mean may explain the poorer agreement. 

As in the case of the analysis of the best S/N spectra and the multi-epoch observations, we provide in column "RV final" of Table~\ref{table.A4} a list of identifiers to separate the outliers of the distribution of final values of RV from stars within the 2$\sigma$ boundaries. This information is used in the next section to determine additional SB1 stars that have not previously been identified based on the available multi-epoch spectra.    

%_ _ _ _ _ _ _ _ _ _ _ _ _ _ _ _ _ _ _ _ _ _ _ _ _ _

\subsubsection{Runaway and binary stars.}
\label{subsubsection.523_rw&bin}

\begin{table*}
  \centering
  \caption[]{Summary of the number of binary stars in the sample (see Table~\ref{table.A4}). For columns "SB1" and "SB2", the percentage shows the fraction with respect to the total number for each SpT. We split the B-type stars into two groups to separate giants from supergiants. Column "Lit." counts the number of binary stars found in the literature. Column "SB1?" counts the sum of the stars labeled "LPV/SB1?" in column "Spec. variability" and "SB1?" in column "Comments" (we note that if a star is labeled both as "LPV/SB1?" and "SB1?", we only count the first). The total number of stars are in column N$_{\rm All}$. Column "\% bin" gives the percentage of total and potential binary stars with respect to the total number of stars.} 
    \label{table.bin}
      \begin{tabular}{cccccccc}
\hline
\hline
\noalign{\smallskip}
SpT & SB1 & SB2 & Lit. & SB1? & N$_{\rm All}$ & \% bin. \\ %
\noalign{\smallskip}
\hline
\noalign{\smallskip}
O             & 0        & 2~(15\%)   & 1 & 1 & 13 & 15 -- 30\% \\ % 
B~I \& II     & 3~(8\%)  & 1~(3\%)    & 1 & 5 & 37 & 10 -- 27\% \\ %
A/F          & 0         & 0          & 0 & 2 & 11 & 0 -- 18\% \\ %
K/M          & 0         & 0          & 0 & 2 & 18 &  0 -- 10\% \\ %
\noalign{\smallskip}
\hline
\noalign{\smallskip}
B~III         & 2~(22\%) & 2~(22\%)   & 0 & 1 &  9 & 45 -- 55\% \\ %
\noalign{\smallskip}
\hline
      \end{tabular}
\end{table*}

The last two columns of Table~\ref{table.rw} summarize the final number and percentage of identified runaways in each of the four SpT groups. We proceed as follows to obtain this. First, we assigned the runaway status to all outliers in proper motion (meaning that the magnitude of any of their individual components, or the total proper motion, deviates by more than 2$\sigma$ from the mean of the corresponding distribution). Then we considered the possibility of identifying additional runaway stars through their RVs. In this case, we decided to only label them (if not detected as SB2) clear runaways if their RV deviates by more than 15~\kms from the mean of the final RV distribution presented in Fig.~\ref{figure.fig10}. We made this decision based on two arguments: the first refers to the result presented in the fifth column of Table~\ref{table.rw}; namely, the number of identified runaways when all outliers in RV are considered that deviate by more than 2$\sigma$ is too large when compared with those detected through PMs. The second argument is based on the results of RV$_{\rm PP}$ that is expected to be produced by intrinsic variability, which can be as high as 10\,--\,20~\kms\ in some cases (see Table~\ref{table.variab}).

These arguments are supported by the fact that it is very unlikely to find a runaway star that is an outlier in RV, but not in at least one of the components of the proper motion. In contrast, as described above, intrinsic variability can lead to single-snapshot RV measurement that can easily deviate by up to 10\,--\,20~\kms (or, equivalently, about 4\,--\,5$\sigma$ in this specific sample of stars). Furthermore, this situation can be even more dramatic for large-amplitude SB1 systems for which only a low number of spectra is available. It is therefore more likely that a star that is not an outlier in proper motion but is an outlier in RV is an SB1 than a runaway. Alternatively, if the deviation in RV is smaller than the typical intrinsic variability corresponding to the associated SpT, it might not even be a binary star.   

A practical example of the latter situation is the BN2~II-III star BD\,+56578, for which we only have three spectra that cover a very short time-span (one day). This star is not an outlier in proper motion, but is an outlier in RV (it deviates by 13$\sigma$ from the mean of the RV distribution). Based on what we have described above, this star should be labeled as a potential binary, a suspicion that is confirmed from the literature (\citeads{2017A&A...598A.108L}).

Following these arguments, all stars in Fig.~\ref{figure.fig10} whose RV measurements deviate by more than 2$\sigma$ (i.e., which lie outside the gray band) and up to 10\,--\,20~\kms and that have not previously been detected as runaways through proper motions (open triangles) or as spectroscopic binaries through multi-epoch spectroscopy (filled circles) are quite likely single pulsating stars.

Overall, we identify a total of 11 runaway stars. The group of stars with a larger number of runaways (45\%) are the O-type stars. This is followed by the B and K/M Sgs, with 5\% each. (We note, however, that the runaway status of the M Sg BD\,+56724 can be a spurious result because it is based on the $Gaia$-DR2 proper motion, which may not be as reliable as for the other stars because of the problems regarding size and variability of the red Sgs.) Last, the lower percentage of runaways is found for the A Sgs, with only 1 or 2\%, depending on whether we trust the $Gaia$-DR2 proper motion of the bright star HD14489, which also has a much larger parallax than the remaining stars in Per\,OB1.

It thus becomes clear again that a high percentage of the O-type stars in the Per~OB1 region can be considered a dynamically distinct group. However, in contrast to previous assumptions (see, e.g., \citeads{2002AJ....124..507W}), the fact that all of them are found within the 2$\sigma$ boundaries of the parallax distribution indicates that they belong to the same grouping as the remaining blue and red supergiants in Per~OB1, and not to a more distant, dispersed association. Although further confirmation is needed, the most likely origin of the O-star runaways is a dynamical kick by a supernova explosion in a previously bounded binary system. This hypothesis is reinforced by the fact that none of the detected runaways are identified as binary systems (and the other way round).  

Table~\ref{table.bin} summarizes the results for the detected binaries, again separated by SpT group, and this time differentiating the B Sgs from the B Gs because they represent the evolutionary descendants of main-sequence stars in two different mass domains. We refer to Sect.~\ref{subsubsection.442_multi} for a description of how the SB1 and SB2 stars where identified. The targets labeled "SB1?" include targets fulfilling any of the two following criteria. On the one hand, stars with five or more spectra for which we cannot clearly decide whether the detected line-profile variability is due to intrinsic variability or orbital motion. On the other hand, following the arguments above, we identified stars as "SB1?" whose RV$_{\rm PP}$ is larger than the typical intrinsic variability expected for their SpT (see Fig.~\ref{figure.fig9}).

The main conclusions from inspection of the results presented in Table~\ref{table.bin} (and Table~\ref{table.rw}) are summarized as follows. First, the percentage of detected spectroscopic binaries decreases toward later SpT, or equivalently as the massive star evolution proceeds. This result agrees with recent findings by \citetads{2017IAUS..329...89B}; \citetads{2019A&A...624A.129P,2020A&A...635A..29P}; Sim\'on-D\'iaz et al. (2020, subm.). When we assume that the detected runaways indicate a past binary evolution, the total percentage of clear binaries (excluding those labeled "SB1?") would decrease from $\sim$60\% to $\sim$15\% when the O star and B~Sg samples are compared, and further below $\sim$5\% when the cooler Sgs are considered. Second, while the decreasing tendency remains in both cases, the exact behavior of the percentage of detected spectroscopic binaries is different depending on whether we also include the stars labeled "SB1?" stars. Therefore it is critical to confirm or dismiss our suspicion that most of the stars with RV$_{\rm PP}$ below 10-15~\kms\ are actually single pulsating stars and not spectroscopic binaries. Access to multi-epoch data for the whole sample of star is therefore crucial to obtain reliable empirical information about the relative percentage of binaries throughout the massive star evolution. Finally, as an aside, the percentage of spectroscopic binaries is much higher among the B Gs than in the B Sgs.

This clearly shows that any further attempt to interpret the empirical properties of this sample of massive stars in an evolutionary context must take into account that a large fraction of the O stars is or likely has been part of a binary or multiple system. In addition, some of the other more evolved targets may also have been affected by binary evolution.

%========================================================================== 

\section{Summary and future prospects.}
\label{section.6_summary}

Our study has provided all the necessary environmental information that will be used in a forthcoming paper, in which we will also incorporate results obtained from a quantitative spectroscopic analysis of the whole sample (including stellar parameters and surface abundances) to perform a complete homogeneous characterization of the physical and evolutionary properties of the massive star population of the Per~OB1 association.

In this paper, we have studied a sample of 88 massive stars located within 4.5\,deg from the center of the Per~OB1 association using high-resolution multi-epoch spectroscopy, and astrometric information from the {\em Gaia} second data release (DR2). 

We have investigated membership of all star in the sample to the Per\,OB1 association, resulting in 70 members, 9 likely members, and another 5 candidates that require further investigation, while the other 4 were considered nonmembers as they belong to IC\,1805.

We have found eight clear and two likely runaway stars, most of them O-type stars. We also identified 5 SB1 and five SB2 stars (these include three and one new binary systems, respectively), plus another 11 potential SB1 stars that we propose are single pulsating stars. 

To obtain these results, we took their parallaxes and proper motions (as compiled from {\em Gaia} DR2) into account, and the RV estimates obtained from the available multi-epoch and/or single snapshot spectra. In addition, we also considered the reliability of the astrometry provided by {\em Gaia} through the RUWE value, the potential decrease in reliability of {\em Gaia} astrometry in the case of the red Sgs because of their large size and photocentric variability, and the expected amplitude of spectroscopic variability produced by stellar pulsations and/or wind variability when spectroscopic binaries are identified based on their RV measurements. 

We have also analyzed some global properties of the sample and obtained averages in parallax, total proper motion, and RV of $\varpi$ = 0.40 $\pm$ 0.07\,mas, $\mu$ = 1.22 $\pm$ 0.26\,mas\,yr$^{-1}$ ($\mu_{\alpha}\cos{\delta}$ = -0.50 $\pm$ 0.48, $\mu_{\delta}$ = -0.99 $\pm$ 0.31), and $-42.9$ $\pm$ 3.5\,\kms. All these results agree relatively well with previous studies based on different stellar samples comprising the Per~OB1 association (some of them focused on the h~and~$\chi$ Persei clusters). 

Generally speaking, no important differences are detected in the distribution of parallaxes, proper motions, and RVs when stars in h~and~$\chi$ Persei or the full sample are considered, which suggests a very extended dynamically interrelated population. However, a few clear outliers in the proper motion and RV distributions are also found. A large fraction of these are O-type stars (almost 50\%). The further analysis of their proper motions and RVs indicates that they are runaway stars, probably resulting from the kick of a supernova explosion in a previously bounded binary system.

Finally, we have found that the percentage of secure binaries decreases from the hotter to the cooler Sgs. In particular, this percentage decreases from 15\% to 10\% when the O star and B~Sg samples are compared (or alternatively, from 60\% to 15\% when we consider the runaway stars as previous binaries), and it practically vanishes in the A/F and K/M Sgs. Further investigation of the potential connection between this result and merging processes that occur during the evolution of massive stars is an interesting direction of future work.

%========================================================================== 

\begin{acknowledgements}

Based on observations made with the Nordic Optical Telescope, operated by NOTSA, and the Mercator Telescope, operated by the Flemish Community, both at the Observatorio de El Roque de los Muchachos (La Palma, Spain) of the Instituto de Astrof\'isica de Canarias. We acknowledge funding from the Spanish Government Ministerio de Ciencia e Innovación through grants PGC-2018-091\,3741-B-C211/C22, SEV 2015-0548, and CEX2019-000920-S and from the Canarian Agency for Research, Innovation and Information Society (ACIISI), of the Canary Islands Government, and the European  Regional Development Fund (ERDF), under grant with reference ProID2017010115. This research made use of the SIMBAD, operated at Centre de Donn\'ees astronomiques de Strasbourg, France, and NASA's Astrophysics Data System. The background images were taken from The STScI Digitized Sky Survey (\href{http://archive.stsci.edu/dss/copyright.html}{Copyright link)}.

\end{acknowledgements}

%%%%%%%%%%%%%%%%%%%%%%%%%%%%%%%%%%%%%%%%%%%%%%%%%

\bibliographystyle{aa} % style aa.bst
\bibliography{aa_mybib} % the references 

%%%%%%%%%%%%%%%%%%%%%%%%%%%%%%%%%%%%%%%%%%%%%%%%%

\appendix

\section{Long tables}

\scriptsize{
\begin{longtable}{llccccccccl}
\label{table.A1}\\ %
\caption[]{Sample of (super-)giant stars within 4.5 degrees around the Perseus double cluster.}\\ %

\hline
\hline
\noalign{\smallskip}
Name    &SpC    &$\alpha$       &$\delta$       &G      &BP-RP  &$\mu_{\alpha}\cos{\delta}$     &$\mu_{\delta}$         &$\varpi$         &RV                             & Notes \\ %
            &           &(J2000)        &(J2000)        &[mag]  &[mag]  &[mas yr$^{-1}$]                      &[mas\,yr$^{-1}$]       &[mas]          &[km\,s$^{-1}$] &       \\ %
\noalign{\smallskip}
\hline
\noalign{\smallskip}
\endfirsthead

\caption[]{Sample of (super-)giant stars within 4.5 degrees around the Perseus double cluster.}\\ %

\hline
\hline
\noalign{\smallskip}
Name    &SpC    &$\alpha$       &$\delta$       &G      &BP-RP  &$\mu_{\alpha}\cos{\delta}$     &$\mu_{\delta}$         &$\varpi$         &RV                             & Notes \\ %
            &           &(J2000)        &(J2000)        &[mag]  &[mag]  &[mas yr$^{-1}$]                      &[mas\,yr$^{-1}$]       &[mas]          &[km\,s$^{-1}$] &       \\ %
\noalign{\smallskip}
\hline
\noalign{\smallskip}
\endhead

\noalign{\smallskip}
\hline
\endfoot

\multicolumn{11}{c}{\textbf{O giants (III), bright giants (II), and supergiants (I)}} \\\noalign{\smallskip}
HD15570   & O4If          & 02:32:49.420 & +61:22:42.09 & 7.9 & 1.1 & -0.612 $\pm$ 0.04  & -0.847 $\pm$ 0.063 & 0.429 $\pm$ 0.035 & -43.4 $\pm$ 3.8 & IC\,1805 \\ %
HD16691   & O4If          & 02:42:52.026 & +56:54:16.46 & 8.5 & 0.7 & -2.052 $\pm$ 0.078 & -3.34  $\pm$ 0.085 & 0.470 $\pm$ 0.043 & -42.8 $\pm$ 7.5 & \\ %
HD14947   & O4.5If        & 02:26:46.990 & +58:52:33.12 & 7.9 & 0.7 & +0.735 $\pm$ 0.083 & -0.764 $\pm$ 0.109 & 0.295 $\pm$ 0.048 & -41.2 $\pm$ 5.9 & \\ %
HD15558$^{e}$& O4.5III(f) & 02:32:42.537 & +61:27:21.58 & 7.8 & 0.8 &(-0.441 $\pm$ 0.097)&(-1.488 $\pm$ 0.15) &(0.496 $\pm$ 0.086)& -83.3           & IC\,1805 \\ %
          &               &              &              &     &     &                    &                    &                   & 101.1           & \\ %
HD14442   & O5n(f)p       & 02:22:10.700 & +59:32:58.89 & 9.1 & 0.6 & -0.765 $\pm$ 0.09  & -1.031 $\pm$ 0.067 & 0.286 $\pm$ 0.034 & -33.1 $\pm$ 5.2 & \\ %
HD17603   & O7.5Ib(f)     & 02:51:47.798 & +57:02:54.47 & 8.2 & 0.9 & +0.307 $\pm$ 0.119 & -0.653 $\pm$ 0.096 & 0.419 $\pm$ 0.045 & -25.6 $\pm$ 2.2 & \\ %
HD13268   & ON8.5IIIn     & 02:11:29.700 & +56:09:31.72 & 8.1 & 0.2 & +0.31  $\pm$ 0.091 & -2.457 $\pm$ 0.091 & 0.621 $\pm$ 0.047 & -106.2$\pm$ 0.8 & \\ %
HD16429   & O9II(n)       & 02:40:44.975 & +61:16:56.07 & 7.6 & 0.9 &(-12.367$\pm$ 0.457)&(-1.265 $\pm$ 0.46)&(-0.179 $\pm$ 0.384)& -55.5 $\pm$ 0.7 & IC\,1805 \\ % Negative plx
HD16832   & O9.2III       & 02:44:12.716 & +56:39:27.25 & 8.7 & 0.6 & +0.107 $\pm$ 0.073 & -1.548 $\pm$ 0.078 & 0.447 $\pm$ 0.042 & -26.2 $\pm$ 0.8 & \\ %
HD15642   & O9.5II-IIIn   & 02:32:56.383 & +55:19:39.06 & 8.5 & 0.1 & -1.372 $\pm$ 0.075 & -3.021 $\pm$ 0.079 & 0.271 $\pm$ 0.041 & -18.7 $\pm$ 7.4 & \\ %
HD13831   & O9.7III       & 02:16:39.218 & +56:44:16.09 & 8.2 & 0.2 & -0.74  $\pm$ 0.083 & -1.54  $\pm$ 0.09  & 0.305 $\pm$ 0.048 & -48.6 $\pm$ 6.9 & \\ %
HD13745   & O9.7II(n)     & 02:15:45.935 & +55:59:46.73 & 7.8 & 0.3 & +0.293 $\pm$ 0.096 & -4.302 $\pm$ 0.12  & 0.471 $\pm$ 0.053 & -18.1 $\pm$ 0.7 & \\ %
HD13022   & O9.7III       & 02:09:30.071 & +58:47:01.56 & 8.7 & 0.5 & +0.075 $\pm$ 0.1   & -0.407 $\pm$ 0.095 & 0.443 $\pm$ 0.047 & -47.7 $\pm$ 1.3 & \\ %

\noalign{\smallskip}
\hline
\noalign{\smallskip}

\multicolumn{11}{c}{\textbf{B giants (III), bright giants (II), and supergiants (I)}} \\\noalign{\smallskip}
HD13402$^{e}$& B0II:+B0   & 02:12:57.257 & +59:32:18.78 & 7.9 & 0.9 & -1.269 $\pm$ 0.069 & -0.479 $\pm$ 0.057 & 0.358 $\pm$ 0.034 & +57.4           & \\ %
          &               &              &              &     &     &                    &                    &                    & -164.8          & \\ %
HD14331   & B0III         & 02:20:51.405 & +55:49:31.34 & 8.4 & 0.2 & -0.023 $\pm$ 0.079 & -1.865 $\pm$ 0.085 & 0.375 $\pm$ 0.039 & -27.5 $\pm$ 1.1 & \\ %
HD16808   & B0.2Ib        & 02:44:06.775 & +58:19:34.58 & 8.4 & 0.8 & +0.144 $\pm$ 0.073 & -0.894 $\pm$ 0.069 & 0.389 $\pm$ 0.038 & -39.0 $\pm$ 0.7 & \\ %
HD13036   & B0.2III       & 02:09:37.838 & +59:37:13.41 & 8.3 & 0.7 & -1.135 $\pm$ 0.072 & -0.755 $\pm$ 0.073 & 0.528 $\pm$ 0.041 & -48.4 $\pm$ 0.8 & \\ %
BD+60493  & B0.5Ia        & 02:30:51.355 & +61:10:39.97 & 8.1 & 1.2 & -1.138 $\pm$ 0.042 & -0.068 $\pm$ 0.068 & 0.474 $\pm$ 0.039 & -26.9 $\pm$ 0.6 & IC\,1805 \\ %
HD13716   & B0.5III       & 02:15:39.407 & +57:45:47.95 & 8.2 & 0.5 & -1.4   $\pm$ 0.086 & -0.995 $\pm$ 0.093 & 0.420 $\pm$ 0.044 & -47.5 $\pm$ 0.6 & \\ %
HD13970$^{e}$& B0.5:III:  & 02:17:46.340 & +56:38:28.92 & 8.5 & 0.3 &(+0.656 $\pm$ 0.261)&(-1.290 $\pm$ 0.180)&(0.768 $\pm$ 0.134)& +118.7          & \\ %
          &                   &              &              &     &     &                    &                    &                   & -189.8          & \\  %
HD14053   & B0.7II        & 02:18:23.048 & +57:00:36.72 & 8.4 & 0.4 &(-0.036 $\pm$ 0.228)&(-0.991 $\pm$ 0.253)&(0.319 $\pm$ 0.123)& -37.0 $\pm$ 0.3 & $h$~Per \\ %
HD13969   & B0.7III       & 02:17:49.843 & +57:05:25.57 & 8.8 & 0.4 &(+0.743 $\pm$ 0.353)&(-2.284 $\pm$ 0.358)&(0.451 $\pm$ 0.194)& -12.3 $\pm$ 0.6 & $h$~Per \\ %
HD13854   & B1Ia-Iab      & 02:16:51.716 & +57:03:18.88 & 6.4 & 0.5 & -0.549 $\pm$ 0.079 & -1.033 $\pm$ 0.102 & 0.352 $\pm$ 0.041 & -41.4 $\pm$ 0.8 & \\ %
HD13659   & B1Ib          & 02:15:06.497 & +56:55:35.49 & 8.5 & 0.7 & -0.708 $\pm$ 0.091 & -1.361 $\pm$ 0.076 & 0.476 $\pm$ 0.039 & -52.6 $\pm$ 0.2 & \\ %
HD15571   & B1Ib          & 02:32:24.779 & +57:25:44.40 & 8.2 & 0.8 &(-0.934 $\pm$ 0.338)&(-1.635 $\pm$ 0.355)&(0.106 $\pm$ 0.160)& -45.4 $\pm$ 0.3 & \\ %
BD+59451  & B1Ib-II       & 02:16:10.776 & +59:41:05.59 & 9.1 & 1.0 & -0.571 $\pm$ 0.076 & -0.773 $\pm$ 0.078 & 0.310 $\pm$ 0.037 & -27.6 $\pm$ 1.2 & \\ %
HD14052   & B1Ib-II       & 02:18:28.157 & +57:12:30.14 & 8.1 & 0.5 & -0.408 $\pm$ 0.093 & -1.193 $\pm$ 0.138 & 0.500 $\pm$ 0.054 & -90.8 $\pm$ 0.3 & $h$~Per \\ %
HD14476   & B1II          & 02:22:16.958 & +57:16:18.96 & 8.7 & 0.6 & -0.807 $\pm$ 0.087 & -1.164 $\pm$ 0.126 & 0.422 $\pm$ 0.050 & -40.6 $\pm$ 0.3 & $\chi$~Per \\ %
BD+56576  & B1III         & 02:22:09.710 & +57:07:02.30 & 9.3 & 0.4 & -0.211 $\pm$ 0.111 & -0.798 $\pm$ 0.099 & 0.326 $\pm$ 0.048 & -84.7 $\pm$ 0.3 & $\chi$~Per \\ %
BD+57520  & B1III         & 02:13:43.610 & +58:26:10.74 & 9.5 & 0.5 &(+0.176 $\pm$ 0.207)&(-0.731 $\pm$ 0.246)&(0.440 $\pm$ 0.109)& -46.0 $\pm$ 0.3 & \\ %
HD13841   & B1.5Ib        & 02:16:46.391 & +57:01:45.67 & 7.3 & 0.4 & -0.561 $\pm$ 0.078 & -1.548 $\pm$ 0.075 & 0.331 $\pm$ 0.040 & -40.2 $\pm$ 0.2 & \\ %
HD14443   & BC1.5Ib       & 02:22:00.576 & +57:08:41.87 & 7.9 & 0.5 & -0.594 $\pm$ 0.073 & -0.947 $\pm$ 0.092 & 0.376 $\pm$ 0.041 & -45.7 $\pm$ 0.2 & $\chi$~Per \\ %
HD16779   & B1.5Ib N-weak & 02:43:38.498 & +57:49:40.81 & 8.6 & 1.1 & +0.126 $\pm$ 0.083 & -1.012 $\pm$ 0.093 & 0.422 $\pm$ 0.046 & -39.6 $\pm$ 0.5 & \\ %
BD+56574  & BC1.5Ib-II    & 02:22:07.383 & +57:06:42.25 & 8.4 & 0.5 & -0.416 $\pm$ 0.076 & -0.651 $\pm$ 0.105 & 0.429 $\pm$ 0.045 & -39.6 $\pm$ 3.1 & $\chi$~Per \\ %
HD14302   & B1.5Ib-II     & 02:20:34.128 & +56:19:47.83 & 8.5 & 0.4 & -0.247 $\pm$ 0.078 & -1.043 $\pm$ 0.095 & 0.376 $\pm$ 0.044 & -36.8 $\pm$ 0.6 & \\ %
BD+56527  & B1.5II        & 02:19:10.448 & +57:07:50.10 & 8.3 & 0.6 & -0.746 $\pm$ 0.103 & -1.064 $\pm$ 0.119 & 0.447 $\pm$ 0.051 & -33.8 $\pm$ 1.1 & $h$~Per \\ %
HD14357   & B1.5II        & 02:21:10.436 & +56:51:56.34 & 8.4 & 0.5 & -0.319 $\pm$ 0.135 & -1.348 $\pm$ 0.121 & 0.471 $\pm$ 0.055 & -37.7 $\pm$ 2.1 & \\ %
HD14520   & B1.5II        & 02:22:43.527 & +57:05:12.48 & 9.1 & 0.5 & -0.357 $\pm$ 0.068 & -0.923 $\pm$ 0.082 & 0.479 $\pm$ 0.038 & -43.1 $\pm$ 1.1 & $\chi$~Per \\ %
BD+57626  &B1.5II-III     & 02:44:56.669 & +57:39:07.80 & 9.7 & 0.9 & +0.174 $\pm$ 0.066 & -0.992 $\pm$ 0.074 & 0.426 $\pm$ 0.036 & -43.9 $\pm$ 0.9 & \\ %
BD+57513  & B1.5IIIn      & 02:12:36.069 & +58:05:54.11 & 9.4 & 0.4 & -0.958 $\pm$ 0.06  & -1.11  $\pm$ 0.08  & 0.355 $\pm$ 0.034 & -45.7 $\pm$ 7.5 & \\ %
HD14143   & B2Ia          & 02:19:13.942 & +57:10:09.23 & 6.5 & 0.8 & -0.347 $\pm$ 0.07  & -1.276 $\pm$ 0.099 & 0.463 $\pm$ 0.038 & -41.4 $\pm$ 0.6 & $h$~Per \\ %
HD14818   & B2Ia          & 02:25:16.029 & +56:36:35.36 & 6.1 & 0.5 & -0.676 $\pm$ 0.06  & -1.355 $\pm$ 0.071 & 0.345 $\pm$ 0.033 & -43.1 $\pm$ 0.6 & \\ %
HD14956   & B2Ia          & 02:26:45.696 & +57:40:45.04 & 7.0 & 1.0 & -0.422 $\pm$ 0.113 & -1.106 $\pm$ 0.112 & 0.285 $\pm$ 0.048 & -36.9 $\pm$ 0.3 & \\ %
HD15690   & B2Iab         & 02:33:32.786 & +57:32:14.75 & 7.8 & 0.9 & -0.531 $\pm$ 0.132 & -1.138 $\pm$ 0.097 & 0.399 $\pm$ 0.046 & -42.4 $\pm$ 0.6 & \\ %
HD13866   & BC2Ib         & 02:16:57.576 & +56:43:07.68 & 7.4 & 0.3 & -0.954 $\pm$ 0.08  & -1.568 $\pm$ 0.102 & 0.411 $\pm$ 0.048 & -47.1 $\pm$ 0.6 & \\ %
BD+56578  & BN2II-III     & 02:22:17.690 & +57:07:24.63 & 9.1 & 0.5 & -0.823 $\pm$ 0.08  & -1.202 $\pm$ 0.086 & 0.282 $\pm$ 0.041 & -87.8 $\pm$ 0.8 & $\chi$~Per \\ %
HD10898   & B2.5Ib        & 01:48:35.030 & +58:27:28.21 & 7.3 & 0.5 & -0.82  $\pm$ 0.061 & -0.772 $\pm$ 0.079 & 0.270 $\pm$ 0.042 & -30.0 $\pm$ 0.5 & \\ %
BD+59387  & B2.5II N-weak & 02:02:37.167 & +60:04:47.23 & 9.4 & 0.9 & -0.92  $\pm$ 0.033 & -0.686 $\pm$ 0.065 & 0.332 $\pm$ 0.033 & -32.4 $\pm$ 0.8 & \\ %
BD+59388  & BN2.5III      & 02:02:37.351 & +59:58:31.93 & 9.4 & 0.9 & -0.952 $\pm$ 0.041 & -0.729 $\pm$ 0.089 & 0.399 $\pm$ 0.042 & -47.2 $\pm$ 0.3 & \\ %
HD14134   & B3Ia          & 02:19:04.452 & +57:08:07.80 & 6.4 & 0.7 & -0.637 $\pm$ 0.078 & -1.294 $\pm$ 0.098 & 0.443 $\pm$ 0.044 & -41.8 $\pm$ 0.6 & $h$~Per \\ %
HD15497   & B6Ia          & 02:31:53.376 & +57:41:51.51 & 6.8 & 1.1 & -0.124 $\pm$ 0.067 & -1.022 $\pm$ 0.065 & 0.333 $\pm$ 0.032 & -44.2 $\pm$ 0.3 & \\ %
HD13267   & B6Iab         & 02:11:29.193 & +57:38:43.96 & 6.3 & 0.5 & -0.54  $\pm$ 0.068 & -0.675 $\pm$ 0.09  & 0.447 $\pm$ 0.040 & -35.0 $\pm$ 0.5 & \\ %
HD17145   & B6Iab         & 02:47:24.247 & +57:40:37.59 & 7.8 & 1.2 & -0.021 $\pm$ 0.092 & -1.358 $\pm$ 0.081 & 0.384 $\pm$ 0.047 & -37.3 $\pm$ 0.6 & \\ %
HD14322   & B8Iab         & 02:20:42.918 & +55:54:32.68 & 6.7 & 0.5 & -0.52  $\pm$ 0.078 & -1.418 $\pm$ 0.083 & 0.219 $\pm$ 0.041 & -33.4 $\pm$ 0.4 & \\ %
HD14542   & B8Iab         & 02:23:00.426 & +57:23:13.22 & 6.8 & 1.0 & -0.766 $\pm$ 0.084 & -0.941 $\pm$ 0.087 & 0.374 $\pm$ 0.043 & -50.9 $\pm$ 0.4 & \\ %
HD15620   & B8Iab         & 02:32:52.876 & +57:55:45.65 & 8.0 & 1.3 & -0.349 $\pm$ 0.077 & -0.837 $\pm$ 0.083 & 0.398 $\pm$ 0.041 & -43.6 $\pm$ 0.4 & \\ %
HDE236995 & B8Ib          & 02:45:03.505 & +58:33:04.56 & 8.3 & 1.0 & -0.024 $\pm$ 0.068 & -1.512 $\pm$ 0.074 & 0.405 $\pm$ 0.036 & -42.4 $\pm$ 0.2 & \\ %
HD17088   & B9Ia          & 02:46:51.450 & +57:44:01.69 & 7.2 & 1.2 & +0.273 $\pm$ 0.08  & -1.046 $\pm$ 0.073 & 0.430 $\pm$ 0.041 & -47.0 $\pm$ 0.7 & \\ %
HD14899   & B9Iab         & 02:26:18.472 & +57:13:41.99 & 7.2 & 0.6 & -1.166 $\pm$ 0.086 & -0.691 $\pm$ 0.106 & 0.535 $\pm$ 0.051 & -41.0 $\pm$ 0.1 & \\ %

\noalign{\smallskip}
\hline
\noalign{\smallskip}

\multicolumn{11}{c}{\textbf{F and A supergiants (I)}} \\\noalign{\smallskip}
HD13744   & A0Iab         & 02:15:58.697 & +58:17:37.02 & 7.3 & 1.1 & -0.854 $\pm$ 0.064 & -0.951 $\pm$ 0.073 & 0.417 $\pm$ 0.038 & -50.7 $\pm$ 0.4 & \\ %
HD12953   & A1Iae         & 02:08:40.579 & +58:25:24.97 & 5.5 & 1.0 & -1.044 $\pm$ 0.117 & -0.638 $\pm$ 0.13  & 0.308 $\pm$ 0.067 & -33.1 $\pm$ 0.3 & \\ %
HD14433   & A1Ia          & 02:21:55.435 & +57:14:34.49 & 6.2 & 0.9 & -0.463 $\pm$ 0.07  & -0.908 $\pm$ 0.075 & 0.382 $\pm$ 0.038 & -47.9 $\pm$ 0.1 & $\chi$~Per \\ %
HD14489   & A1Ia          & 02:22:21.435 & +55:50:44.35 & 5.1 & 0.7 & +0.374 $\pm$ 0.23  & -1.798 $\pm$ 0.205 & 0.790 $\pm$ 0.114 & -18.7 $\pm$ 0.1 & \\ %
HD14535   & A1Ia          & 02:22:53.497 & +57:14:42.54 & 7.2 & 1.1 & -0.25  $\pm$ 0.076 & -0.667 $\pm$ 0.079 & 0.420 $\pm$ 0.040 & -43.5 $\pm$ 0.5 & $\chi$~Per \\ %
HD11831   & A2Ia          & 01:58:03.814 & +60:23:27.93 & 7.7 & 1.3 & -0.843 $\pm$ 0.027 & -0.379 $\pm$ 0.052 & 0.303 $\pm$ 0.033 & -39.2 $\pm$ 0.1 & \\ %
HD16778   & A2Ia-Iab      & 02:43:53.691 & +59:49:21.94 & 7.3 & 1.4 & +0.307 $\pm$ 0.049 & -0.662 $\pm$ 0.065 & 0.444 $\pm$ 0.034 & -42.9 $\pm$ 0.1 & \\ %
HD13476   & A3Iab         & 02:13:41.611 & +58:33:38.10 & 6.2 & 0.9 & -0.85  $\pm$ 0.066 & -0.65  $\pm$ 0.063 & 0.357 $\pm$ 0.035 & -39.6 $\pm$ 0.4 & \\ %
HD15316   & A3Iab         & 02:29:58.556 & +57:49:14.57 & 6.9 & 1.2 & -0.252 $\pm$ 0.067 & -0.961 $\pm$ 0.084 & 0.454 $\pm$ 0.036 & -49.9 $\pm$ 0.4 & \\ %
HD17378   & A6Ia          & 02:49:30.737 & +57:05:03.55 & 5.9 & 1.3 & +0.828 $\pm$ 0.105 & -1.235 $\pm$ 0.082 & 0.477 $\pm$ 0.041 & -40.1 $\pm$ 0.3 & \\ %
HD12842   & F3Ib          & 02:07:46.336 & +58:39:58.68 & 8.2 & 1.2 & -0.041 $\pm$ 0.082 & -0.106 $\pm$ 0.077 & 0.527 $\pm$ 0.047 & -24.1 $\pm$ 0.1 & \\ %

\noalign{\smallskip}
\hline
\noalign{\smallskip}

\multicolumn{11}{c}{\textbf{M supergiants (I)}} \\\noalign{\smallskip}
HD14580   & K3Ib-II       & 02:23:24.110 & +57:12:43.04 & 7.3 & 2.6 & -0.12  $\pm$ 0.081 & -0.66  $\pm$ 0.11  & 0.530 $\pm$ 0.046 & -45.4 $\pm$ 0.2 & $\chi$~Per \\ %
BD+59372  & K4Ib-II       & 01:59:39.666 & +60:15:01.94 & 8.1 & 2.7 & -0.842 $\pm$ 0.04  & -0.722 $\pm$ 0.078 & 0.570 $\pm$ 0.040 & -39.4 $\pm$ 0.1 & \\ %
HD14330   & K5-M0Iab-Ib   & 02:20:59.645 & +57:09:29.96 & 6.9 & 2.6 & -0.696 $\pm$ 0.075 & -1.223 $\pm$ 0.11  & 0.469 $\pm$ 0.041 & -44.9 $\pm$ 0.1 & $\chi$~Per \\ %
HD236979  & K5-M0Iab-Ib   & 02:38:25.420 & +57:02:46.20 & 6.5 & 2.8 & -0.119 $\pm$ 0.099 & -1.391 $\pm$ 0.101 & 0.360 $\pm$ 0.055 & -41.1 $\pm$ 0.1 & \\ %
BD+56595  & K5-M0Ib       & 02:23:11.065 & +57:11:57.98 & 7.0 & 2.6 & -0.308 $\pm$ 0.083 & -0.92  $\pm$ 0.09  & 0.475 $\pm$ 0.041 & -44.3 $\pm$ 0.2 & $\chi$~Per \\ %
BD+57530A & K5-M0Ib       & 02:17:08.232 & +58:31:46.96 & 7.7 & 2.9 & -0.893 $\pm$ 0.088 & -1.317 $\pm$ 0.086 & 0.446 $\pm$ 0.047 & -52.0 $\pm$ 0.1 & \\ %
HD14242   & M0Iab-Ib      & 02:20:22.463 & +59:40:16.91 & 7.0 & 2.8 & -0.711 $\pm$ 0.079 & -0.953 $\pm$ 0.07  & 0.329 $\pm$ 0.039 & -34.7 $\pm$ 0.1 & \\ %
HD14404   & M0Ib          & 02:21:42.410 & +57:51:46.15 & 6.7 & 2.7 & -0.788 $\pm$ 0.101 & -1.328 $\pm$ 0.115 & 0.565 $\pm$ 0.049 & -45.9 $\pm$ 0.1 & \\ %
HD13136   & M1Iab-Ib      & 02:10:15.784 & +56:33:32.66 & 6.6 & 2.8 & +0.045 $\pm$ 0.105 & -0.978 $\pm$ 0.135 & 0.622 $\pm$ 0.052 & -41.7 $\pm$ 0.2 & \\ %
HD14142   & M1Iab-Ib      & 02:19:21.877 & +58:57:40.35 & 7.2 & 3.0 & -1.274 $\pm$ 0.104 & -0.543 $\pm$ 0.084 & 0.345 $\pm$ 0.052 & -46.2 $\pm$ 0.1 & \\ %
HD236915  & M1Ib          & 01:58:28.911 & +59:16:08.78 & 7.0 & 2.9 & -1.461 $\pm$ 0.071 & -0.76  $\pm$ 0.106 & 0.493 $\pm$ 0.045 & -48.7 $\pm$ 0.1 & \\ %
HD14270   & M2Iab-Ib      & 02:20:29.003 & +56:59:35.24 & 6.7 & 3.0 & -0.066 $\pm$ 0.113 & -1.423 $\pm$ 0.138 & 0.487 $\pm$ 0.056 & -43.6 $\pm$ 0.2 & \\ %
HD14826   & M2Iab-Ib      & 02:25:21.860 & +57:26:14.14 & 6.6 & 3.0 & -0.254 $\pm$ 0.108 & -1.559 $\pm$ 0.153 & 0.552 $\pm$ 0.063 & -44.2 $\pm$ 0.1 & \\ %
BD+56512  & M3Iab         & 02:18:53.291 & +57:25:16.76 & 7.3 & 3.4 & -0.527 $\pm$ 0.16  & -1.106 $\pm$ 0.195 & 0.695 $\pm$ 0.088 & -36.0 $\pm$ 0.1 & \\ %
HD14469   & M3.5Iab       & 02:22:06.894 & +56:36:14.90 & 5.9 & 3.0 & -0.617 $\pm$ 0.129 & -1.49  $\pm$ 0.171 & 0.346 $\pm$ 0.080 & -44.8 $\pm$ 0.3 & \\ %
HD14488   & M4Iab         & 02:22:24.295 & +57:06:34.10 & 6.4 & 3.3 & -0.371 $\pm$ 0.137 & -0.931 $\pm$ 0.165 & 0.674 $\pm$ 0.082 & -43.6 $\pm$ 0.3 & $\chi$~Per \\ %
BD+56724  & M4-M5Ia-Iab   & 02:50:37.893 & +56:59:00.30 & 7.2 & 4.0 & +0.243 $\pm$ 0.17  & -1.991 $\pm$ 0.168 & 0.825 $\pm$ 0.082 & -42.2 $\pm$ 0.3 & \\ %
HD14528$^{d}$& M5-M6Iab   & 02:22:51.710 & +58:35:11.45 & 7.8 & 4.3 & -0.49  $\pm$ 0.23  & -1.19  $\pm$ 0.20  & 0.413 $\pm$ 0.017 & -39.5 $\pm$ 0.4 & \\ %

\noalign{\smallskip}
\hline
\noalign{\smallskip}

\multicolumn{11}{c}{\textbf{Interesting stars which have not been yet observed. Collected from \citetads{1978ApJS...38..309H}.}} \\\noalign{\smallskip}
HD15785  & B1Iab  & 02:34:48.052 & +60:33:07.45 & 8.1 & 0.8 & +0.737 $\pm$ 0.039 & +0.381 $\pm$ 0.080 & 0.268 $\pm$ 0.039 & \\
HD16310  & B1II:  & 02:39:23.029 & +59:03:59.44 & 7.9 & 0.9 & +0.071 $\pm$ 0.087 & -0.718 $\pm$ 0.085 & 0.355 $\pm$ 0.047 & \\
HD16243  & B2II:  & 02:38:38.081 & +57:49:01.16 & 8.0 & 0.9 & -0.308 $\pm$ 0.076 & -1.408 $\pm$ 0.076 & 0.478 $\pm$ 0.040 & \\
HD13658  & M:Ib   & 02:15:13.297 & +58:08:32.32 & 7.6 & 2.7 & -0.508 $\pm$ 0.072 & -1.066 $\pm$ 0.083 & 0.485 $\pm$ 0.039 & \\
BD+57647 & M2Ia   & 02:51:03.948 & +57:51:19.92 & 7.5 & 3.6 & -0.184 $\pm$ 0.158 & -1.252 $\pm$ 0.143 & 0.562 $\pm$ 0.072 & \\
BD+58501 & M2Iab  & 02:39:50.440 & +59:35:51.30 & 7.5 & 3.6 & -0.382 $\pm$ 0.136 & -0.760 $\pm$ 0.146 & 0.626 $\pm$ 0.068 & \\
BD+54444 & M4Ib+  & 02:03:09.358 & +55:13:56.62 & 6.5 & 3.2 & -1.232 $\pm$ 0.114 & -1.856 $\pm$ 0.126 & 0.491 $\pm$ 0.066 & \\

\noalign{\smallskip}
\hline
\noalign{\smallskip}

\multicolumn{11}{c}{\textbf{Interesting stars which have not been yet observed. Collected from \citeads{1992A&AS...94..211G}.}} \\\noalign{\smallskip}
HD14422   & B1Ie     & 02:21:50.811 & +57:23:11.64 & 8.5 & 0.9 & -0.429 $\pm$ 0.096 & -1.084 $\pm$ 0.087 & 0.439 $\pm$ 0.046 & \\
BD+59461  & B1II     & 02:19:22.019 & +59:43:38.49 & 9.9 & 0.7 & -0.375 $\pm$ 0.075 & -0.629 $\pm$ 0.081 & 0.270 $\pm$ 0.035 & \\
HD236954  & B3Ib-II  & 02:13:37.333 & +59:10:14.80 & 9.2 & 0.9 & -0.348 $\pm$ 0.064 & -0.643 $\pm$ 0.075 & 0.391 $\pm$ 0.035 & \\
HD14050   & B5II     & 02:18:35.182 & +58:35:21.24 & 9.1 & 0.6 & +0.233 $\pm$ 0.083 & -0.417 $\pm$ 0.074 & 0.950 $\pm$ 0.039 & \\
IRC+60091 & M1Ib     & 02:27:23.988 & +60:40:47.83 & 8.5 & 4.8 & -0.824 $\pm$ 0.104 & -2.201 $\pm$ 0.170 & 1.345 $\pm$ 0.109 & \\

\noalign{\smallskip}
\hline
\noalign{\smallskip}

\multicolumn{11}{c}{\textbf{Interesting stars which have not been yet observed. Collected from \citeads{2010ApJS..186..191C}}} \\\noalign{\smallskip}
BD+56553  & B0.4I  & 02:21:07.700 & +57:10:00.45 & 9.4 & 0.5 & -0.157 $\pm$ 0.068 & -1.006 $\pm$ 0.089 & 0.435 $\pm$ 0.036 & & $\chi$~Per \\
BD+56501  & B0.5I  & 02:18:29.834 & +57:09:03.17 & 9.3 & 0.4 & -0.625 $\pm$ 0.086 & -1.555 $\pm$ 0.097 & 0.472 $\pm$ 0.046 & & $h$~Per \\
BD+56528  & B0.5I  & 02:19:10.681 & +57:01:29.86 & 9.8 & 0.5 & -0.566 $\pm$ 0.081 & -1.436 $\pm$ 0.100 & 0.493 $\pm$ 0.044 & & $h$~Per \\
BD+56584  & B0.7I  & 02:22:29.851 & +57:12:28.84 & 9.4 & 0.5 & -0.781 $\pm$ 0.068 & -1.072 $\pm$ 0.084 & 0.491 $\pm$ 0.037 & & $\chi$~Per \\
HD14661   & B1.5I  & 02:24:02.236 & +57:21:14.13 & 9.0 & 0.8 & -0.422 $\pm$ 0.077 & -1.574 $\pm$ 0.094 & 0.367 $\pm$ 0.042 & & \\

\end{longtable}
}

\begin{list}{}{}
\small{
\item[$^{a}$] Sky coordinates retrieved from {\em Gaia} DR2.
\item[$^{b}$] Parallaxes retrieved from {\em Gaia} DR2 are corrected from a -0.03\,mas zero-point offset (see \citetads{2018A&A...616A...2L}).
\item[$^{c}$] The proper motion and parallax of stars whose {\em Gaia} DR2 astrometry RUEW value is higher than 1.4 are given in parentheses.
\item[$^{d}$] Proper motion and parallax of HD\,14528 is taken from \citetads{2010ApJ...721..267A} (see Sect. \ref{subsection.23_gaiaobs}).
\item[$^{e}$] Stars identified as SB2. The the radial velocity measurements for HD\,15558 refer to the main component (top line), and second component (bottom line), as measured using the line O~{\sc iii} $\lambda$5592.25\,{\AA}. For HD\,13402, and HD\,13970 the Si~{\sc iii} $\lambda$4552.62\,{\AA} was used instead.

}
\end{list}

%%%%%%%%%%%%%%%%%%%%%%%%%%%%%%%%%%%%%%%%%%%%%%%%%

\begin{longtable}{llcccc}
\label{table.A2}\\ %
\caption[]{Results from the multi-epoch RV measurements for stars with five or more spectra, excluding confirmed SB1 and SB2 stars (see also Table~\ref{table.A3}).}\\ %

\hline
\hline
\noalign{\smallskip}
Name    &SpC            &Time span      &Number of      &$\overline{RV}$        &RV$_{PP}$      \\ %
                &                       &[days]         &spectra        &[km\,s$^{-1}$]         &[km\,s$^{-1}$] \\ %
\noalign{\smallskip}
\hline
\noalign{\smallskip}
\endfirsthead

\caption[]{Results from the multi-epoch radial velocity measurements for stars with five or more spectra, excluding confirmed SB1/SB2 (see also Table~\ref{table.A3}).}\\ %

\hline
\hline
\noalign{\smallskip}
Name    &SpC            &Time span      &Number of      &$\overline{RV}$        &RV$_{PP}$      \\ %
                &                       &[days]         &spectra        &[km\,s$^{-1}$]         &[km\,s$^{-1}$] \\ %
\noalign{\smallskip}
\hline
\noalign{\smallskip}
\endhead

\noalign{\smallskip}
\hline
\endfoot

\multicolumn{6}{c}{\textbf{O-type stars}} \\\noalign{\smallskip}
HD15570 & O4If           & 2266 & 6 & -52.6 $\pm$ 8.0 & 21.0 $\pm$ 5.1  \\ %
HD15642 & O9.5II-IIIn    & 2293 & 6 & -22.7 $\pm$ 7.1 & 22   $\pm$ 13   \\ %

\noalign{\smallskip}
\hline
\noalign{\smallskip}

\multicolumn{6}{c}{\textbf{B-type stars}} \\\noalign{\smallskip}
HD14053 & B0.7II   & 1176 & 8  & -33.9 $\pm$ 2.5 &  8.10 $\pm$ 0.51 \\ %
HD13854 & B1Ia-Iab & 1552 & 14 & -38.0 $\pm$ 4.5 & 16.13 $\pm$ 0.81 \\ %
HD14357 & B1.5II   & 3229 & 8  & -40.2 $\pm$ 3.7 & 11.0  $\pm$ 5.2  \\ %
HD14143 & B2Ia     & 1897 & 14 & -43.3 $\pm$ 2.8 &  9.15 $\pm$ 0.86 \\ %
HD14818 & B2Ia     & 1451 & 19 & -42.1 $\pm$ 1.9 &  7.22 $\pm$ 0.92 \\ %
HD14134 & B3Ia     & 1550 & 12 & -42.5 $\pm$ 1.1 &  3.67 $\pm$ 0.59 \\ %
HD13267 & B6Iab    & 818  & 13 & -36.1 $\pm$ 2.5 & 10.01 $\pm$ 0.73 \\ %
HD14542 & B8Iab    & 1556 & 11 & -51.8 $\pm$ 3.1 &  9.62 $\pm$ 0.52 \\ %
HD14322 & B8Iab    & 2128 & 15 & -34.1 $\pm$ 1.2 &  4.60 $\pm$ 0.67 \\ %

\noalign{\smallskip}
\hline
\noalign{\smallskip}

\multicolumn{6}{c}{\textbf{A-type stars}} \\\noalign{\smallskip}
HD12953 & A1Iae    & 2867 & 19 & -29.7 $\pm$ 3.4 & 10.43 $\pm$ 0.32 \\ %
HD14489 & A1Ia     & 2867 & 25 & -17.9 $\pm$ 1.5 &  5.41 $\pm$ 0.22 \\ %

\noalign{\smallskip}
\hline
\noalign{\smallskip}

\multicolumn{6}{c}{\textbf{M-type stars}} \\\noalign{\smallskip}
HD14270 & M2Iab-Ib & 2865 & 15 & -43.7 $\pm$ 0.3 & 1.3 $\pm$ 0.23 \\ %
HD13136 & M1Iab-Ib & 2865 & 16 & -42.0 $\pm$ 0.8 & 2.0 $\pm$ 0.21 \\ %

\end{longtable}

%%%%%%%%%%%%%%%%%%%%%%%%%%%%%%%%%%%%%%%%%%%%%%%%%

\begin{longtable}{llcccc}
\label{table.A3}\\ %
\caption[]{Results from the multi-epoch RV measurements for all confirmed SB1 stars.}\\ %

\hline
\hline
\noalign{\smallskip}
Name    &SpC            &Time span      &Number of      &$\overline{RV}$        &RV$_{PP}$      \\ %
                &                       &[days]         &spectra        &[km\,s$^{-1}$]         &[km\,s$^{-1}$] \\ %
\noalign{\smallskip}
\hline
\noalign{\smallskip}
\endfirsthead

\caption[]{Results from the radial velocities measurements using the multi-epoch spectra of all the confirmed SB1/SB2.}\\ %

\hline
\hline
\noalign{\smallskip}
Name    &SpC            &Time span      &Number of      &$\overline{RV}$        &RV$_{PP}$      \\ %
                &                       &[days]         &spectra        &[km\,s$^{-1}$]         &[km\,s$^{-1}$] \\ %
\noalign{\smallskip}
\hline
\noalign{\smallskip}
\endhead

\noalign{\smallskip}
\hline
\endfoot

HD13969  & B0.7III    &  111 & 3 & -26   $\pm$ 10  & 23.68 $\pm$ 0.76 \\ %
HD14052  & B1Ib-II    & 1100 & 4 & -70   $\pm$ 36  & 82.01 $\pm$ 0.43 \\ %
HD14476  & B1II       & 3265 & 3 & -54.5 $\pm$ 9.9 & 21.51 $\pm$ 0.42 \\ %
BD+56576 & B1III      & 3346 & 2 & -30   $\pm$ 54  &108.50 $\pm$ 0.46 \\ %
\noalign{\smallskip}

\end{longtable}

%%%%%%%%%%%%%%%%%%%%%%%%%%%%%%%%%%%%%%%%%%%%%%%%%

\scriptsize{
\begin{longtable}{lcccccccl}
\label{table.A4}\\ %
\caption[]{Membership and features of investigated stars.}\\ %

\hline
\hline
\noalign{\smallskip}
Name & $\varpi$ & $\mu$ & RV best & RV multi & RV final & Spec. variability & Member$^{c}$ & Comments$^{d}$ \\ %
\noalign{\smallskip}
\hline
\noalign{\smallskip}
\endfirsthead

\caption[]{Membership and features of investigated stars.}\\ %

\hline
\hline
\noalign{\smallskip}
Name & $\varpi$ & $\mu$ & RV best & RV multi & RV final & Spec. variability & Member$^{c}$ & Comments$^{d}$ \\ %
\noalign{\smallskip}
\hline
\noalign{\smallskip}
\endhead

\noalign{\smallskip}
\hline
\endfoot

\multicolumn{9}{c}{\textbf{O giants (III), bright giants (II), and supergiants (I)}} \\\noalign{\smallskip}
HD15570 &  $\bullet$ &  $\bullet$ & $\bullet$ & $\bullet$ & $\circ$   &     & N & IC\,1805          \\ % Const. (Hol18)
HD16691 &  $\bullet$ &  $\times$  & $\bullet$ & $\bullet$ & $\circ$   &     & Y & Runaway (PM)      \\ % LPV (Hol18)
HD14947 &  $\bullet$ &  $\bullet$ & $\bullet$ & $\bullet$ & $\bullet$ &     & Y &                   \\ % LPV, WVe (Hol18)
HD15558 & ($\bullet$)& ($\bullet$)& $\bullet$ & $\bullet$ &($\circ$)  & SB2 & N & RUWE, SB2 (MA19), IC\,1805 \\ %
HD14442 &  $\bullet$ &  $\bullet$ & $\bullet$ & --        & $\bullet$ &     & Y &                   \\ %
HD17603 &  $\bullet$ &  $\bullet$ & $\bullet$ & --        &($\circ$)  &SB1 (lit.)&Y& SB1 (Hol18)    \\ %
HD13268 &  $\circ$   &  $\times$  & $\circ$   & $\circ$   & $\times$  &     & Y$^{e}$ & Runaway (PM,RV)\\ % Const. (Hol20)
HD16429 & ($\times$) & ($\times$) & $\bullet$ & $\bullet$ &($\times$) & SB2 & N & RUWE, Triple sys. (MA19), IC\,1805 \\ %
HD16832 &  $\bullet$ &  $\bullet$ & $\bullet$ & --        & $\circ$   &     & Y & SB1?              \\ % Const. (Hol18)
HD15642 &  $\bullet$ &  $\times$  & $\bullet$ & $\bullet$ & $\times$  &     & Y & Runaway (PM,RV)   \\ % LPV (Hol20)
HD13831 &  $\bullet$ &  $\bullet$ & $\bullet$ & --        & $\bullet$ &     & Y &                   \\ % BCEP (VSI)
HD13745 &  $\bullet$ &  $\times$  & $\bullet$ & $\bullet$ & $\times$  &     & Y & Runaway (PM,RV)   \\ % LPV (Hol20), BCEP (VSI)
HD13022 &  $\bullet$ &  $\circ$   & $\bullet$ & --        & $\circ$   &     & Y & Runaway (PM)      \\ % Const. (Hol20)

\noalign{\smallskip}
\hline
\noalign{\smallskip}

\multicolumn{9}{c}{\textbf{B giants (III), bright giants (II), and supergiants (I)}} \\\noalign{\smallskip}
HD13402 &  $\bullet$ &  $\bullet$ & $\bullet$ & $\times$  &($\times$) & SB2 & Y &                   \\ % Binary ejected
HD14331 &  $\bullet$ &  $\circ$   & $\circ$   & --        & $\circ$   &     & Y & Runaway (PM, RV)  \\ %
HD16808 &  $\bullet$ &  $\bullet$ & $\bullet$ & --        & $\bullet$ &     & Y &                   \\ %
HD13036 &  $\bullet$ &  $\bullet$ & $\bullet$ & --        &($\circ$)  &LPV/SB1?&Y&                  \\ % Be?
BD+60493&  $\bullet$ &  $\bullet$ & $\circ$   & --        & $\circ$   &     & N & IC\,1805, SB1?    \\ %
HD13716 &  $\bullet$ &  $\bullet$ & $\bullet$ & $\bullet$ &($\bullet$)& SB2 & Y & E (VSI), Eclip. Var. (Zac85) \\ %
HD13970 & ($\times$) & ($\bullet$)& $\times$  & --        &($\circ$)  & SB2 & C & RUWE, ELL (VSI)   \\ %
HD14053 & ($\bullet$)& ($\bullet$)& $\bullet$ & $\bullet$ & $\circ$   &     & P & RUWE              \\ % BCEP (VSI), ACYG (Lau17)
HD13969 & ($\bullet$)& ($\times$) & $\times$  & --        &($\circ$)  & SB1 & P & RUWE              \\ % 
HD13854 &  $\bullet$ &  $\bullet$ & $\bullet$ & $\bullet$ & $\bullet$ &LPV/SB1?&Y&                  \\ % ACYG (VSI, Lau17)
HD13659 &  $\bullet$ &  $\bullet$ & $\circ$   & --        & $\bullet$ &     & Y &                   \\ %
HD15571 & ($\times$) & ($\circ$)  & $\bullet$ & --        & $\bullet$ &     & C & RUWE              \\ % ACYG (Lau17)
BD+59451&  $\bullet$ &  $\bullet$ & $\circ$   & --        & $\circ$   &     & Y & SB1?              \\ %
HD14052 &  $\bullet$ &  $\bullet$ & $\times$  & $\times$  & $\times$  & SB1 & Y & SB1 (Hel73)       \\ % Binary ejected
HD14476 &  $\bullet$ &  $\bullet$ & $\bullet$ & --        &($\circ$)  & SB1 & Y &                   \\ %
BD+56576&  $\bullet$ &  $\bullet$ & $\times$  & --        &($\circ$)  & SB1 & Y & EA (VSI), EB (Lau17) \\ %
BD+57520& ($\bullet$)& ($\bullet$)& $\bullet$ & --        & $\bullet$ &     & P & RUWE              \\ % 
HD13841 &  $\bullet$ &  $\bullet$ & $\bullet$ & --        & $\bullet$ &     & Y &                   \\ % ACYG (Lau17)
HD14443 &  $\bullet$ &  $\bullet$ & $\bullet$ & $\bullet$ & $\bullet$ &     & Y &                   \\ % ACYG (Lau17)
HD16779 &  $\bullet$ &  $\bullet$ & $\bullet$ & --        & $\bullet$ &     & Y &                   \\ %
BD+56574&  $\bullet$ &  $\bullet$ & $\bullet$ & --        & $\bullet$ &     & Y &                   \\ %
HD14302 &  $\bullet$ &  $\bullet$ & $\bullet$ & --        & $\bullet$ &     & Y &                   \\ %
BD+56527&  $\bullet$ &  $\bullet$ & $\bullet$ & --        & $\bullet$ &     & Y &                   \\ %
HD14357 &  $\bullet$ &  $\bullet$ & $\bullet$ & $\bullet$ & $\bullet$ &     & Y &                   \\ %
HD14520 &  $\bullet$ &  $\bullet$ & $\bullet$ & --        & $\bullet$ &     & Y &                   \\ %
BD+57626&  $\bullet$ &  $\bullet$ & $\bullet$ & --        & $\bullet$ &     & Y &                   \\ %
BD+57513&  $\bullet$ &  $\bullet$ & $\bullet$ & --        & $\bullet$ &     & Y &                   \\ %
HD14143 &  $\bullet$ &  $\bullet$ & $\bullet$ & $\bullet$ & $\bullet$ &     & Y &                   \\ % ACYG (Lau17)
HD14818 &  $\bullet$ &  $\bullet$ & $\bullet$ & $\bullet$ & $\bullet$ &     & Y &                   \\ % ACYG (VSI)
HD14956 &  $\bullet$ &  $\bullet$ & $\bullet$ & --        &($\circ$)  &SB1 (lit.)&Y& SB1 (Hel73)    \\ % ACYG (VSI, Lau17)
HD15690 &  $\bullet$ &  $\bullet$ & $\bullet$ & $\bullet$ & $\bullet$ &     & Y &                   \\ %
HD13866 &  $\bullet$ &  $\circ$   & $\bullet$ & $\bullet$ & $\bullet$ &     & Y & Runaway (PM)      \\ % BCEP (VSI)
BD+56578&  $\bullet$ &  $\bullet$ & $\times$  & --        &($\times$) &EB (lit.)&Y& ELL (VSI), EB (Lau17), SB1 \\ % Be
HD10898 &  $\bullet$ &  $\bullet$ & $\circ$   & $\bullet$ & $\circ$   &     & Y &                   \\ %
BD+59387&  $\bullet$ &  $\bullet$ & $\bullet$ & --        & $\circ$   &     & Y &                   \\ %
BD+59388&  $\bullet$ &  $\bullet$ & $\bullet$ & --        & $\bullet$ &     & Y &                   \\ %
HD14134 &  $\bullet$ &  $\bullet$ & $\bullet$ & $\bullet$ & $\bullet$ &     & Y &                   \\ % IA (VSI), ACYG (Lau17)
HD15497 &  $\bullet$ &  $\bullet$ & $\bullet$ & --        & $\bullet$ &     & Y &                   \\ % ACYG (VSI)
HD13267 &  $\bullet$ &  $\bullet$ & $\bullet$ & $\bullet$ & $\bullet$ &LPV/SB1?&Y&                  \\ %
HD17145 &  $\bullet$ &  $\bullet$ & $\bullet$ & --        & $\bullet$ &     & Y &                   \\ %
HD14322 &  $\circ$   &  $\bullet$ & $\bullet$ & $\bullet$ & $\circ$   &     & C &                   \\ % ACYG (Lau17)
HD14542 &  $\bullet$ &  $\bullet$ & $\circ$   & $\circ$   & $\circ$   &LPV/SB1?&Y&                  \\ % ACYG (VSI, Lau17)
HD15620 &  $\bullet$ &  $\bullet$ & $\bullet$ & --        & $\bullet$ &     & Y &                   \\ % ACYG (Lau17)
HDE236995& $\bullet$ &  $\bullet$ & $\bullet$ & --        & $\bullet$ &     & Y &                   \\ %
HD17088 &  $\bullet$ &  $\bullet$ & $\bullet$ & $\bullet$ & $\bullet$ &     & Y &                   \\ %
HD14899 &  $\times$  &  $\bullet$ & $\bullet$ & --        & $\bullet$ &     & Y &                   \\ %

\noalign{\smallskip}
\hline
\noalign{\smallskip}

\multicolumn{9}{c}{\textbf{A supergiants (I)}} \\\noalign{\smallskip}
HD13744 &  $\bullet$ &  $\bullet$ & $\bullet$ & $\bullet$ & $\bullet$ &     & Y &                   \\ %
HD12953 &  $\bullet$ &  $\bullet$ & $\bullet$ & $\circ$   & $\circ$   &LPV/SB1?&Y& G$_{mag}$\,=\,5.5\\ % ACYG (VSI)
HD14433 &  $\bullet$ &  $\bullet$ & $\bullet$ & --        & $\bullet$ &     & Y &                   \\ % ACYG (Lau17)
HD14489 & ($\circ$)  & ($\circ$)  & $\circ$   & $\circ$   & $\times$  &     & C & G$_{mag}$\,=\,5.1, Runaway? \\ % High err Plx/PM, ACYG (VSI)
HD14535 &  $\bullet$ &  $\bullet$ & $\bullet$ & $\bullet$ & $\bullet$ &     & Y &                   \\ % ACYG (VSI, Lau17)
HD11831 &  $\bullet$ &  $\bullet$ & $\bullet$ & --        & $\bullet$ &     & Y &                   \\ %
HD16778 &  $\bullet$ &  $\bullet$ & $\bullet$ & --        & $\bullet$ &     & Y &                   \\ %
HD13476 &  $\bullet$ &  $\bullet$ & $\bullet$ & --        & $\bullet$ &     & Y &                   \\ %
HD15316 &  $\bullet$ &  $\bullet$ & $\bullet$ & $\bullet$ & $\bullet$ &     & Y &                   \\ %
HD17378 &  $\bullet$ &  $\bullet$ & $\bullet$ & --        &($\bullet$)&LPV/SB1?&Y&                  \\ % ACYG (VSI), G=5.9
HD12842 &  $\bullet$ &  $\times$  & $\circ$   & --        & $\circ$   &     & Y & Runaway (PM)      \\ % 

\noalign{\smallskip}
\hline
\noalign{\smallskip}

\multicolumn{9}{c}{\textbf{Yellow and red supergiants (I)}} \\\noalign{\smallskip}
HD14580 &  $\bullet$ &  $\bullet$  $\bullet$ & --        & $\bullet$ & & P & $\varpi$, $\mu$ affected by size \\ %
BD+59372& ($\circ$)  & ($\bullet$)& $\bullet$ & --        & $\bullet$ & & P & $\varpi$, $\mu$ affected by size \\ %
HD14330 &  $\bullet$ &  $\bullet$ & $\bullet$ & --        & $\bullet$ & & Y &                       \\ %
HD236979&  $\bullet$ &  $\bullet$ & $\bullet$ & --        & $\bullet$ & & Y &                       \\ %
BD+56595&  $\bullet$ &  $\bullet$ & $\bullet$ & --        & $\bullet$ & & Y &                       \\ %
BD+57530A& $\bullet$ &  $\bullet$ & $\circ$   & --        & $\circ$   & & Y & SB1?                  \\ %
HD14242 &  $\bullet$ &  $\bullet$ & $\circ$   & --        & $\circ$   & & Y & SB1?                  \\ %
HD14404 & ($\circ$)  & ($\bullet$)& $\bullet$ & --        & $\bullet$ & & P & $\varpi$, $\mu$ affected by size \\ %
HD13136 & ($\circ$)  & ($\bullet$)& $\bullet$ & $\bullet$ & $\bullet$ & & P & $\varpi$, $\mu$ affected by size \\ % Angsize 5.1mas
HD14142 &  $\bullet$ &  $\bullet$ & $\bullet$ & --        & $\bullet$ & & Y &                       \\ %
HD236915&  $\bullet$ &  $\bullet$ & $\circ$   & --        & $\bullet$ & & Y &                       \\ %
HD14270 &  $\bullet$ &  $\bullet$ & $\bullet$ & $\bullet$ & $\bullet$ & & Y &                       \\ %
HD14826 & ($\circ$)  & ($\bullet$)& $\bullet$ & --        & $\bullet$ & & P & $\varpi$, $\mu$ affected by size \\ %
BD+56512& ($\circ$)  & ($\bullet$)& $\circ$   & --        & $\bullet$ & & P & $\varpi$, $\mu$ affected by size \\ % High err Plx/PM
HD14469 &  $\bullet$ &  $\bullet$ & $\bullet$ & --        & $\bullet$ & & Y &                       \\ % G=5.9, High err Plx/PM
HD14488 & ($\circ$)  & ($\bullet$)& $\bullet$ & --        & $\bullet$ & & P & $\varpi$, $\mu$ affected by size \\ % High err Plx/PM
BD+56724& ($\times$) & ($\circ$)  & $\bullet$ & --        & $\bullet$ & & C & $\varpi$, $\mu$ affected by size. Runaway (PM)? \\ % High err Plx/PM
HD14528 &  $\bullet$ &  $\bullet$ & $\bullet$ & --        & $\bullet$ & & Y &                       \\ % High err PM (Asaki 2010)

\end{longtable}
}

\begin{list}{}{}
\small{
\item[$^{a}$] For each parameter in columns \textit{$\varpi$}, \textit{$\mu$}, \textit{RV best,} and \textit{RV multi}: filled circles indicate stars within 2~$\sigma$ from the mean of the distribution, empty circles indicate outliers by more than 2~$\sigma$, and crosses indicate outliers by more than 3~$\sigma$. For \textit{RV best}, the 2~$\sigma$ was computed separately for O, B, A/F, and K/M spectral types. For \textit{RV multi,} the 2~$\sigma$ was computed independently for each spectral type, taking stars with more than one spectrum into account, but only applying the boundaries stars with more than three spectra.
\item[$^{b}$] For columns \textit{$\varpi$} and \textit{$\mu$}, stars in parentheses have astrometric issues that are indicated in the column \textit{Comments}. For the column \textit{RV final}, stars in parentheses correspond to stars that are binaries or potential binaries with fewer than four spectra and a time span shorter than 500 days (see Sec.~\ref{subsection.51_memb}).
\item[$^{c}$] Y: members of Per~OB1; L: likely member of Per~OB1; C: candidate to be a member of Per~OB1; N: nonmember of Per~OB1.
\item[$^{d}$] Runaway: runaway stars according to proper motion alone (PM), and also according to RV (PM+RV); Runaway/SB1?: stars with more than four spectra and a time span longer than 500 days that are outliers only in RV; Runaway?: same as previously, but they are already identified as binaries; SB1?: stars with four or fewer spectra and a time span shorter than 500 days that are outliers only in RV;
RUWE: stars with poor astrometry because the {\em Gaia} RUWE value is higher than 1.4; Runaway (PM/RV): runaway stars according to proper motion (PM) or/and radial velocity (RV); EA: $\beta$ Persei-type (Algol) eclipsing systems; EB: eclipsing binaries; ELL: close binary systems with ellipsoidal components;
References: Hel73: \citeads{1973ApJ...184..167A}; Zac85: \citeads{1985AbaOB..58..313Z}; Lau17: \citeads{2017A&A...598A.108L};
Hol18: \citeads{2018A&A...613A..65H}; MA19: \citeads{2019A&A...626A..20M}; Hol20: \citeads{2020arXiv200505446H}; VSI: The International Variable Star Index.
\item[$^{e}$] Although this star has a higher parallax, it is considered a member because of its high negative RV. See Sect.~\ref{subsection.51_memb}).
}
\end{list}

%%%%%%%%%%%%%%%%%%%%%%%%%%%%%%%%%%%%%%%%%%%%%%%%%

\begin{longtable}{llccc}
\label{table.rvstable}\\ %
\caption[]{All RV measurements for individual epochs in all the stars in the sample.}\\ %

\hline
\hline
\noalign{\smallskip}
Name    &SpC    &Number         & MBJD  &RV                     \\ %
                &           &of lines   &       &[km\,s$^{-1}$] \\ %
\noalign{\smallskip}
\hline
\noalign{\smallskip}
\endfirsthead

\caption[]{All RV measurements..}\\ %

\hline
\hline
\noalign{\smallskip}
Name    &SpC    &Number         & MBJD  &RV                     \\ %
                &           &of lines   &       &[km\,s$^{-1}$] \\ %
\noalign{\smallskip}
\hline
\endhead

\noalign{\smallskip}
\hline
\endfoot

\multicolumn{4}{c}{\textbf{O-type stars}} \\\noalign{\smallskip}
HD15570  & O4If      &  3 & 54777.0 & -43.4 $\pm$ 3.8 \\ %
...      & ...       &  3 & 54777.0 & -41.2 $\pm$ 4.6 \\ %
...      & ...       &  3 & 55575.9 & -54.3 $\pm$ 3.5 \\ %
...      & ...       &  3 & 55815.2 & -53.7 $\pm$ 7.3 \\ %
...      & ...       &  2 & 56227.1 & -61.2 $\pm$ 6.2 \\ %
...      & ...       &  4 & 57043.8 & -62.2 $\pm$ 2.0 \\ %
HD16691  & O4If      &  4 & 55576.8 & -42.8 $\pm$ 7.5 \\ %
...      & ...       &  2 & 56287.0 & -73.7 $\pm$ 1.2 \\ %
...      & ...       &  4 & 58017.2 & -42.3 $\pm$ 6.8 \\ %
...      & ...       &  4 & 58017.2 & -41.1 $\pm$ 7.7 \\ %
HD14947  & O4.5If    &  5 & 54779.0 & -41.6 $\pm$ 6.2 \\ %
...      & ...       &  6 & 55814.2 & -36.1 $\pm$ 4.4 \\ %
...      & ...       &  4 & 56226.1 & -54.2 $\pm$ 4.2 \\ %
...      & ...       &  6 & 56285.0 & -41.2 $\pm$ 5.9 \\ %
HD15558$^{a}$& O4.5III(f) & 7 & 54777.0 & -29.5 $\pm$ 7.7 \\ %
...      & ...       &  6 & 54777.0 & -23.1 $\pm$ 6.9 \\ %
...      & ...       &  7 & 54777.9 & -28.0 $\pm$ 7.5 \\ %
...      & ...       &  6 & 54778.0 & -21.0 $\pm$ 7.8 \\ %
...      & ...       &  7 & 54778.0 & -24.0 $\pm$ 6.9 \\ %
...      & ...       &  7 & 54778.0 & -26.7 $\pm$ 7.5 \\ %
...      & ...       &  6 & 54779.0 & -26.7 $\pm$ 6.2 \\ %
...      & ...       &  6 & 56226.1 & -39.8 $\pm$ 4.7 \\ %
...      & ...       &  4 & 56285.0 & -78.5 $\pm$ 6.6 \\ %
HD14442  & O5n(f)p   &  3 & 57660.2 & -37.4 $\pm$ 11.0 \\ %
...      & ...       &  2 & 57752.0 & -33.1 $\pm$ 5.2 \\ %
...      & ...       &  5 & 58839.0 & -52.0 $\pm$ 11.6 \\ %
HD17603  & O7.5Ib(f) &  8 & 55575.9 & -25.6 $\pm$ 2.2 \\ %
...      & ...       &  8 & 56226.1 & -31.9 $\pm$ 3.4 \\ %
...      & ...       &  8 & 56287.1 & -29.8 $\pm$ 1.9 \\ %
HD13268  & ON8.5IIIn &  7 & 55573.8 & -109.8 $\pm$ 8.4 \\ %
...      & ...       &  6 & 56320.9 & -95.2 $\pm$ 8.7 \\ %
...      & ...       &  5 & 56321.9 & -99.6 $\pm$ 3.1 \\ %
...      & ...       &  2 & 58035.0 & -106.2 $\pm$ 0.8 \\ %
HD16429  & O9II(n)   &  8 & 55448.2 & -61.2 $\pm$ 3.0 \\ %
...      & ...       &  6 & 55814.1 & -55.5 $\pm$ 0.7 \\ %
...      & ...       &  9 & 56285.0 & -59.5 $\pm$ 2.6 \\ %
...      & ...       &  5 & 56595.0 & -67.3 $\pm$ 3.8 \\ %
HD16832  & O9.2III   & 10 & 55813.1 & -23.6 $\pm$ 0.8 \\ %
...      & ...       & 11 & 55816.2 & -26.2 $\pm$ 0.8 \\ %
...      & ...       & 11 & 56287.0 & -22.8 $\pm$ 0.8 \\ %
HD15642  &O9.5II-IIIn&  6 & 55815.2 & -27.8 $\pm$ 10.8 \\ %
...      & ...       &  5 & 56287.1 & -26.6 $\pm$ 9.7 \\ %
...      & ...       &  3 & 58017.2 & -9.3 $\pm$ 11.7 \\ %
...      & ...       &  3 & 58017.2 & -22.8 $\pm$ 2.5 \\ %
...      & ...       &  5 & 58035.1 & -18.7 $\pm$ 7.4 \\ %
...      & ...       &  6 & 58108.8 & -31.0 $\pm$ 4.5 \\ %
HD13831  & O9.7III   &  5 & 56596.1 & -47.2 $\pm$ 2.0 \\ %
...      & ...       &  5 & 58838.1 & -36.5 $\pm$ 4.4 \\ %
...      & ...       &  6 & 58839.0 & -48.6 $\pm$ 6.9 \\ %
HD13745  & O9.7II(n) &  9 & 55449.2 & -9.2 $\pm$ 2.7 \\ %
...      & ...       &  6 & 55576.9 & -18.1 $\pm$ 0.7 \\ %
...      & ...       &  9 & 55814.1 & -17.9 $\pm$ 3.6 \\ %
...      & ...       &  9 & 56226.0 & -12.4 $\pm$ 2.3 \\ %
HD13022  & O9.7III   &  4 & 55813.2 & -50.2 $\pm$ 0.2 \\ %
...      & ...       &  9 & 55816.2 & -47.7 $\pm$ 1.3 \\ %
...      & ...       & 10 & 56287.0 & -53.1 $\pm$ 1.0 \\ %

\noalign{\smallskip}
\hline
\noalign{\smallskip}

\multicolumn{4}{c}{\textbf{B-type stars}} \\\noalign{\smallskip}
HD13402$^{a}$ & B0II:+B0 &  22 & 55494.1 & +60.4 $\pm$ 2.0 \\ %
...      & ...       & 14 & 55494.1 & +57.3 $\pm$ 1.1 \\ %
...      & ...       & 16 & 58837.9 & -31.3 $\pm$ 0.6 \\ %
...      & ...       & 16 & 58838.8 & -33.8 $\pm$ 0.6 \\ %
...      & ...       & 16 & 58840.0 & -34.4 $\pm$ 0.7 \\ %
HD14331  & B0III     & 19 & 55572.9 & -27.5 $\pm$ 1.1 \\ %
...      & ...       & 18 & 58838.1 & -26.2 $\pm$ 0.6 \\ %
...      & ...       & 22 & 58839.0 & -25.4 $\pm$ 1.0 \\ %
HD16808  & B0.2Ib    & 18 & 57662.2 & -39.0 $\pm$ 0.7 \\ %
HD13036  & B0.2III   & 20 & 55576.9 & -62.2 $\pm$ 1.2 \\ %
...      & ...       & 19 & 58837.9 & -49.7 $\pm$ 0.7 \\ %
...      & ...       & 19 & 58838.9 & -48.4 $\pm$ 0.8 \\ %
BD+60493 & B0.5Ia    & 14 & 57662.2 & -26.9 $\pm$ 0.6 \\ %
...      & ...       & 16 & 58840.1 & -31.4 $\pm$ 0.9 \\ %
HD13716  & B0.5III   & 16 & 58355.1 & -35.0 $\pm$ 1.2 \\ %
...      & ...       & 17 & 58804.0 & -27.6 $\pm$ 1.5 \\ %
...      & ...       & 18 & 58837.8 & -53.2 $\pm$ 0.5 \\ %
...      & ...       & 17 & 58838.0 & -47.5 $\pm$ 0.6 \\ %
HD13970$^{a}$ & B0.5:III: &  2 & 57661.2 & -61.0 $\pm$ 12.3 \\ %
...      & ...       &  2 & 58354.2 &  -8.6 $\pm$ 3.8 \\ %
HD14053  & B0.7II    & 19 & 57662.1 & -31.4 $\pm$ 0.1 \\ %
...      & ...       & 19 & 58354.1 & -35.5 $\pm$ 0.2 \\ %
...      & ...       & 22 & 58728.2 & -37.0 $\pm$ 0.3 \\ %
...      & ...       & 18 & 58804.0 & -35.1 $\pm$ 0.2 \\ %
...      & ...       & 22 & 58815.2 & -28.9 $\pm$ 0.4 \\ %
...      & ...       & 22 & 58836.8 & -35.9 $\pm$ 0.3 \\ %
...      & ...       & 21 & 58838.1 & -33.5 $\pm$ 0.2 \\ %
...      & ...       & 19 & 58839.0 & -34.4 $\pm$ 0.2 \\ %
HD13969  & B0.7III   & 19 & 58728.2 & -12.3 $\pm$ 0.6 \\ %
...      & ...       & 23 & 58838.0 & -36.0 $\pm$ 0.5 \\ %
...      & ...       & 21 & 58839.0 & -30.6 $\pm$ 0.5 \\ %
HD13854  & B1Ia-Iab  & 23 & 55493.3 & -42.5 $\pm$ 0.6 \\ %
...      & ...       & 23 & 55572.8 & -41.4 $\pm$ 0.8 \\ %
...      & ...       & 19 & 56227.0 & -35.7 $\pm$ 0.3 \\ %
...      & ...       & 19 & 56909.0 & -35.4 $\pm$ 0.3 \\ %
...      & ...       & 19 & 56909.2 & -30.8 $\pm$ 0.4 \\ %
...      & ...       & 18 & 56910.0 & -36.9 $\pm$ 0.3 \\ %
...      & ...       & 19 & 56910.2 & -37.2 $\pm$ 0.6 \\ %
...      & ...       & 17 & 56911.0 & -39.0 $\pm$ 0.4 \\ %
...      & ...       & 19 & 56911.2 & -36.4 $\pm$ 0.4 \\ %
...      & ...       & 20 & 56952.2 & -46.9 $\pm$ 0.7 \\ %
...      & ...       & 20 & 56952.9 & -40.1 $\pm$ 0.8 \\ %
...      & ...       & 21 & 57041.9 & -30.9 $\pm$ 0.4 \\ %
...      & ...       & 19 & 57043.9 & -35.3 $\pm$ 0.2 \\ %
...      & ...       & 22 & 57045.8 & -43.8 $\pm$ 0.5 \\ %
HD13659  & B1Ib      & 12 & 57658.2 & -48.6 $\pm$ 0.2 \\ %
...      & ...       & 15 & 57659.2 & -48.2 $\pm$ 0.5 \\ %
...      & ...       & 15 & 58839.1 & -52.6 $\pm$ 0.2 \\ %
HD15571  & B1Ib      & 21 & 56596.1 & -45.4 $\pm$ 0.3 \\ %
BD+59451 & B1Ib-II   & 14 & 57663.2 & -27.6 $\pm$ 1.2 \\ %
HD14052  & B1Ib-II   & 22 & 55494.1 & -90.8 $\pm$ 0.3 \\ %
...      & ...       & 18 & 55494.1 & -90.6 $\pm$ 0.2 \\ %
...      & ...       & 23 & 55494.1 & -90.7 $\pm$ 0.3 \\ %
...      & ...       & 21 & 56594.1 &  -8.8 $\pm$ 0.3 \\ %
HD14476  & B1II      & 20 & 55573.9 & -40.6 $\pm$ 0.3 \\ %
...      & ...       & 20 & 58838.0 & -60.8 $\pm$ 0.2 \\ %
...      & ...       & 22 & 58838.9 & -62.1 $\pm$ 0.3 \\ %
BD+56576 & B1III     & 18 & 55494.2 & -84.7 $\pm$ 0.3 \\ %
...      & ...       &  2 & 58840.0 & +23.8 $\pm$ 0.4 \\ %
BD+57520 & B1III     & 20 & 57663.1 & -46.0 $\pm$ 0.3 \\ %
...      & ...       & 19 & 58840.1 & -45.2 $\pm$ 0.3 \\ %
HD13841  & B1.5Ib    & 23 & 55494.0 & -43.3 $\pm$ 0.3 \\ %
...      & ...       & 22 & 56596.1 & -40.0 $\pm$ 0.4 \\ %
...      & ...       & 20 & 58837.8 & -40.2 $\pm$ 0.2 \\ %
HD14443  & BC1.5Ib   & 21 & 55493.1 & -45.7 $\pm$ 0.2 \\ %
...      & ...       & 23 & 55493.1 & -45.3 $\pm$ 0.3 \\ %
...      & ...       & 21 & 55493.2 & -45.3 $\pm$ 0.4 \\ %
...      & ...       & 27 & 57662.2 & -41.5 $\pm$ 0.6 \\ %
HD16779  & B1.5Ib N-weak & 21 & 57663.1 & -39.6 $\pm$ 0.5 \\ %
BD+56574 & BC1.5Ib-II & 11 & 55490.0 & -39.6 $\pm$ 3.1 \\ %
...      & ...       &  7 & 55490.0 & -43.7 $\pm$ 2.8 \\ %
...      & ...       & 12 & 55490.0 & -44.1 $\pm$ 3.5 \\ %
HD14302  & B1.5Ib-II & 24 & 55576.9 & -36.8 $\pm$ 0.6 \\ %
...      & ...       & 19 & 58838.1 & -38.0 $\pm$ 0.4 \\ %
...      & ...       & 18 & 58839.0 & -38.5 $\pm$ 0.3 \\ %
BD+56527 & B1.5II    & 24 & 58728.2 & -33.8 $\pm$ 1.1 \\ %
...      & ...       & 18 & 58837.8 & -38.6 $\pm$ 1.1 \\ %
...      & ...       & 19 & 58838.9 & -39.8 $\pm$ 0.8 \\ %
HD14357  & B1.5II    & 19 & 55574.8 & -37.7 $\pm$ 2.1 \\ %
...      & ...       & 14 & 58351.1 & -43.1 $\pm$ 2.8 \\ %
...      & ...       & 17 & 58352.2 & -35.4 $\pm$ 2.2 \\ %
...      & ...       & 20 & 58353.1 & -34.9 $\pm$ 4.1 \\ %
...      & ...       & 16 & 58354.1 & -43.3 $\pm$ 3.6 \\ %
...      & ...       & 14 & 58355.2 & -41.5 $\pm$ 1.8 \\ %
...      & ...       & 19 & 58802.2 & -46.0 $\pm$ 3.1 \\ %
...      & ...       & 16 & 58804.0 & -40.2 $\pm$ 1.6 \\ %
HD14520  & B1.5II    & 20 & 57662.2 & -43.1 $\pm$ 1.1 \\ %
BD+57626 & B1.5II-III& 17 & 57658.2 & -43.9 $\pm$ 0.9 \\ %
...      & ...       & 21 & 58840.1 & -31.4 $\pm$ 1.7 \\ %
BD+57513 & B1.5IIIn  &  8 & 55573.8 & -45.7 $\pm$ 7.5 \\ %
...      & ...       & 14 & 58840.1 & -42.1 $\pm$ 6.9 \\ %
HD14143  & B2Ia      & 22 & 55147.0 & -38.8 $\pm$ 0.6 \\ %
...      & ...       & 26 & 55493.2 & -47.8 $\pm$ 0.6 \\ %
...      & ...       & 25 & 55493.2 & -47.9 $\pm$ 0.6 \\ %
...      & ...       & 26 & 55874.1 & -41.4 $\pm$ 0.6 \\ %
...      & ...       & 24 & 56909.0 & -43.4 $\pm$ 0.6 \\ %
...      & ...       & 21 & 56909.2 & -43.3 $\pm$ 0.4 \\ %
...      & ...       & 21 & 56910.1 & -44.5 $\pm$ 0.5 \\ %
...      & ...       & 25 & 56910.2 & -45.2 $\pm$ 0.6 \\ %
...      & ...       & 27 & 56911.0 & -41.2 $\pm$ 0.6 \\ %
...      & ...       & 27 & 56911.2 & -39.1 $\pm$ 0.5 \\ %
...      & ...       & 20 & 56952.2 & -42.6 $\pm$ 0.4 \\ %
...      & ...       & 21 & 56953.0 & -40.7 $\pm$ 0.3 \\ %
...      & ...       & 24 & 57042.0 & -44.0 $\pm$ 0.4 \\ %
...      & ...       & 24 & 57044.0 & -46.9 $\pm$ 0.5 \\ %
HD14818  & B2Ia      & 24 & 54776.0 & -38.2 $\pm$ 0.8 \\ %
...      & ...       & 24 & 54776.0 & -40.9 $\pm$ 0.9 \\ %
...      & ...       & 20 & 54776.0 & -40.8 $\pm$ 0.3 \\ %
...      & ...       & 22 & 54777.0 & -41.9 $\pm$ 0.3 \\ %
...      & ...       & 25 & 54777.0 & -42.2 $\pm$ 0.4 \\ %
...      & ...       & 23 & 54777.0 & -41.9 $\pm$ 0.3 \\ %
...      & ...       & 24 & 54777.0 & -42.5 $\pm$ 0.3 \\ %
...      & ...       & 25 & 54777.1 & -42.8 $\pm$ 0.4 \\ %
...      & ...       & 23 & 54777.1 & -42.8 $\pm$ 0.3 \\ %
...      & ...       & 27 & 54778.0 & -45.0 $\pm$ 0.8 \\ %
...      & ...       & 22 & 54778.0 & -43.3 $\pm$ 0.6 \\ %
...      & ...       & 21 & 54778.0 & -42.5 $\pm$ 0.5 \\ %
...      & ...       & 25 & 54778.0 & -43.1 $\pm$ 0.6 \\ %
...      & ...       & 24 & 54779.0 & -39.8 $\pm$ 0.6 \\ %
...      & ...       & 24 & 54779.0 & -39.6 $\pm$ 0.6 \\ %
...      & ...       & 20 & 55488.3 & -44.6 $\pm$ 0.5 \\ %
...      & ...       & 24 & 55493.1 & -45.4 $\pm$ 0.5 \\ %
...      & ...       & 25 & 55574.8 & -39.4 $\pm$ 0.6 \\ %
...      & ...       & 20 & 56227.1 & -43.8 $\pm$ 0.3 \\ %
HD14956  & B2Ia      & 22 & 55572.8 & -32.2 $\pm$ 0.6 \\ %
...      & ...       & 21 & 57661.2 & -37.7 $\pm$ 0.4 \\ %
...      & ...       & 20 & 58837.8 & -36.9 $\pm$ 0.3 \\ %
HD15690  & B2Iab     & 24 & 56595.0 & -43.5 $\pm$ 0.4 \\ %
...      & ...       & 24 & 57662.2 & -41.5 $\pm$ 0.4 \\ %
...      & ...       & 24 & 58354.1 & -42.4 $\pm$ 0.6 \\ %
...      & ...       & 22 & 58838.1 & -40.2 $\pm$ 0.3 \\ %
HD13866  & BC2Ib     & 25 & 55494.0 & -47.1 $\pm$ 0.6 \\ %
...      & ...       & 20 & 55494.0 & -47.6 $\pm$ 0.3 \\ %
...      & ...       & 20 & 56227.1 & -45.6 $\pm$ 0.3 \\ %
...      & ...       & 22 & 58837.8 & -46.1 $\pm$ 0.5 \\ %
BD+56578 & BN2II-III & 17 & 55494.2 & -89.7 $\pm$ 0.7 \\ %
...      & ...       & 12 & 55494.2 & -89.5 $\pm$ 0.4 \\ %
...      & ...       & 18 & 55494.2 & -87.8 $\pm$ 0.8 \\ %
HD10898  & B2.5Ib    & 23 & 58352.1 & -31.1 $\pm$ 0.5 \\ %
...      & ...       & 20 & 58802.1 & -29.1 $\pm$ 0.6 \\ %
...      & ...       & 25 & 58804.0 & -30.1 $\pm$ 0.5 \\ %
...      & ...       & 28 & 58804.8 & -29.8 $\pm$ 0.6 \\ %
BD+59387 & B2.5II N-weak & 20 & 57681.1 & -33.3 $\pm$ 1.3 \\ %
...      & ...       & 21 & 57681.1 & -38.0 $\pm$ 1.8 \\ %
...      & ...       & 21 & 58839.1 & -32.4 $\pm$ 0.8 \\ %
BD+59388 & BN2.5III  & 28 & 57659.2 & -47.2 $\pm$ 0.3 \\ %
HD14134  & B3Ia      & 37 & 55493.2 & -41.8 $\pm$ 0.6 \\ %
...      & ...       & 30 & 55874.0 & -41.2 $\pm$ 0.3 \\ %
...      & ...       & 32 & 56909.0 & -42.8 $\pm$ 0.4 \\ %
...      & ...       & 27 & 56909.2 & -41.2 $\pm$ 0.3 \\ %
...      & ...       & 35 & 56910.1 & -42.3 $\pm$ 0.5 \\ %
...      & ...       & 31 & 56910.2 & -42.5 $\pm$ 0.4 \\ %
...      & ...       & 34 & 56911.0 & -41.6 $\pm$ 0.5 \\ %
...      & ...       & 33 & 56911.2 & -41.3 $\pm$ 0.5 \\ %
...      & ...       & 35 & 56952.2 & -43.0 $\pm$ 0.5 \\ %
...      & ...       & 32 & 56953.0 & -43.7 $\pm$ 0.4 \\ %
...      & ...       & 35 & 57042.0 & -44.9 $\pm$ 0.5 \\ %
...      & ...       & 36 & 57044.0 & -43.1 $\pm$ 0.7 \\ %
HD15497  & B6Ia      & 24 & 55488.0 & -43.7 $\pm$ 0.3 \\ %
...      & ...       & 25 & 55488.0 & -44.2 $\pm$ 0.3 \\ %
...      & ...       & 29 & 57660.2 & -47.8 $\pm$ 0.4 \\ %
HD13267  & B6Iab     & 31 & 56227.0 & -35.0 $\pm$ 0.5 \\ %
...      & ...       & 29 & 56909.0 & -36.6 $\pm$ 0.3 \\ %
...      & ...       & 35 & 56909.2 & -38.1 $\pm$ 0.6 \\ %
...      & ...       & 34 & 56910.0 & -40.2 $\pm$ 0.5 \\ %
...      & ...       & 36 & 56910.2 & -39.3 $\pm$ 0.6 \\ %
...      & ...       & 32 & 56911.0 & -37.3 $\pm$ 0.4 \\ %
...      & ...       & 31 & 56911.2 & -38.1 $\pm$ 0.4 \\ %
...      & ...       & 30 & 56952.1 & -35.2 $\pm$ 0.4 \\ %
...      & ...       & 29 & 56952.9 & -34.3 $\pm$ 0.4 \\ %
...      & ...       & 32 & 57042.0 & -36.3 $\pm$ 0.7 \\ %
...      & ...       & 32 & 57042.8 & -30.1 $\pm$ 0.5 \\ %
...      & ...       & 28 & 57044.0 & -34.7 $\pm$ 0.4 \\ %
...      & ...       & 34 & 57045.8 & -34.7 $\pm$ 0.4 \\ %
HD17145  & B6Iab     & 25 & 55489.0 & -37.4 $\pm$ 0.6 \\ %
...      & ...       & 24 & 55489.1 & -37.2 $\pm$ 0.6 \\ %
...      & ...       & 24 & 55489.2 & -37.7 $\pm$ 0.5 \\ %
HD14322  & B8Iab     & 36 & 56227.0 & -33.4 $\pm$ 0.4 \\ %
...      & ...       & 33 & 56909.0 & -33.4 $\pm$ 0.3 \\ %
...      & ...       & 35 & 56909.2 & -33.7 $\pm$ 0.3 \\ %
...      & ...       & 33 & 56910.1 & -35.6 $\pm$ 0.3 \\ %
...      & ...       & 34 & 56910.3 & -35.3 $\pm$ 0.3 \\ %
...      & ...       & 39 & 56911.0 & -34.6 $\pm$ 0.5 \\ %
...      & ...       & 32 & 56911.2 & -33.5 $\pm$ 0.3 \\ %
...      & ...       & 36 & 56952.2 & -35.5 $\pm$ 0.4 \\ %
...      & ...       & 33 & 56953.0 & -34.8 $\pm$ 0.3 \\ %
...      & ...       & 37 & 57044.0 & -33.0 $\pm$ 0.4 \\ %
...      & ...       & 39 & 58351.1 & -33.8 $\pm$ 0.6 \\ %
...      & ...       & 32 & 58352.1 & -33.7 $\pm$ 0.5 \\ %
...      & ...       & 34 & 58353.1 & -31.9 $\pm$ 0.4 \\ %
...      & ...       & 29 & 58354.1 & -32.4 $\pm$ 0.4 \\ %
...      & ...       & 37 & 58355.1 & -36.5 $\pm$ 0.5 \\ %
HD14542  & B8Iab     & 37 & 55487.9 & -50.1 $\pm$ 0.5 \\ %
...      & ...       & 37 & 55488.0 & -50.6 $\pm$ 0.5 \\ %
...      & ...       & 37 & 56227.1 & -50.9 $\pm$ 0.4 \\ %
...      & ...       & 36 & 56909.0 & -53.7 $\pm$ 0.4 \\ %
...      & ...       & 35 & 56909.2 & -53.9 $\pm$ 0.3 \\ %
...      & ...       & 37 & 56910.2 & -55.6 $\pm$ 0.3 \\ %
...      & ...       & 39 & 56911.0 & -54.9 $\pm$ 0.5 \\ %
...      & ...       & 41 & 56911.3 & -54.1 $\pm$ 0.6 \\ %
...      & ...       & 37 & 56952.3 & -46.7 $\pm$ 0.5 \\ %
...      & ...       & 36 & 56953.0 & -46.0 $\pm$ 0.4 \\ %
...      & ...       & 38 & 57044.0 & -53.4 $\pm$ 0.4 \\ %
HD15620  & B8Iab     & 33 & 57663.1 & -43.6 $\pm$ 0.4 \\ %
HDE236995& B8Ib      & 29 & 55489.2 & -42.3 $\pm$ 0.2 \\ %
...      & ...       & 24 & 55489.2 & -42.4 $\pm$ 0.2 \\ %
...      & ...       & 25 & 55489.2 & -42.4 $\pm$ 0.2 \\ %
HD17088  & B9Ia      & 33 & 55488.2 & -47.0 $\pm$ 0.7 \\ %
...      & ...       & 34 & 55488.2 & -46.7 $\pm$ 0.7 \\ %
...      & ...       & 35 & 55488.2 & -46.7 $\pm$ 0.7 \\ %
...      & ...       & 33 & 57658.2 & -45.9 $\pm$ 0.4 \\ %
HD14899  & B9Iab     & 30 & 57660.2 & -43.9 $\pm$ 0.1 \\ %
...      & ...       & 29 & 58838.1 & -41.0 $\pm$ 0.1 \\ %
...      & ...       & 29 & 58839.0 & -42.1 $\pm$ 0.1 \\ %

\noalign{\smallskip}
\hline
\noalign{\smallskip}

\multicolumn{4}{c}{\textbf{A-type stars}} \\\noalign{\smallskip}
HD13744  & A0Iab     & 35 & 55488.2 & -50.8 $\pm$ 0.4 \\ %
...      & ...       & 38 & 55488.2 & -50.0 $\pm$ 0.4 \\ %
...      & ...       & 36 & 55488.2 & -50.7 $\pm$ 0.4 \\ %
...      & ...       & 29 & 57663.1 & -48.2 $\pm$ 0.1 \\ %
HD12953  & A1Iae     & 38 & 55487.9 & -32.3 $\pm$ 0.5 \\ %
...      & ...       & 33 & 55487.9 & -32.9 $\pm$ 0.5 \\ %
...      & ...       & 31 & 56641.0 & -32.5 $\pm$ 0.2 \\ %
...      & ...       & 32 & 58046.1 & -25.9 $\pm$ 0.2 \\ %
...      & ...       & 33 & 58047.1 & -25.2 $\pm$ 0.2 \\ %
...      & ...       & 31 & 58049.2 & -25.3 $\pm$ 0.3 \\ %
...      & ...       & 32 & 58049.2 & -25.3 $\pm$ 0.3 \\ %
...      & ...       & 33 & 58049.9 & -26.2 $\pm$ 0.4 \\ %
...      & ...       & 36 & 58051.0 & -27.4 $\pm$ 0.5 \\ %
...      & ...       & 38 & 58051.9 & -27.7 $\pm$ 0.5 \\ %
...      & ...       & 34 & 58055.0 & -31.2 $\pm$ 0.4 \\ %
...      & ...       & 32 & 58055.9 & -33.1 $\pm$ 0.3 \\ %
...      & ...       & 25 & 58077.9 & -35.1 $\pm$ 0.2 \\ %
...      & ...       & 28 & 58078.9 & -35.6 $\pm$ 0.2 \\ %
...      & ...       & 27 & 58351.2 & -33.2 $\pm$ 0.2 \\ %
...      & ...       & 32 & 58352.1 & -31.2 $\pm$ 0.3 \\ %
...      & ...       & 31 & 58353.2 & -29.3 $\pm$ 0.3 \\ %
...      & ...       & 32 & 58354.2 & -28.0 $\pm$ 0.4 \\ %
...      & ...       & 38 & 58355.2 & -26.8 $\pm$ 0.4 \\ %
HD14433  & A1Ia      & 33 & 55488.0 & -42.9 $\pm$ 0.4 \\ %
...      & ...       & 32 & 56642.0 & -47.1 $\pm$ 0.1 \\ %
...      & ...       & 36 & 58837.8 & -47.9 $\pm$ 0.1 \\ %
HD14489  & A1Ia      & 35 & 55487.9 & -16.1 $\pm$ 0.4 \\ %
...      & ...       & 33 & 55488.0 & -15.9 $\pm$ 0.4 \\ %
...      & ...       & 31 & 56641.0 & -18.5 $\pm$ 0.1 \\ %
...      & ...       & 33 & 57660.2 & -19.5 $\pm$ 0.1 \\ %
...      & ...       & 32 & 58046.1 & -17.8 $\pm$ 0.1 \\ %
...      & ...       & 33 & 58047.1 & -18.7 $\pm$ 0.1 \\ %
...      & ...       & 33 & 58047.1 & -18.6 $\pm$ 0.1 \\ %
...      & ...       & 31 & 58047.8 & -19.2 $\pm$ 0.1 \\ %
...      & ...       & 29 & 58049.2 & -17.6 $\pm$ 0.1 \\ %
...      & ...       & 31 & 58049.9 & -17.4 $\pm$ 0.1 \\ %
...      & ...       & 35 & 58050.9 & -18.4 $\pm$ 0.2 \\ %
...      & ...       & 33 & 58050.9 & -18.4 $\pm$ 0.1 \\ %
...      & ...       & 31 & 58052.2 & -21.2 $\pm$ 0.1 \\ %
...      & ...       & 31 & 58054.1 & -21.3 $\pm$ 0.1 \\ %
...      & ...       & 32 & 58055.0 & -19.1 $\pm$ 0.2 \\ %
...      & ...       & 32 & 58056.0 & -16.1 $\pm$ 0.1 \\ %
...      & ...       & 30 & 58077.9 & -18.7 $\pm$ 0.1 \\ %
...      & ...       & 26 & 58078.0 & -18.4 $\pm$ 0.1 \\ %
...      & ...       & 29 & 58078.9 & -17.4 $\pm$ 0.1 \\ %
...      & ...       & 31 & 58351.2 & -17.6 $\pm$ 0.1 \\ %
...      & ...       & 34 & 58351.2 & -17.5 $\pm$ 0.1 \\ %
...      & ...       & 31 & 58352.1 & -16.7 $\pm$ 0.1 \\ %
...      & ...       & 28 & 58353.2 & -16.3 $\pm$ 0.1 \\ %
...      & ...       & 33 & 58354.2 & -15.9 $\pm$ 0.2 \\ %
...      & ...       & 34 & 58355.2 & -15.9 $\pm$ 0.2 \\ %
HD14535  & A1Ia      & 36 & 55488.1 & -43.5 $\pm$ 0.5 \\ %
...      & ...       & 35 & 55488.1 & -44.5 $\pm$ 0.6 \\ %
...      & ...       & 35 & 55488.2 & -44.0 $\pm$ 0.5 \\ %
...      & ...       & 37 & 57662.2 & -48.8 $\pm$ 0.3 \\ %
HD11831  & A2Ia      & 30 & 57659.2 & -39.2 $\pm$ 0.1 \\ %
HD16778  & A2Ia-Iab  & 29 & 57660.2 & -38.1 $\pm$ 0.1 \\ %
...      & ...       & 28 & 58839.2 & -42.0 $\pm$ 0.1 \\ %
...      & ...       & 31 & 58840.0 & -42.9 $\pm$ 0.1 \\ %
HD13476  & A3Iab     & 37 & 55488.0 & -39.6 $\pm$ 0.4 \\ %
...      & ...       & 29 & 57661.2 & -41.0 $\pm$ 0.1 \\ %
...      & ...       & 30 & 58837.8 & -36.3 $\pm$ 0.1 \\ %
HD15316  & A3Iab     & 37 & 55488.1 & -49.9 $\pm$ 0.4 \\ %
...      & ...       & 39 & 55488.1 & -50.0 $\pm$ 0.5 \\ %
...      & ...       & 36 & 55488.1 & -49.9 $\pm$ 0.4 \\ %
...      & ...       & 30 & 57660.2 & -47.8 $\pm$ 0.2 \\ %
HD17378  & A6Ia      & 39 & 55488.0 & -43.5 $\pm$ 0.4 \\ %
...      & ...       & 36 & 56642.8 & -40.1 $\pm$ 0.3 \\ %
...      & ...       & 32 & 58838.1 & -35.5 $\pm$ 0.3 \\ %
HD12842  & F3Ib      & 15 & 57660.2 & -24.2 $\pm$ 0.1 \\ %
...      & ...       & 16 & 57662.1 & -24.2 $\pm$ 0.1 \\ %
...      & ...       & 14 & 58839.1 & -24.1 $\pm$ 0.1 \\ %

\noalign{\smallskip}
\hline
\noalign{\smallskip}

\multicolumn{4}{c}{\textbf{M-type stars}} \\\noalign{\smallskip}
HD14580  & K3Ib-II   & 28 & 55490.2 & -45.4 $\pm$ 0.2 \\ %
BD+59372 & K4Ib-II   & 24 & 58838.9 & -39.4 $\pm$ 0.1 \\ %
HD14330  &K5-M0Iab-Ib& 24 & 55489.9 & -47.5 $\pm$ 0.1 \\ %
...      & ...       & 24 & 55489.9 & -47.4 $\pm$ 0.1 \\ %
...      & ...       & 26 & 58839.1 & -44.9 $\pm$ 0.1 \\ %
HD236979 &K5-M0Iab-Ib& 25 & 55490.0 & -41.1 $\pm$ 0.1 \\ %
BD+56595 & K5-M0Ib   & 28 & 55489.9 & -44.3 $\pm$ 0.2 \\ %
...      & ...       & 21 & 58840.1 & -41.8 $\pm$ 0.1 \\ %
BD+57530A& K5-M0Ib   & 25 & 58351.2 & -52.0 $\pm$ 0.2 \\ %
...      & ...       & 21 & 58840.1 & -55.7 $\pm$ 0.1 \\ %
HD14242  & M0Iab-Ib  & 25 & 55490.2 & -34.7 $\pm$ 0.1 \\ %
HD14404  & M0Ib      & 26 & 55489.9 & -49.4 $\pm$ 0.2 \\ %
...      & ...       & 24 & 58839.1 & -45.7 $\pm$ 0.1 \\ %
...      & ...       & 22 & 58840.0 & -45.9 $\pm$ 0.1 \\ %
HD13136  & M1Iab-Ib  & 27 & 55489.2 & -42.9 $\pm$ 0.2 \\ %
...      & ...       & 27 & 55489.9 & -43.0 $\pm$ 0.2 \\ %
...      & ...       & 23 & 58046.1 & -41.1 $\pm$ 0.1 \\ %
...      & ...       & 22 & 58047.2 & -41.0 $\pm$ 0.1 \\ %
...      & ...       & 26 & 58049.9 & -41.3 $\pm$ 0.1 \\ %
...      & ...       & 22 & 58051.0 & -41.3 $\pm$ 0.1 \\ %
...      & ...       & 25 & 58051.0 & -41.4 $\pm$ 0.2 \\ %
...      & ...       & 24 & 58051.9 & -41.4 $\pm$ 0.1 \\ %
...      & ...       & 26 & 58055.9 & -41.7 $\pm$ 0.1 \\ %
...      & ...       & 24 & 58078.0 & -41.4 $\pm$ 0.1 \\ %
...      & ...       & 29 & 58078.9 & -41.2 $\pm$ 0.2 \\ %
...      & ...       & 27 & 58351.2 & -42.9 $\pm$ 0.2 \\ %
...      & ...       & 27 & 58352.1 & -42.7 $\pm$ 0.2 \\ %
...      & ...       & 25 & 58353.2 & -42.9 $\pm$ 0.2 \\ %
...      & ...       & 26 & 58354.2 & -43.0 $\pm$ 0.2 \\ %
...      & ...       & 27 & 58355.2 & -42.6 $\pm$ 0.2 \\ %
HD14142  & M1Iab-Ib  & 27 & 58352.2 & -46.2 $\pm$ 0.1 \\ %
HD236915 & M1Ib      & 28 & 55490.2 & -48.7 $\pm$ 0.1 \\ %
HD14270  & M2Iab-Ib  & 28 & 55489.9 & -44.6 $\pm$ 0.2 \\ %
...      & ...       & 26 & 58046.1 & -43.4 $\pm$ 0.2 \\ %
...      & ...       & 26 & 58047.2 & -43.4 $\pm$ 0.2 \\ %
...      & ...       & 26 & 58049.9 & -43.5 $\pm$ 0.2 \\ %
...      & ...       & 28 & 58049.9 & -43.4 $\pm$ 0.2 \\ %
...      & ...       & 24 & 58051.1 & -43.7 $\pm$ 0.2 \\ %
...      & ...       & 23 & 58052.2 & -43.9 $\pm$ 0.1 \\ %
...      & ...       & 26 & 58055.9 & -43.9 $\pm$ 0.2 \\ %
...      & ...       & 25 & 58078.1 & -43.9 $\pm$ 0.1 \\ %
...      & ...       & 23 & 58079.0 & -44.1 $\pm$ 0.1 \\ %
...      & ...       & 27 & 58351.2 & -43.9 $\pm$ 0.2 \\ %
...      & ...       & 26 & 58352.1 & -43.6 $\pm$ 0.2 \\ %
...      & ...       & 26 & 58353.2 & -43.6 $\pm$ 0.2 \\ %
...      & ...       & 26 & 58354.2 & -43.5 $\pm$ 0.2 \\ %
...      & ...       & 27 & 58355.2 & -43.3 $\pm$ 0.2 \\ %
HD14826  & M2Iab-Ib  & 28 & 55490.2 & -44.2 $\pm$ 0.1 \\ %
BD+56512 & M3Iab     & 23 & 58351.2 & -36.0 $\pm$ 0.1 \\ %
...      & ...       & 25 & 58840.0 & -39.3 $\pm$ 0.1 \\ %
HD14469  & M3.5Iab   & 23 & 55489.9 & -44.9 $\pm$ 0.2 \\ %
...      & ...       & 23 & 56227.0 & -44.8 $\pm$ 0.3 \\ %
...      & ...       & 25 & 58839.1 & -48.6 $\pm$ 0.2 \\ %
HD14488  & M4Iab     & 22 & 55490.2 & -43.6 $\pm$ 0.3 \\ %
BD+56724 &M4-M5Ia-Iab& 22 & 55490.1 & -42.2 $\pm$ 0.3 \\ %
...      & ...       & 21 & 55490.1 & -42.2 $\pm$ 0.3 \\ %
...      & ...       & 21 & 55490.1 & -42.2 $\pm$ 0.2 \\ %
HD14528  & M5-M6Iab  & 18 & 58352.1 & -39.5 $\pm$ 0.4 \\ %
...      & ...       & 28 & 58839.2 & -38.3 $\pm$ 0.7 \\ %
...      & ...       & 24 & 58840.0 & -37.8 $\pm$ 0.5 \\ %

\end{longtable}

\begin{list}{}{}
\small{
\item[$^{a}$] Stars identified as SB2. The values in this table only refer to the main component. 
}
\end{list}

\end{document}